\documentclass[prd,aps,showpacs,tightenlines,preprintnumbers,10pt, reprint,nofootinbib,superscriptaddress,floatfix,twocolumn,eqsecnum]{revtex4-2}

\usepackage[utf8]{inputenc} 

\usepackage{acronym}
\usepackage{amsmath}
\usepackage{amsthm}  % theorems etc
\usepackage{amssymb}
\usepackage{mathtools}
\usepackage{phaistos}
\usepackage{graphicx}
\usepackage[section]{placeins}
\usepackage{IEEEtrantools}

\usepackage[dvipsnames]{xcolor}   % for \textcolor

\usepackage{physics} % /dv /pdv
\usepackage{IEEEtrantools}  % equations
\usepackage{xspace}
\usepackage{cancel}
\usepackage[normalem]{ulem} % strikeout 

\usepackage{tikz}
\usepackage[compat=1.1.0]{tikz-feynman}
\newcommand{\twoVertexDistance}{0.28}
\newcommand{\insertionDistance}{0.5}
\newcommand{\be}{\begin{equation}}
\newcommand{\ee}{\end{equation}}
\newcommand{\ba}{\begin{eqnarray}}
\newcommand{\ea}{\end{eqnarray}}

\tikzset{graviton/.style={decorate, decoration={snake, amplitude=.4mm, segment length=1.5mm, pre length=.5mm, post length=.5mm}, double}}
\usepackage{array}
\newcolumntype{C}[1]{>{\centering\let\newline\\\arraybackslash\hspace{0pt}}m{#1}}

\usepackage{cellspace}
\usepackage{comment}
\usepackage{epsfig}
\usepackage{multirow} 
\usepackage{color}
\usepackage[export]{adjustbox} %To adjust subplots at the top.
\usepackage{floatrow}
\usepackage[font=small,labelfont=bf,justification=Justified]{caption}
\usepackage{float}
\usepackage{ragged2e}
\usepackage{subcaption}
\usepackage{makecell}
\usepackage[titles]{tocloft}
\usepackage{titlesec}
\usepackage{bm}

\titleformat{\section}
  {\centering\normalfont\fontsize{10}{15}\bfseries}{\thesection}{1em}{}
\titleformat{\subsection}
  {\centering\normalfont\fontsize{10}{15}\bfseries}{\thesubsection}{1em}{}

\usepackage{verbatim}
\usepackage{tabularx}
\usepackage[pdfencoding=auto, psdextra]{hyperref}

\usepackage{units}
\usepackage[T1]{fontenc}
\usepackage[normalem]{ulem} %For crossing out words.
\usepackage{slashed}
\usepackage{bbold}
\usepackage{orcidlink}
\usepackage{soul}

\usepackage{academicons}
\definecolor{orcidlogocol}{HTML}{A6CE39}
\newcommand{\orcid}[1]{\href{https://orcid.org/#1}{\textcolor[HTML]{A6CE39}{\aiOrcid}}}

\hypersetup{
    colorlinks,
    linkcolor={red!60!black},
    citecolor={blue!50!black},
    urlcolor={blue!80!black}
}
\allowdisplaybreaks

\tikzset{
    scalarmoredash/.style={decorate, draw=black, dashed, dash pattern=on 0.5pt off 0.5pt}
}
\tikzset{
    scalarlessdash/.style={decorate, draw=black, dashed, dash pattern=on 6pt off 2pt}
 }
\tikzset{
    potentialScalar/.style={decorate, draw=black, dashed, dash pattern=on 4pt off 2pt, double}
}
\tikzset{
    radiationScalar/.style={decorate, draw=black, dashed, dash pattern=on 4pt off 2pt}
}
\tikzset{
    radiationGraviton/.style={photon}
}
\tikzset{
    potentialGraviton/.style={decorate, decoration={snake, amplitude=.4mm, segment length=2mm, pre length=.5mm, post length=.5mm}, double}
}

\tikzset{
    potentialGraviton/.style={decorate, decoration={snake, amplitude=.4mm, segment length=2mm, pre length=.5mm, post length=.5mm}, double}
}

\tikzset{
    phistyle/.style={decorate, draw=blue!80!black, dashed, dash pattern=on 2pt off 1pt}
}

\tikzset{
    varphistyle/.style={decorate, draw=red!80!black}
}

\newcommand{\phistyle}{phistyle}
\newcommand{\varphistyle}{varphistyle}
\advance\cftsecnumwidth 0.5em\relax
\advance\cftsubsecindent 0.5em\relax
\advance\cftsubsecnumwidth 0.5em\relax

\newcommand{\rd}{{\rm d}}
\newcommand{\mpl}{M_\mathrm{Pl}}

\begin{document}
%\DeclareUnicodeCharacter{2212}{-}

\preprint{WUCG-25-02}

%%%%%%%%%%%%%%%%%%%%%%%% Title page %%%%%

\title{Tidal Love Numbers of Neutron Stars in Horndeski Theories} 

\author{Robin Fynn Diedrichs\,\orcidlink{0009-0007-3998-4608}}
\email{diedrichs@itp.uni-frankfurt.de} 
\affiliation{Institute for Theoretical Physics, Goethe University, 60438 Frankfurt am Main, Germany}

\author{Shinji Tsujikawa\,\orcidlink{0000-0002-9378-2229}}
\email{tsujikawa@waseda.jp} 
\affiliation{Department of Physics, Waseda University, 3-4-1 Okubo, Shinjuku, Tokyo 169-8555, Japan}

\author{Kent Yagi\,\orcidlink{0000-0002-0642-5363}}
\email{ky5t@Virginia.edu} 
\affiliation{Department of Physics, University of Virginia, Charlottesville, Virginia 22904, USA}

\date{\today}
\begin{abstract}
Precision measurements of the gravitational wave signal from compact binary inspirals allow us to constrain the internal structure of those objects via physical parameters such as the tidal Love numbers. In scalar-tensor theories, one typically finds new types of Love numbers that are usually not considered or simply absent in General Relativity, which further allows us to constrain deviations from General Relativity.
Building upon previous results, we present the linear perturbation equations necessary to calculate static and 
even-parity tidal Love numbers in Horndeski theories, the most general scalar-tensor theories with second-order field equations of motion. We further focus on the quadrupolar Love numbers and demonstrate how these can be extracted from the asymptotic expansion of the perturbation fields. We find that there is a potential ambiguity in extracting the Love numbers in this way, which we resolve by performing supplementary calculations in the effective field theory framework. We show that, in the case of scalar-tensor theories, the tidal Love numbers are not directly given by the $1/r^3$ term in the asymptotic expansion of 
the perturbation fields, as there is an additional contribution to this term independent of the Love numbers. We calculate such a contribution for a minimally coupled scalar field and also for the Damour-Esposito-Farèse model. 
For the latter, we find that the Love numbers can differ by 
$\mathcal{O}(1 \sim 10)\,\%$, if this additional contribution is not taken into account.
\end{abstract}

\maketitle

\begingroup
\small
\tableofcontents
\addtocontents{toc}{\vspace{-8pt}}
\endgroup

%%%%%%%%%%%%%%%%%%%%%%%%%%%%%%%%%%%%%%%%
\section{Introduction \label{sec:intro}} 

Data obtained by the gravitational wave (GW) detector network LIGO/Virgo/KAGRA is crucial to perform tests of General Relativity (GR) in the high curvature regime (see, e.g.,~\cite{LIGOScientific:2018dkp, LIGOScientific:2021sio,Berti:2015itd,Yunes:2016jcc,Barack:2018yly,Berti:2018cxi,Takeda:2023wqn, Langlois:2017dyl,Jana:2018djs,Niu:2021nic,Yunes:2024lzm}). These detectors have measured the inspiral GW signal from compact binary systems, typically assumed to consist of 
black holes (BHs) and neutron stars (NSs). 
To extract information on the constituents from the GW signal, it is necessary to provide waveforms that encode information about these objects. These waveforms are typically constructed with a so-called post-Newtonian expansion~\cite{Blanchet:2013haa, Galley:2009px,Porto:2016pyg,Foffa:2019yfl,Kalin:2020fhe,Dlapa:2021npj,Goldberger:2009qd,Goldberger:2012kf,Goldberger:2022ebt,Goldberger:2022rqf, Cardoso:2008gn,Kuntz:2019zef,Liu:2020moh,Poddar:2021yjd,Higashino:2022izi,Yunes:2024lzm}. 
One of the key parameters for constructing the waveforms is the quadrupolar tidal deformability, also known as the quadrupolar Love number, which encodes information about the internal structure of the objects~\cite{Hinderer:2007mb}. As such, understanding the Love numbers is crucial in extracting information on the objects and, hence, constraining nuclear physics~\cite{LIGOScientific:2018hze, LIGOScientific:2021sio, LIGOScientific:2021qlt, LIGOScientific:2018cki, LIGOScientific:2017vwq} as 
well as deviations from GR~\cite{Cardoso:2017cfl,Saffer:2021gak,Katagiri:2024fpn, Takeda:2023wqn, Creci:2024wfu, Cayuso:2024ppe, Diedrichs:2023foj, Quartin:2023tpl, Motaharfar:2025typ, Kobayashi:2025bdh, Cano:2025zyk, Motaharfar:2025ihv}.

It has been shown that several long-standing problems, 
such as the unknown nature of dark components in today's Universe, 
can be explained by modifying GR~\cite{Copeland:2006wr,DeFelice:2010aj,Clifton:2011jh,Joyce:2014kja,Koyama:2015vza,Heisenberg:2018vsk,Ishak:2018his}. While there are a plethora of 
ways to modify GR, we will focus here on modifications that introduce an additional scalar degree of freedom with the constraint that the overall system obeys second-order field equations of motion. 
These models are known as Horndeski theories~\cite{Horndeski:1974wa}, which include, e.g., the Damour-Esposito-Farèse (DEF)~\cite{Damour:1993hw, Damour:1996ke} model or scalar
Gauss-Bonnet gravity~\cite{Kanti:1995vq,Kanti:1997br,Torii:1996yi}. 
Both of these models can reproduce GR in the low-curvature regime. 
Still, in the high-curvature regime, they lead to effects, such as spontaneous scalarization, which distinguish them from GR (see~Ref.~\cite{Doneva:2022ewd} and references therein). In Ref.~\cite{Kase:2021mix}, the authors derived 
the general linear perturbation equations of motion
in Horndeski theories in the presence of a perfect fluid 
(see also Refs.~\cite{Kobayashi:2012kh,Kobayashi:2014wsa} 
for BH perturbations in Horndeski gravity).

In this work, we will extend the above-mentioned study, 
Ref.~\cite{Kase:2021mix}, by providing the general equations that govern static and even-parity perturbations, which thus allow the computation of Love numbers for NSs in Horndeski theories. 
It was noted in Ref.~\cite{Pani:2015hfa} 
(see also Ref.~\cite{Gralla:2017djj}) that there exists a possible ambiguity when extracting the tidal Love numbers: While the coefficient of the $1/r^3$ term in the asymptotic expansion of perturbation fields is considered to solely encode the information on a tidally-induced quadrupole moment, it is possible that additional terms that are independent of the tidal Love numbers also contribute at the same order in $r$. While in pure GR, it turns out that this is not the case --- as long as one uses Schwarzschild coordinates combined with the Regge-Wheeler gauge --- in scalar-tensor theory, we find that additional terms do appear, which have not been considered in the previous literature (see, e.g., \cite{Pani:2014jra, Brown:2022kbw, Creci:2023cfx}). Leveraging an effective field theory (EFT) approach, we calculate these terms first for the case of a minimally coupled scalar field and then for a model of spontaneous 
scalarization proposed by Damour and 
Esposito-Farèse (DEF) \cite{Damour:1993hw}. 
We further numerically compute the Love numbers in the DEF model 
and demonstrate that the result can have a noticeable difference 
if one does not correctly extract the Love numbers.

This paper is structured as follows. 
In Sec.~\ref{sec:horndeskiAction}, we present the equations of motion describing the perturbations, and subsequently, in Sec.~\ref{sec:pp_action}, we will use the EFT formalism to define the Love numbers. 
In Sec.~\ref{sec:minimally_coupled_scalar}, we investigate the case of a minimally coupled scalar field, and next, we turn to the DEF model in Sec.~\ref{sec:DEF_model}. We briefly comment on 
scalar-Gauss-Bonnet gravity in Sec.~\ref{sec:SGB}, 
and finally, we conclude this work 
in Sec.~\ref{sec:conclusions}.

Unless otherwise stated, we utilize natural units in which $\hbar = G = c = 1$. We choose the metric signature to be $(-, +, +, +)$ and define the reduced Planck mass as $\mpl = 1 / \sqrt{8 \pi}$.

%%%%%%%%%%%%%%%%%%%%%%%%%%%%%%%
\section{Horndeski Theories \label{sec:horndeskiAction}}
%%%%%%%%%%%%%%%%%%%%%%%%%%%%%%%

In this section, we revisit the background 
and perturbation equations of motion on the 
static and spherically symmetric background 
in Horndeski theories \cite{Horndeski:1974wa}.  
We will closely follow Ref.~\cite{Kase:2021mix} 
in which the issue of linear perturbations of 
relativistic stars was studied in detail. 
The Horndeski's action is 
given by \cite{Horndeski:1974wa,Kobayashi:2011nu}
\begin{align}
\label{eq::horndeskiActionTotal}
{\cal S} = \int {\rm d}^4 x \sqrt{-g}\, 
{\cal L}_\mathrm{H} + {\cal S}_m(g_{\mu\nu})\,,
\end{align}
where $g$ is the determinant of the metric 
tensor $g_{\mu \nu}$, and 
\begin{align}
\label{eq::horndeski_action}
{\cal L}_\mathrm{H} = &\ G_2 - G_3 \square \phi + G_4 R + G_5 G_{\mu\nu} \nabla^\mu \nabla^\nu \phi \notag\\
&+ G_{4,X} \Big[ (\square \phi)^2 - (\nabla_\mu \nabla_\nu \phi) (\nabla^\mu \nabla^\nu \phi) \Big] \notag \\ &- \frac{1}{6} G_{5,X} \Big[ (\square \phi)^3 - 3 (\square \phi) (\nabla_\mu \nabla_\nu \phi) (\nabla^\mu \nabla^\nu \phi) \notag \\
&\hspace{1.5cm} {} + 2 (\nabla^\mu \nabla_\alpha \phi) (\nabla^\alpha \nabla_\beta \phi) (\nabla^\beta \nabla_\mu \phi)\Big].
\end{align}
Here, $R$ is the Ricci scalar, $G_{\mu\nu}$ is 
the Einstein tensor, and 
$G_{2,3,4,5}$ are functions of $\phi$ and 
$X=-\nabla^\mu \phi \nabla_\mu \phi/2$, 
with $\nabla_{\mu}$ being a covariant 
derivative operator and $\square \equiv \nabla_{\mu}\nabla^{\mu}$.
To simplify the notation, we denote a partial derivative with respect to a certain quantity 
with a comma in the subscript, e.g.,
$G_{2,\phi} = \partial G_2/\partial \phi$ and 
$G_{4, X}=\partial G_4/\partial X$.

For the matter sector described by 
the action ${\cal S}_m$, we consider a perfect 
fluid minimally coupled to gravity.
This is described by the Schutz-Sorkin 
action \cite{Schutz:1977df,Brown:1992kc,DeFelice:2009bx},
\begin{align}
{\cal S}_{m} =  -\int {\rm d}^{4}x 
\left[ \sqrt{-g}\,\rho(n)
+ J^{\mu} (\partial_{\mu} \ell+{\cal A}_i\partial_{\mu}{\cal B}^i)\right]\,,
\label{Sm}
\end{align}
where $\rho$ is the matter density that 
depends on its number density $n$, 
and $J^\mu$ is related to the fluid 
four-velocity $u^\mu$ via
\begin{align}
u^\mu = \frac{J^\mu}{n \sqrt{-g}}\,.
\label{umu}
\end{align}
Further, $\ell$, $\mathcal{A}_i$ and $\mathcal{B}^i$ (where $i=1,2,3$) 
are the Lagrange multipliers. 
Varying this action with respect to the metric, 
the resulting energy-momentum tensor takes 
the usual form \cite{Amendola:2020ldb},
\begin{align}
T_{\mu\nu} = \left( \rho + P \right) 
u_\mu u_\nu + P g_{\mu\nu}\,,
\end{align}
where $\rho$ and $P=n \rho_{,n}-\rho$ 
are the density and pressure, respectively. 

In Horndeski theories, the propagation speed 
$c_\mathrm{GW}$ of gravitational waves on a homogenous and 
isotropic cosmological background generally 
deviates from the speed of light \cite{Kobayashi:2011nu,DeFelice:2011bh}. 
In a subclass of Horndeski theories where the 
conditions $G_{4,X} = G_5 = 0$ are satisfied, 
we have $c_\mathrm{GW}=1$.
In this work, we primarily focus on models with $c_\mathrm{GW}=1$, but we also briefly consider the case of scalar-Gauss-Bonnet gravity, where neither 
neither $G_{4,X}$ nor $G_5$ vanish.
For the computation of the tidal deformation of NSs, 
we treat the scalar field as time-independent, with its effects manifesting in the local region of the Universe.
In other words, this scalar field is not necessarily related to the cosmological dynamics of dark energy.

\subsection{Background Equations \label{sec:background}}

The line element describing the static and 
spherically symmetric background is given by 
\begin{equation}
{\rm d} s^2 = -f(r) {\rm d}t^2 
+ h^{-1}(r) {\rm d}r^2 + r^2 
\rd \Omega^2\,,
\label{metric}
\end{equation}
where $f$ and $h$ are functions of 
the areal distance $r$, and 
$\rd \Omega^2={\rm d}\theta^2 + \sin^2 \theta 
{\rm d} \varphi^2$.
On this background, we consider a scalar field 
$\phi$ that is a function of $r$ alone. 
In the Schutz-Sorkin action (\ref{Sm}), we take 
the configurations $J^{\mu}=[\sqrt{-g}\,N(r),0,0,0]$ 
and ${\cal A}_i=0$, where $N(r)$ is related to 
the fluid number density $n$, as 
$n(r)=f(r)^{1/2} N(r)$.
{}From Eq.~(\ref{umu}), the fluid four-velocity
has the components $u^{\mu}=[f(r)^{-1/2},0,0,0]$.

Varying the action \eqref{eq::horndeski_action} with respect to $g_{\mu \nu}$ and utilizing 
the metric ansatz \eqref{metric}, we obtain \cite{Kase:2013uja,Kase:2021mix} 
\begin{widetext}
\begin{eqnarray}
\label{eq::horndeski_bg_equations}
{\cal E}_{00}&\equiv&
\left(A_1+\frac{A_2}{r}+\frac{A_3}{r^2}\right)\phi''
+\left(\frac{\phi'}{2h}A_1+\frac{A_4}{r}+\frac{A_5}{r^2}\right)h'+A_6+\frac{A_7}{r}+\frac{A_8}{r^2}=\rho\,,\label{back1}\\
{\cal E}_{11}&\equiv&
-\left(\frac{\phi'}{2h}A_1+\frac{A_4}{r}+\frac{A_5}{r^2}\right) \frac{hf'}{f}
+A_9-\frac{2\phi'}{r}A_1-\frac{1}{r^2}\left[\frac{\phi'}{2h}A_2+(h-1)A_4\right]=P
\,,\label{back2}\\
{\cal E}_{22}&\equiv&
\left[\left\{A_2+\frac{(2h-1)\phi'A_3+2hA_5}{h\phi' r}\right\}\frac{f'}{4f}+A_1+\frac{A_2}{2r}\right]\phi''
+\frac{1}{4f}\left(2hA_4-\phi'A_2+\frac{2hA_5-\phi'A_3}{r}\right)\left(f''-\frac{f'^2}{2f}\right)\notag\\
&&+\left[A_4+\frac{2h(2h+1)A_5-\phi'A_3}{2h^2r}\right]\frac{f'h'}{4f}
+\left(\frac{A_7}{4}+\frac{A_{10}}{r}\right)\frac{f'}{f}
+\left(\frac{\phi'}{h}A_1+\frac{A_4}{r}\right)\frac{h'}{2}+A_6+\frac{A_7}{2r}
=-P\,,\label{back3}
\end{eqnarray}
\end{widetext}
where a prime denotes the derivative with respect 
to $r$ and the explicit forms of the functions 
$A_1$ to $A_{10}$ are given in Appendix~\ref{app::backgroundEquations}.
Additionally, the continuity equation 
in the fluid sector, $\nabla^\mu T_{\mu\nu}=0$, 
leads to 
\begin{equation}
{\cal E}_{P} \equiv 
P'+\frac{f'}{2f} \left( \rho+P \right)=0\,.
\label{back4}
\end{equation}
The equation of motion for the scalar field 
can be obtained by varying the action \eqref{eq::horndeski_action} with respect to $\phi$. 
This equation is equivalent to
\be
{\cal E}_{\phi} \equiv -\frac{2}{\phi'} 
\left[ \frac{f'}{2f}{\cal E}_{00}
+{\cal E}'_{11}+\left( \frac{f'}{2f}
+\frac{2}{r} \right){\cal E}_{11}
+\frac{2}{r}{\cal E}_{22}
+{\cal E}_{P} \right].
\label{back5}
\ee
With a given equation of state $\rho=\rho (P)$, 
the four Eqs.~(\ref{back1}), (\ref{back2}), 
(\ref{back4}), and (\ref{back5})
define a set of differential equations for the quantities $f$, $h$, $P$, and $\phi$. 
Thus, upon specifying the functions $G_j$ and boundary conditions, we can determine these quantities as functions of $r$ by integrating 
the four background equations mentioned above.

\subsection{Perturbation Equations \label{sec:peturbation_equations}}

The tidal deformability is encoded in the behavior of linear perturbations around the static and 
spherically symmetric NS background. 
Since the equations of motion for them  
were derived in Ref.~\cite{Kase:2021mix}, we will briefly revisit the essential steps. 
We focus on the perturbations in the even-parity sector, which are the dominant contributions to the tidal deformability of relativistic stars. 
We are also interested in a binary system where 
its orbital frequency is much smaller than 
the mode frequency of stars (such as the 
$f$-mode frequency). 
In this case, the perturbations can be 
dealt as static ones \cite{Flanagan:2007ix,Hinderer:2007mb}, 
so that all the perturbed fields do 
not depend on time $t$.

We first split the metric into the form 
$g_{\mu\nu} = \bar{g}_{\mu\nu} + h_{\mu\nu}$, 
where $\bar{g}_{\mu\nu}$ is the background value  
and $h_{\mu\nu}$ corresponds to perturbations. 
As was done in Ref.~\cite{Regge:1957td}, we further decompose 
metric perturbations into multipole moments by using 
the spherical harmonics $Y_{lm}$, such that 
\begin{align}
\label{eq::metric_perturb_definition}
h_{\mu\nu} =
\sum_{l,m} \begin{pmatrix}
f H_0 & H_1 & h_0 \nabla_a \\
H_1 & h^{-1} H_2 & h_1 \nabla_a \\
h_0 \nabla_a & h_1 \nabla_a & K g_{ab} + r^2 G  \nabla_a \nabla_b
\end{pmatrix} Y_{lm}(\theta, \varphi)\,,
\end{align}
where the seven perturbed fields 
$H_0$, $H_1$, $H_2$, $h_0$, $h_1$, $K$ and $G$ 
depend on $r$ alone under the static approximation. 
We note that the operator $\nabla_a$, where 
$a=(\theta, \varphi)$, denotes the covariant derivative on the two-sphere in spherical coordinates.
The scalar field is decomposed similarly as
\begin{equation}
\phi = \bar{\phi}(r) + \sum_{l,m} \delta \phi(r) Y_{lm} (\theta, \varphi) \,,
\end{equation}
where $\bar{\phi}(r)$ is the background 
scalar field. In the following, 
we will mostly drop the overhead 
bar from $\bar{\phi}(r)$ for brevity, unless 
it is crucial to explicitly distinguish it from the full field.

For the fluid sector, the perturbed quantities 
associated with the Schutz-Sorkin action 
(\ref{Sm}) were discussed in Ref.~\cite{Kase:2021mix}. 
Here, we skip the details and 
introduce the following several key 
perturbed quantities:
\ba
\delta n &=& \sum_{l,m} 
\frac{\delta \rho (r)}{\rho_{,n}(r)}
Y_{lm}(\theta, \varphi)\,,\\
u_r &=& \sum_{l,m} 
\delta u_r (r) Y_{lm}(\theta, \varphi)\,,\\
u_a &=& \sum_{l,m} 
v (r) \nabla_{a} Y_{lm}(\theta, \varphi)\,.
\ea
Here $\delta \rho (r)$, $\delta u_r (r)$, 
and $v(r)$ correspond to the density perturbation, 
the radial component of $u_{\mu}$, and 
the velocity potential, respectively.

In Ref.~\cite{Kase:2021mix}, the authors derived 
the second-order action of even-parity perturbations 
${\cal S}^{(2)}$ by integrating the action 
with respect to $\theta$ and $\varphi$. 
The linear perturbation equations of motion follow 
by varying the second-order action with respect 
to each perturbed field. They are given in Ref.~\cite{Kase:2021mix} 
in a gauge-ready form, 
i.e., without fixing particular gauge conditions. 
In the following, we choose 
the Regge-Wheeler gauge~\cite{Regge:1957td}, 
which is characterized by the conditions 
\be
h_0=0\,,\qquad h_1=0\,,\qquad G=0\,.
\label{gauge}
\ee
We also focus on static perturbations, so that 
all of the time derivatives are dropped from 
the perturbation equations of motion.

First of all, the perturbation equation that follows 
from the variation of ${\cal S}^{(2)}$ 
with respect to $H_1$ reads
\begin{equation}
\left[ 2 L b_1 - \frac{r^2(\rho + P) \sqrt{h}}{\sqrt{f}} 
\right] H_1 + r^2 (\rho + P) \sqrt{h}\, \delta u_r = 0 \,,
\label{eq::elimH1}
\end{equation}
where 
\be
L \equiv l(l+1)\,,
\ee
and the definition of $b_1$ is given 
in Appendix \ref{app::evenParityPerturbations}.
From the variation of the second-order action 
with respect to one of the components of the 
Lagrange multiplier field ${\cal A}_{\mu}$, 
we obtain
\begin{equation}
H_1 = \sqrt{f} \delta u_r\,,
\label{H1ur}
\end{equation}
see Eq.~(4.24) of Ref.~\cite{Kase:2021mix}.
Inserting Eq.~(\ref{H1ur}) into 
Eq.~\eqref{eq::elimH1}, it follows that 
\begin{equation}
2 L b_1 H_1 = 0\,.
\end{equation}
Since, in general, $b_1 \neq 0$, 
we have that 
\be
H_1=0\,.
\label{H1con}
\ee
Under this equality together with the gauge 
conditions (\ref{gauge}), we find that 
all of the off-diagonal components of $h_{\mu \nu}$ in 
Eq.~\eqref{eq::metric_perturb_definition} 
vanish for static perturbations.

The perturbation equations of motion 
following from the variation of ${\cal S}^{(2)}$ 
with respect to $H_0$, $H_2$, $h_1$, $K$, $G$, 
and $\delta \phi$ are given in Eqs.~(4.30), (4.32), 
(4.34), (4.35), (4.36), and (4.37) 
of Ref.~\cite{Kase:2021mix} in a gauge-ready form.
Under the gauge choice (\ref{gauge}) together 
with the condition (\ref{H1con}), we just need 
to set $h_0=h_1=G=H_1=0$ in these equations. 
Then, we have 
\begin{widetext}
\ba
\label{eq::pert_eq_first}
0 &=& a_1 \delta \phi'' + a_2 \delta \phi' + a_3 H_2' + (a_5 + L a_6) \delta \phi + (a_7 + L a_8) H_2 + \frac{r^2 \sqrt{f}}{2\sqrt{h}} \delta \rho - 2 g_2 K'' - 2 g_{13} K' + (L - 2) k_1 K\,, 
\label{pereq1}\\
0 &=& c_2 \delta \phi' + (c_3 + L c_4) \delta \phi + 2 c_6 H_2 - a_3 H_0' + (L a_8 + a_7 - a_3') H_0 - 2 g_{14} K' + k_2 (L - 2) K\,, 
\label{pereq2}\\
0 &=& d_2 \delta \phi' + d_3 \delta \phi - a_4 H_0' + (a_9 - a_4') H_0 + c_5 H_2 - g_{16} K'\,, 
\label{pereq3}\\
0 &=& -2 g_2 H_0'' - 4 {g_4}  K'' - 4 g_4' K' + 2 g_{12} \delta \phi'' + 2 (g_{13} - 2 g_2') H_0' + 2 g_{14} H_2' + 2 (g_{15} + g_{12}') \delta \phi'\,, 
\nonumber \\
& &+\left[ (L - 2) k_3 + 2 g_{15}' \right] \delta \phi + \left[ (L - 2) k_1 - 2 g_2'' + 2 {g_{13}'} \right] H_0 + \left[ (L - 2) k_2 + 2 g_{14}' \right] H_2\,,
\label{pereq4}\\
0 &=& (L - 2) \left( k_1 H_0 + k_2 H_2 
+ k_3 \delta \phi \right)\,, 
\label{pereq5}\\
0 &=& -2 e_2 \delta \phi'' + 2(e_3 + L e_4) \delta \phi + a_1 H_0'' + (2a_1' - a_2) H_0' + (a_1'' - a_2' + a_5 + L a_6) H_0 - c_2 H_2' \nonumber \\
& & - (c_2' - c_3 - L c_4) H_2 -2 e_2' \delta \phi' + 2 g_{12} K'' - 2 (g_{15} - g_{12}') K' + k_3 (L - 2) K\,,
\label{pereq6}
\ea
\end{widetext}
where the coefficients $a_1$, etc.~are presented 
in Appendix~\ref{app::evenParityPerturbations} 
(which are the same as those used 
in Ref.~\cite{Kase:2021mix}).
Also, the matter density perturbation obeys
\begin{equation}
\delta \rho = \frac{\rho + P}{2 c_m^2}H_0\,,
\label{eq::pert_eq_last}
\end{equation}
where $c_m^2 \equiv n \rho_{,nn}/\rho_{,n}$ is 
the squared sound speed of the fluid (see 
Eq.~(4.22) in Ref.~\cite{Kase:2021mix}).
Thus, the perturbation Eqs.~(\ref{pereq1})--(\ref{pereq6}) 
contain the four fields 
$H_0$, $H_2$, $K$, $\delta \phi$ and their 
radial derivatives.

We are interested in the computation 
of tidal Love numbers for the multipole mode,
\be
l=2\,,
\ee
in which case $L=6$. 
Then, for $k_2 \neq 0$, Eq.~(\ref{pereq5}) gives 
\be
H_2=-\frac{k_1}{k_2}H_0
-\frac{k_3}{k_2}\delta \phi\,,
\ee
which relates $H_2$ with $H_0$ and
$\delta \phi$. 
Solving Eqs.~(\ref{pereq2})--(\ref{pereq3}) 
for $K$ and $K'$, we can express them in terms 
of $H_0$, $\delta \phi$, and their 
radial derivatives. 
Then, we take the $r$ derivatives of $K'$ 
to eliminate the $K$-dependent terms in 
Eqs.~(\ref{pereq1}) and (\ref{pereq6}). 
Solving them for $H_0''$ 
and $\delta \phi''$, we obtain the second-order 
differential equations for $H_0$ and $\delta \phi$
coupled to each other. 
Thus, the system of even-parity static perturbations 
in Horndeski theories reduces to 
that of the two fields $H_0$ and $\delta \phi$. 
For given boundary conditions of them around 
$r=0$, we can integrate their field equations of 
motion outward to obtain 
$H_0$ and $\delta \phi$ as functions of $r$.
In Sec.~\ref{sec:minimally_eft}, we will see that 
the asymptotic behavior of these perturbations 
at radial infinity allows us to compute 
the tidal Love numbers.

\section{Point-Particle Action \label{sec:pp_action}}

If one is only interested in the behavior of the fields at large distances (distances 
greater than the radius of the considered object), 
then instead of solving the background and perturbation equations of motion explicitly, 
it is possible to work with an EFT approach by replacing the finite-sized object with 
a point particle (PP) and, hence, the matter action 
in Eq.~\eqref{eq::horndeskiActionTotal} with a PP action~\cite{Goldberger:2004jt,Goldberger:2007hy}. Quantities such as the mass and charge of an object are then captured by \textit{Wilsonian coefficients}, and the external fields are generated by corresponding \textit{operators} that live on the worldline of the PP. 

When constructing the PP action, we only need 
to consider operators respecting symmetries 
that we wish to be present at large distances, 
such as diffeomorphism invariance. 
Specifically, in pure GR, one would thus 
find that the PP action can be written as
\begin{align}
{\cal S}_\mathrm{pp} = \int {\rm d} \tau 
\left( M + c_R R + d_R u^\mu u^\nu R_{\mu\nu} + \dots \right)\,,
\end{align}
where ${\rm d} \tau = \sqrt{- {\rm d} x_\mu {\rm d} x^\mu}$ 
is the proper time interval of the considered 
object, $u^\mu$ is its four-velocity, $R_{\mu \nu}$ is 
the Ricci tensor, and $c_R$ and $d_R$ denote Wilsonian coefficients. Any operator that vanishes 
by the equations of motion can be removed from 
the PP action under a redefinition of 
the field variables. 
Hence, we do not need to consider operators 
that are built from $R$ and $R_{\mu\nu}$. 
The first non-trivial operator is thus built from the Riemann tensor $R_{\mu\alpha\nu\beta}$, 
which encodes the quadrupolar tidal deformability of the object~\cite{Hinderer:2007mb}.

In scalar-tensor theories, we are free to 
take additional operators into account, including the scalar field 
and its derivatives. For example, we could thus modify the PP action to be
\begin{align}
{\cal S}_\mathrm{pp} = 
-\int {\rm d} \tau \left( M + Q \frac{\phi}{\mpl} + p \frac{\phi^2}{\mpl^2} + \dots \right)\,,
\label{PP0}
\end{align}
i.e.,~adding powers of the scalar field $\phi$. 
The Wilsonian coefficient corresponding to the operator linear in $\phi$ can be identified as 
a scalar charge of the object, as it generates 
a term that asymptotically falls off with $1/r$ at large distances in the scalar field (see below). 
We can also add nonlinear operators, such as 
$p \phi^2$, which can be interpreted as 
an induced monopole due to an external field~\cite{Huang:2018pbu}.

In this work, we are interested in the 
quadrupolar even-parity Love numbers, which 
we define in the PP action as 
\ba
\label{eq::pp_action_tidal_part}
{\cal S}_\mathrm{pp} &\supset& 
{\cal S}_\mathrm{pp}^\lambda = 
\int {\rm d} \tau \bigg( \frac{\lambda_{hh}}{4} 
E_{\mu\nu} E^{\mu\nu} + \frac{\lambda_{h\phi}}{2 \mpl} 
E^{\mu\nu} \nabla_\mu \nabla_\nu \phi 
\nonumber \\
& &\qquad \qquad \qquad
+\frac{\lambda_{\phi\phi}}{4 \mpl^2} 
\nabla_\mu \nabla_\nu \phi \nabla^\mu \nabla^\nu \phi 
\bigg) \,,
\ea
where $E_{\mu\nu} \equiv 
u^\alpha u^\beta R_{\mu\alpha\nu\beta}$. 
We will use ${\cal S}_\mathrm{pp}^\lambda$ 
to denote the part of the PP action that encodes 
the quadrupolar Love numbers and 
${\cal S}_\mathrm{pp}^\mathrm{(bg)}$ to 
represent the background PP action, containing 
the mass and scalar charge.
We can interpret $\lambda_{hh}$ as the strength 
of the gravitational tidal response due to an external
gravitational field. Likewise, $\lambda_{\phi\phi}$ 
quantifies the strength of the scalar tidal field 
that develops due to an external scalar field. 
As noted in Refs.~\cite{Bernard:2019yfz,Creci:2023cfx}, 
we can additionally have mixed modes, i.e., a scalar tidal 
field developing due to an external gravitational field 
and the opposite situation, a tidal gravitational field 
developing due to an external scalar field. 
Note that both of these cases are quantified 
by the same coefficient $\lambda_{h\phi}$. 

To obtain the values of the tidal Love numbers, 
it is necessary to compute the asymptotic field using 
the PP action, as well as exploiting the full 
equations of motion and match both in the limit 
that $r \rightarrow \infty$. 
In GR, it turns out that there is an exact solution 
to the perturbative field outside the object, and 
it is, hence, possible to match both expansions 
at the surface of the object \cite{Hinderer:2007mb} 
instead of asymptotic infinity. In the following, 
we will demonstrate how this matching can be done 
for a minimally coupled scalar field.

%%%%%%%%%%%%%%%%%%%%%%%%%%%%%%%%%%%%%%%%%
\section{Minimally Coupled Scalar Field 
\label{sec:minimally_coupled_scalar}} 
%%%%%%%%%%%%%%%%%%%%%%%%%%%%%%%%%%%%%%%%%

We first consider a massless and real scalar field 
$\phi$ minimally coupled to gravity which does not 
feature any further couplings. 
We thus augment the bulk action with 
a kinetic term of the field $\phi$, 
such that
\ba
\label{eq:S_bulk}
\hspace{-0.7cm}
{\cal S}_\mathrm{bulk}
&=& 
{\cal S}_\mathrm{EH} + {\cal S}_\phi \nonumber \\
\hspace{-0.7cm}
&=& 
\int {\rm d}^4 x \sqrt{-g} \left( \frac{\mpl^2}{2} R
- \frac{1}{2} \nabla_\mu \phi \nabla^\mu \phi \right),
\label{Sbulk}
\ea
where ${\cal S}_\mathrm{bulk}$ does not 
contain a contribution from the fluid. 
This action can be obtained from the 
Horndeski's action by choosing
\begin{align}
G_2 = X\,,\quad 
G_3 = 0\,,\quad 
G_4 = \frac{\mpl^2}{2}\,,\quad 
G_5 = 0\,.
\end{align}
In the following, we will describe the asymptotic 
behavior of the background and perturbed fields 
by first using their equations of motion and 
then by resorting to the EFT approach. 
Subsequently, we can match both approaches to obtain 
an expression that allows us to extract the tidal Love numbers. 
The results presented in this section are quite generic and 
apply to any scalar-tensor theories whose Einstein-frame action 
reduces to Eq.~\eqref{eq:S_bulk} in the absence 
of the matter action. 
As an example, we consider their application 
to the DEF model in Sec.~\ref{sec:DEF_model}.

\subsection{Asymptotic Expansion 
\label{sec:minimally_asymp_expansion}} 

We expand the background fields $f$, $h$, 
and $\phi$ around $r = \infty$, as
\be 
f = \sum_{i = 0}^\infty f_i r^{-i}\,,
\quad 
h = \sum_{i = 0}^\infty h_i r^{-i}\,,
\quad 
\phi = \sum_{i = 0}^\infty \phi_i r^{-i}\,,
\label{eq::bg_ansatz}
\ee
where $f_i$, $h_i$, and $\phi_i$ are constants.
Imposing asymptotic flatness, we have that 
$f_0=1$ and $h_0=1$. The ADM mass $M$ 
of the object is related to $h_1$, as 
$h_1=-2M$. 
Outside the compact object ($\rho=0=P$), 
the background Eqs.~(\ref{back1}), 
(\ref{back2}), and (\ref{back5}) give
\ba
& &
h'=\frac{1-h}{r}-\frac{rh \phi'^2}{2\mpl^2}\,,
\label{back1D}\\
& &
\frac{f'}{f}-\frac{h'}{h}=\frac{r \phi'^2}
{\mpl^2}\,,
\label{back2D}\\
& &
\phi''+\frac{h+1}{rh}\phi'=0\,.
\label{back3D}
\ea
Plugging the ansatz \eqref{eq::bg_ansatz} 
into Eqs.~(\ref{back1D})--(\ref{back3D}), 
we find that all of the coefficients are 
determined by $\phi_0$, $\phi_1$, and $M$. 
Up to order $r^{-4}$, the expanded 
solutions are given by 
\ba
\label{eq::minimally_coupled_bg_expansion}
f &=& 1-\frac{2M}{r}+\frac{M \phi_1^2}
{6\mpl^2 r^3}+\frac{M^2 \phi_1^2}
{3 \mpl^2 r^4}+{\cal O}(r^{-5})\,,\\
h &=& 1-\frac{2M}{r}+\frac{\phi_1^2}
{2\mpl^2 r^2}+\frac{M \phi_1^2}{2\mpl^2 r^3}
+\frac{2M^2 \phi_1^2}{3 \mpl^2 r^4} \nonumber \\
& &
+{\cal O}(r^{-5})\,,
\label{hlexpan}\\
\phi &=& \phi_0 + \frac{\phi_1}{r} 
+ \frac{M \phi_1}{r^2}
+\frac{\phi_1( 16 M^2 \mpl^2-\phi_1^2)}
{12 \mpl^2 r^3} \nonumber \\
& &
+\frac{M \phi_1 (6 M^2 \mpl^2-\phi_1^2)}
{3\mpl^2 r^4}+{\cal O}(r^{-5})\,.
\label{phiLexpan}
\ea
Since the coefficients $\phi_0$, $\phi_1$, 
and $M$ remain unconstrained by the background equations, 
they have to be obtained by fitting the above expressions 
to numerical solutions for Eqs.~(\ref{back1D})--(\ref{back3D}). 
For this purpose, we have calculated the background expansions 
up to order $r^{-9}$ and present the results in Appendix~\ref{asymp_coeffs_minimally_coupled}.

Likewise, at large distances, we expand the perturbed 
fields $H_0$ and $\delta \phi$ in the forms
\be
H_0 = \sum_{i=-2}^{\infty} H_{0,i} r^{-i}\,,
\qquad 
\delta \phi = 
\sum_{i=-2}^{\infty} \delta \phi_{i} r^{-i}\,,
\label{eq::minimally_coupled_perturb_expansion}
\ee
where $H_{0,i}$ and $\delta \phi_{i}$ 
are constants. Outside the object, 
the field equations of motion for 
$H_0$ and $\delta \phi$ following from 
Eqs.~(\ref{pereq1})--(\ref{pereq6}) 
are given by 
\begin{widetext}
\begin{align}
%\hspace{-1.0cm}
\raisetag{1.3cm}
&
H_0''+\frac{h+1}{rh}H_0'
-\frac{4 [1+ h^2 + (L - 2)h] \mpl^4 
- 4 h r^2 \phi'^2 (h - 1) \mpl^2 
+ r^4 h^2 \phi'^4}
{4 \mpl^4 r^2 h^2}H_0
-\frac{\phi'[2\mpl^2 (h-1)-r^2 h 
\phi'^2]}{\mpl^4 rh} \delta \phi=0,
\label{perH0} \\
%\hspace{-1.0cm}
&
\delta \phi''+\frac{h+1}{rh} \delta \phi'
-\frac{2r^2 h \phi'^2+L \mpl^2}
{\mpl^2 r^2 h}\delta \phi
+\frac{\phi'[2(1-h)\mpl^2+r^2 h \phi'^2]}
{2\mpl^2 rh}H_0=0.
\label{perdp}
\end{align}
\end{widetext}
Inserting the ansatz \eqref{eq::minimally_coupled_perturb_expansion} 
together with the expansion for the background fields into the perturbation 
Eqs.~(\ref{perH0})~and~(\ref{perdp}), we find that 
all of the coefficients in the expansion 
\eqref{eq::minimally_coupled_perturb_expansion} 
are determined by $H_{0,-2}$, $H_{0,3}$, 
$\delta \phi_{-2}$, and $\delta \phi_3$. 
Specifically, up to order $r^{-3}$, we find
\begin{align}
\label{eq::minimal_asymp_expansion_H0}
H_0(r) &= H_{0,-2} r^2 - 2 H_{0,-2} M r + \frac{\phi_1 (H_{0,-2} \phi_1
-2\delta \phi_{-2}M)}{3 \mpl^2} \notag\\
&\phantom{{} = {}}
+\frac{H_{0,-2} M \phi_1^2}
{6 \mpl^2 r} + \frac{H_{0,-2} M^2 \phi_1^2}
{3 \mpl^2 r^2}+\frac{H_{0, 3}}{r^3} \notag\\
&\phantom{{} = {}} +\mathcal{O}\left( r^{-4} \right) \,,\\
\label{eq::minimal_asymp_expansion_dphi}
\delta\phi(r) &= \delta\phi_{-2} r^2 
- 2\delta\phi_{-2}M r 
+\frac{M(2M \delta\phi_{-2}-H_{0,-2}\phi_1)}{3} \notag\\
&\phantom{{} = {}} + \frac{\delta\phi_{-2} M \phi_1^2}{6 \mpl^2 r} 
+\frac{\delta\phi_{-2} M^2 \phi_1^2}
{3 \mpl^2 r^2}+ \frac{\delta\phi_3}{r^3} \notag\\
&\phantom{{} = {}} + \mathcal{O}\left( r^{-4} \right) \,. 
\end{align}
We have derived these coefficients up to order $r^{-8}$ 
and present their values in 
Appendix \ref{asymp_coeffs_minimally_coupled}.

Because of the linear nature of the perturbation equations, we can at this point already conclude that the coefficients $H_{0,3}$ and $\delta \phi_3$ need to behave as
\ba
\label{eq::perturb_coeff_naiv_H03}
H_{0,3} &=& c_{hh} H_{0,-2} 
+ c_{h\phi} \delta \phi_{-2}\,,\\
\delta \phi_3 &=& c_{\phi\phi} \delta \phi_{-2} 
+ c_{\phi h} H_{0,-2}\,,   
\label{eq::perturb_coeff_naiv_dphi03}
\ea
where $c_{hh}$, $c_{h\phi}$, $c_{\phi\phi}$, 
and $c_{\phi h}$ are real coefficients, which we will connect to the tidal Love numbers in the next section by leveraging the EFT approach.

\subsection{EFT Side \label{sec:minimally_eft}} 

Although the previous calculations have been performed 
in Schwarzschild coordinates and the Regge-Wheeler gauge, 
we here follow Ref.~\cite{Kol:2011vg} 
and utilize isotropic coordinates for the EFT calculations, 
since it significantly simplifies
the analysis. To distinguish both coordinate systems more easily, we denote all quantities in isotropic coordinates with a hat, e.g., 
the metric is represented with $\hat{g}_{\mu\nu}$. 
Additionally, we use the same metric decomposition 
as in Ref.~\cite{Kol:2011vg}, which is given by\footnote{We note that the metric of a spherically symmetric and static spacetime can always be expressed in terms of two radial functions.
Namely, spherical symmetry, time-translation invariance, and parity invariance reduce the ten a priori independent components of the metric to just two \cite{Schutz:1985jx}.}
\ba
\hat{g}_{00} &=& -e^{\sqrt{2} 
\hat{\varphi}(\hat{x}^i)/\mpl},\nonumber \\
\hat{g}_{0i} &=& 0, \nonumber \\
\hat{g}_{ij} &=& e^{-\sqrt{2} 
\hat{\varphi}(\hat{x}^i)/\mpl}\,
\hat{\psi}(\hat{x}^i)\,\delta_{ij}\,,
\label{eq::eft_metric_decomposition}
\ea
with $\hat{\varphi}$ and $\hat{\psi}$ are 
scalar fields that are functions of 
spatial coordinates $\hat{x}^i$. 
Then, up to boundary terms, the bulk 
action (\ref{Sbulk}) reduces to
\ba
\label{eq::minimal_bulk_action_reduced}
{\cal S}_\mathrm{bulk} &=& \int {\rm d}^4x 
\bigg[ \frac{\mpl^2}{4} \hat{\psi}^{-3/2} 
(\partial \hat{\psi})^2
-\frac{1}{2} \hat{\psi}^{1/2} 
(\partial \hat{\varphi})^2  \nonumber \\
& &\qquad \quad 
-\frac{1}{2} \hat{\psi}^{1/2} 
(\partial \hat{\phi})^2 \bigg]\,,
\label{Sbulk2}
\ea
where $(\partial \hat{\psi})^2 
\equiv (\partial_{i}\hat{\psi})(\partial_{i}\hat{\psi})$, 
etc. To make the field $\hat{\psi}$ canonical, 
we further define
\be
\hat{\psi} = 1 +  
\frac{\sqrt{2}\hat{\sigma}}{\mpl}\,,
\ee
so that the bulk action (\ref{Sbulk2}) 
is expressed as
\ba
\label{eq::minimal_bulk_action_reduced2}
{\cal S}_\mathrm{bulk} &=& \int {\rm d}^4x 
\bigg[ \frac{1}{2} \hat{\psi}^{-3/2} 
(\partial \hat{\sigma})^2
-\frac{1}{2} \hat{\psi}^{1/2} 
(\partial \hat{\varphi})^2  \nonumber \\
& &\qquad \quad 
-\frac{1}{2} \hat{\psi}^{1/2} 
(\partial \hat{\phi})^2 \bigg]\,.
\label{Sbulk3}
\ea
Upon expanding $\hat{\sigma}$ and $\hat{\varphi}$ 
around zero, we can directly read off the propagator 
for the $\hat{\varphi}$ field:
\begin{align}
\langle \hat{\varphi}(x) \hat{\varphi}(y) \rangle = 
- \int \frac{\rd^4 k}{(2\pi)^4} \frac{i e^{ik(x - y)}}
{k^2 + i \epsilon}\,,
\label{propagator}
\end{align}
where we used the notations
$kx=k_{\mu}x^{\mu}=k_0 t+{\bm k} \cdot {\bm x}$
and $k^2=k_{\mu}k^{\mu}$, with 
the four momentum $k_{\mu}=(k_0, {\bm k})$.
For the $\hat{\sigma}$ and $\hat{\phi}$ fields, 
we also have the propagators analogous to 
Eq.~(\ref{propagator}).

\paragraph*{\bf Background.}
Under the above-defined metric decomposition, 
the background PP action (\ref{PP0}) simplifies to
\begin{equation}
{\cal S}_\mathrm{pp}^\mathrm{(bg)} = 
-\int \rd t \, e^{\sqrt{2} \hat{\varphi} / (2 \mpl)} 
\left( M + Q \frac{\hat{\phi}}{\mpl} \right)\,.
\label{pp1}
\end{equation}
We further simplify Eq.~(\ref{pp1}) by  
considering only the linear terms 
in the fields, such that
\begin{equation}
{\cal S}_\mathrm{pp}^\mathrm{(bg)} = 
-\int \rd t \, \left( M \frac{\hat{\varphi}}
{\sqrt{2} \mpl} + Q \frac{\hat{\phi}}{\mpl} 
\right) \,.
\label{Spp1}
\end{equation}
This truncation is justified since one can show that 
non-linear terms can only contribute to the resulting fields 
when external fields are present (see the discussion in Sec.~IIC 
of Ref.~\cite{Goldberger:2004jt} and Sec.~IVB 
of Ref.~\cite{Diedrichs:2023foj}). 
Here, we only consider the impact of an external quadrupolar field, whose interaction with the worldline is separately defined later in this section.\footnote{In essence, while a constant term in the PP action would just vanish upon applying the Euler-Lagrange equations, other non-linear terms would either generate (1) a monopolar perturbation due to an external field, (2) a quantum loop diagram which carries a power of $\hbar$ and is thus of no interest to us as we are only considering classical observables, or (3) it would result in a diagram that contains a scale-less divergent integral, which vanishes upon using dimensional regularization.}

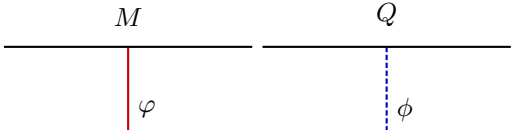
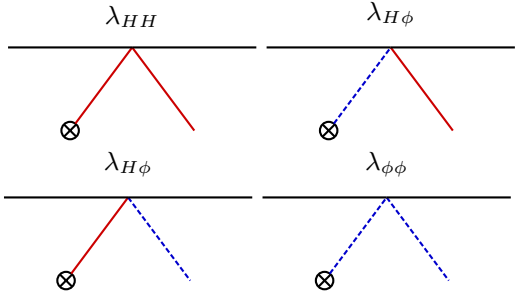
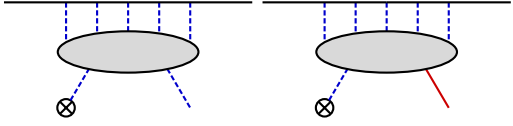
\begin{figure}
\centering
\begin{subfigure}[b]{\textwidth}
\begin{tikzpicture}[thick, scale = 1.1, every node/.style={transform shape}]
\begin{feynman}
            % set up the worldline
            \vertex (a1);
            \vertex [right=of a1] (a2);
            \vertex [right=of a2] (a3);
            
            \vertex (a11) at ($(a2)!\twoVertexDistance!(a1)$);
            \vertex (a12) at ($(a2)!\twoVertexDistance!(a3)$);
            
            % set up some vertices for the bottom (invisible) line
            \vertex [below=of a1] (b1);
            \vertex [right=of b1] (b2);
            \vertex [right=of b2] (b3);
            
            % \node[crossed dot, minimum size=6pt] (b11) at ($(b2)!\twoVertexDistance!(b1)$);
            \vertex (b12) at ($(b2)!\twoVertexDistance!(b3)$);
            
            \vertex (c1) at ($(b2)!0.33!(a2)$);
            \vertex (c2) at ($(a2)!0.33!(b2)$);

            \diagram* {
            (a2) -- [\varphistyle, edge label=$\varphi$, near end] (c1),
            % (a11) -- [\varphistyle] (c2),
            % (a12) -- [\varphistyle] (c2),
            % (c2) -- [\sigmastyle, edge label=$\sigma$] (c1),
            % (c1) -- [\varphistyle] (b11),
            % (c1) -- [\varphistyle, edge label=$\varphi$, near end] (b12),
            (a1) -- [thick] (a3) % worldline
            };
            
            \vertex [above=0.4em of a2] {$M$};
            \end{feynman}
        \end{tikzpicture}
        \begin{tikzpicture}[thick, scale = 1.1, every node/.style={transform shape}]
            \begin{feynman}
            % set up the worldline
            \vertex (a1);
            \vertex [right=of a1] (a2);
            \vertex [right=of a2] (a3);
            
            \vertex (a11) at ($(a2)!\twoVertexDistance!(a1)$);
            \vertex (a12) at ($(a2)!\twoVertexDistance!(a3)$);
            
            % set up some vertices for the bottom (invisible) line
            \vertex [below=of a1] (b1);
            \vertex [right=of b1] (b2);
            \vertex [right=of b2] (b3);
            
            % \node[crossed dot, minimum size=6pt] (b11) at ($(b2)!\twoVertexDistance!(b1)$);
            \vertex (b12) at ($(b2)!\twoVertexDistance!(b3)$);
            
            \vertex (c1) at ($(b2)!0.33!(a2)$);
            \vertex (c2) at ($(a2)!0.33!(b2)$);

            \diagram* {
            (a2) -- [\phistyle, edge label=$\phi$, near end] (c1),
            % (a11) -- [\varphistyle] (c2),
            % (a12) -- [\varphistyle] (c2),
            % (c2) -- [\sigmastyle, edge label=$\sigma$] (c1),
            % (c1) -- [\varphistyle] (b11),
            % (c1) -- [\varphistyle, edge label=$\varphi$, near end] (b12),
            (a1) -- [thick] (a3) % worldline
            };
            
            \vertex [above=0.4em of a2] {$Q$};
            \end{feynman}
        \end{tikzpicture}
    \vspace*{3mm}
    \subcaption{Leading-order contributions to the background metric (left) and scalar field (right) by coupling to the mass and scalar charge of the object, respectively.}
    \label{fig::eft_background_LO_diagrams}
    \end{subfigure}\vspace*{3mm}
    \begin{subfigure}[b]{\textwidth}
        \begin{tikzpicture}[thick, scale = 1.1, every node/.style={transform shape}]
            \begin{feynman}
            % set up the worldline
            \vertex (a1);
            \vertex [right=of a1] (a2);
            \vertex [right=of a2] (a3);

            \vertex (a11) at ($(a2)!\insertionDistance!(a1)$);
            \vertex (a12) at ($(a2)!\insertionDistance!(a3)$);
            
            % set up some vertices for the bottom (invisible) line
            \vertex [below=of a1] (b1);
            \vertex [right=of b1] (b2);
            \vertex [right=of b2] (b3);
            
            % \node[crossed dot, minimum size=6pt] (b11) at ($(b2)!\twoVertexDistance!(b1)$);
            \vertex (b12) at ($(b2)!\insertionDistance!(b3)$);
            \vertex (b11) at ($(b2)!\insertionDistance!(b1)$);
            
            \vertex (c1) at ($(b2)!0.33!(a2)$);
            \vertex (c2) at ($(a2)!0.33!(b2)$);

            \node[crossed dot, minimum size=6pt] (c11) at ($(b11)!0.33!(a11)$);
            \vertex (c12) at ($(b12)!0.33!(a12)$);

            \diagram* {
            (c11) -- [\varphistyle] (a2),
            (a2) -- [\varphistyle] (c12),
            % (a11) -- [\varphistyle] (c2),
            % (a12) -- [\varphistyle] (c2),
            % (c2) -- [\sigmastyle, edge label=$\sigma$] (c1),
            % (c1) -- [\varphistyle] (b11),
            % (c1) -- [\varphistyle, edge label=$\varphi$, near end] (b12),
            (a1) -- [thick] (a2) -- [thick] (a3) % worldline
            };
            
            \vertex [above=0.4em of a2] {$\lambda_{HH}$};
            
            \end{feynman}
        \end{tikzpicture}
        \begin{tikzpicture}[thick, scale = 1.1, every node/.style={transform shape}]
            \begin{feynman}
            % set up the worldline
            \vertex (a1);
            \vertex [right=of a1] (a2);
            \vertex [right=of a2] (a3);

            \vertex (a11) at ($(a2)!\insertionDistance!(a1)$);
            \vertex (a12) at ($(a2)!\insertionDistance!(a3)$);
            
            % set up some vertices for the bottom (invisible) line
            \vertex [below=of a1] (b1);
            \vertex [right=of b1] (b2);
            \vertex [right=of b2] (b3);
            
            % \node[crossed dot, minimum size=6pt] (b11) at ($(b2)!\twoVertexDistance!(b1)$);
            \vertex (b12) at ($(b2)!\insertionDistance!(b3)$);
            \vertex (b11) at ($(b2)!\insertionDistance!(b1)$);
            
            \vertex (c1) at ($(b2)!0.33!(a2)$);
            \vertex (c2) at ($(a2)!0.33!(b2)$);

            \node[crossed dot, minimum size=6pt] (c11) at ($(b11)!0.33!(a11)$);
            \vertex (c12) at ($(b12)!0.33!(a12)$);

            \diagram* {
            (c11) -- [\phistyle] (a2),
            (a2) -- [\varphistyle] (c12),
            % (a11) -- [\varphistyle] (c2),
            % (a12) -- [\varphistyle] (c2),
            % (c2) -- [\sigmastyle, edge label=$\sigma$] (c1),
            % (c1) -- [\varphistyle] (b11),
            % (c1) -- [\varphistyle, edge label=$\varphi$, near end] (b12),
            (a1) -- [thick] (a2) -- [thick] (a3) % worldline
            };

            \vertex [above=0.4em of a2] {$\lambda_{H\phi}$};
            
            \end{feynman}
        \end{tikzpicture}
        
        \begin{tikzpicture}[thick, scale = 1.1, every node/.style={transform shape}]
            \begin{feynman}
            % set up the worldline
            \vertex (a1);
            \vertex [right=of a1] (a2);
            \vertex [right=of a2] (a3);

            \vertex (a11) at ($(a2)!\insertionDistance!(a1)$);
            \vertex (a12) at ($(a2)!\insertionDistance!(a3)$);
            
            % set up some vertices for the bottom (invisible) line
            \vertex [below=of a1] (b1);
            \vertex [right=of b1] (b2);
            \vertex [right=of b2] (b3);
            
            % \node[crossed dot, minimum size=6pt] (b11) at ($(b2)!\twoVertexDistance!(b1)$);
            \vertex (b12) at ($(b2)!\insertionDistance!(b3)$);
            \vertex (b11) at ($(b2)!\insertionDistance!(b1)$);
            
            \vertex (c1) at ($(b2)!0.33!(a2)$);
            \vertex (c2) at ($(a2)!0.33!(b2)$);

            \node[crossed dot, minimum size=6pt] (c11) at ($(b11)!0.33!(a11)$);
            \vertex (c12) at ($(b12)!0.33!(a12)$);

            \diagram* {
            (c11) -- [\varphistyle] (a2),
            (a2) -- [\phistyle] (c12),
            % (a11) -- [\varphistyle] (c2),
            % (a12) -- [\varphistyle] (c2),
            % (c2) -- [\sigmastyle, edge label=$\sigma$] (c1),
            % (c1) -- [\varphistyle] (b11),
            % (c1) -- [\varphistyle, edge label=$\varphi$, near end] (b12),
            (a1) -- [thick] (a2) -- [thick] (a3) % worldline
            };

            \vertex [above=0.4em of a2] {$\lambda_{H\phi}$};
            
            \end{feynman}
        \end{tikzpicture}
        \begin{tikzpicture}[thick, scale = 1.1, every node/.style={transform shape}]
            \begin{feynman}
            % set up the worldline
            \vertex (a1);
            \vertex [right=of a1] (a2);
            \vertex [right=of a2] (a3);

            \vertex (a11) at ($(a2)!\insertionDistance!(a1)$);
            \vertex (a12) at ($(a2)!\insertionDistance!(a3)$);
            
            % set up some vertices for the bottom (invisible) line
            \vertex [below=of a1] (b1);
            \vertex [right=of b1] (b2);
            \vertex [right=of b2] (b3);
            
            % \node[crossed dot, minimum size=6pt] (b11) at ($(b2)!\twoVertexDistance!(b1)$);
            \vertex (b12) at ($(b2)!\insertionDistance!(b3)$);
            \vertex (b11) at ($(b2)!\insertionDistance!(b1)$);
            
            \vertex (c1) at ($(b2)!0.33!(a2)$);
            \vertex (c2) at ($(a2)!0.33!(b2)$);

            \node[crossed dot, minimum size=6pt] (c11) at ($(b11)!0.33!(a11)$);
            \vertex (c12) at ($(b12)!0.33!(a12)$);

            \diagram* {
            (c11) -- [\phistyle] (a2),
            (a2) -- [\phistyle] (c12),
            % (a11) -- [\varphistyle] (c2),
            % (a12) -- [\varphistyle] (c2),
            % (c2) -- [\sigmastyle, edge label=$\sigma$] (c1),
            % (c1) -- [\varphistyle] (b11),
            % (c1) -- [\varphistyle, edge label=$\varphi$, near end] (b12),
            (a1) -- [thick] (a2) -- [thick] (a3) % worldline
            };

            \vertex [above=0.4em of a2] {$\lambda_{\phi\phi}$};
            
            \end{feynman}
        \end{tikzpicture}
    \vspace*{3mm}
    \subcaption{Leading-order contributions of 
    tidal Love numbers to the asymptotic expansion of the perturbation fields. 
    The crossed circle denotes the external field.}\label{fig::tidalLOcontribution}
    \end{subfigure}\vspace*{5mm}
    \begin{subfigure}[b]{\textwidth}
        \begin{tikzpicture}[thick, scale = 1.1, every node/.style={transform shape}]
            \begin{feynman}[every blob={/tikz/fill=gray!30,/tikz/inner sep=2pt}]
            % set up the worldline
            \vertex (a1);
            \vertex [right=of a1] (a2);
            \vertex [right=of a2] (a3);

            \vertex (a11) at ($(a2)!\insertionDistance!(a1)$);
            \vertex (a12) at ($(a2)!\insertionDistance!(a3)$);
            
            % set up some vertices for the bottom (invisible) line
            \vertex [below=of a1] (b1);
            \vertex [right=of b1] (b2);
            \vertex [right=of b2] (b3);
            
            % \node[crossed dot, minimum size=6pt] (b11) at ($(b2)!\twoVertexDistance!(b1)$);
            \vertex (b12) at ($(b2)!\insertionDistance!(b3)$);
            \vertex (b11) at ($(b2)!\insertionDistance!(b1)$);
            
            \vertex (c1) at ($(b2)!0.33!(a2)$);
            \vertex (c2) at ($(a2)!0.33!(b2)$);

            \node[crossed dot, minimum size=6pt] (c11) at ($(b11)!0.15!(a11)$);
            \vertex (c12) at ($(b12)!0.15!(a12)$);
            
            \vertex (c111) at ($(c11)!0.5!(a2)$);
            \vertex (c222) at ($(a2)!0.5!(c12)$);

            \vertex (d1) at ($(a11)!0.45!(b11)$);
            \vertex (d2) at ($(a12)!0.45!(b12)$);

            \vertex (e1) at ($(a11)!0.5!(a2)$);
            \vertex (e2) at ($(a2)!0.5!(a12)$);
            
            \vertex [below=of e1] (f1);
            \vertex [below=of e2] (f2);

            \vertex (g1) at ($(e1)!0.45!(f1)$);
            \vertex (g2) at ($(e2)!0.45!(f2)$);
            
            \diagram* {
            (c11) -- [\phistyle] (c111),
            (c222) -- [\phistyle] (c12),
            (a2) -- [\phistyle] (c2),
            (a11) -- [\phistyle] (d1),
            (a12) -- [\phistyle] (d2),
            (e1) -- [\phistyle] (g1),
            (e2) -- [\phistyle] (g2),
            % (a11) -- [\varphistyle] (c2),
            % (a12) -- [\varphistyle] (c2),
            % (c2) -- [\sigmastyle, edge label=$\sigma$] (c1),
            % (c1) -- [\varphistyle] (b11),
            % (c1) -- [\varphistyle, edge label=$\varphi$, near end] (b12),
            (a1) -- [thick] (a2) -- [thick] (a3) % worldline
            };
            
            \vertex[blob,shape=ellipse,minimum height=5mm,minimum width=17mm] at ($(a2)!0.4!(b2)$) (V) {};

            \end{feynman}
        \end{tikzpicture}
        \begin{tikzpicture}[thick, scale = 1.1, every node/.style={transform shape}]
            \begin{feynman}[every blob={/tikz/fill=gray!30,/tikz/inner sep=2pt}]
            % set up the worldline
            \vertex (a1);
            \vertex [right=of a1] (a2);
            \vertex [right=of a2] (a3);

            \vertex (a11) at ($(a2)!\insertionDistance!(a1)$);
            \vertex (a12) at ($(a2)!\insertionDistance!(a3)$);
            
            % set up some vertices for the bottom (invisible) line
            \vertex [below=of a1] (b1);
            \vertex [right=of b1] (b2);
            \vertex [right=of b2] (b3);
            
            % \node[crossed dot, minimum size=6pt] (b11) at ($(b2)!\twoVertexDistance!(b1)$);
            \vertex (b12) at ($(b2)!\insertionDistance!(b3)$);
            \vertex (b11) at ($(b2)!\insertionDistance!(b1)$);
            
            \vertex (c1) at ($(b2)!0.33!(a2)$);
            \vertex (c2) at ($(a2)!0.33!(b2)$);

            \node[crossed dot, minimum size=6pt] (c11) at ($(b11)!0.15!(a11)$);
            \vertex (c12) at ($(b12)!0.15!(a12)$);
            
            \vertex (c111) at ($(c11)!0.5!(a2)$);
            \vertex (c222) at ($(a2)!0.5!(c12)$);

            \vertex (d1) at ($(a11)!0.45!(b11)$);
            \vertex (d2) at ($(a12)!0.45!(b12)$);

            \vertex (e1) at ($(a11)!0.5!(a2)$);
            \vertex (e2) at ($(a2)!0.5!(a12)$);
            
            \vertex [below=of e1] (f1);
            \vertex [below=of e2] (f2);

            \vertex (g1) at ($(e1)!0.45!(f1)$);
            \vertex (g2) at ($(e2)!0.45!(f2)$);
            
            \diagram* {
            (c11) -- [\phistyle] (c111), %external incoming perturbation
            (c222) -- [\varphistyle] (c12), %outgoing field
            (a2) -- [\phistyle] (c2),
            (a11) -- [\phistyle] (d1),
            (a12) -- [\phistyle] (d2),
            (e1) -- [\phistyle] (g1),
            (e2) -- [\phistyle] (g2),
            % (a11) -- [\varphistyle] (c2),
            % (a12) -- [\varphistyle] (c2),
            % (c2) -- [\sigmastyle, edge label=$\sigma$] (c1),
            % (c1) -- [\varphistyle] (b11),
            % (c1) -- [\varphistyle, edge label=$\varphi$, near end] (b12),
            (a1) -- [thick] (a2) -- [thick] (a3) % worldline
            };
            
            \vertex[blob,shape=ellipse,minimum height=5mm,minimum width=17mm] at ($(a2)!0.4!(b2)$) (V) {};

            \end{feynman}
        \end{tikzpicture}
    \vspace*{3mm}
\subcaption{Two exemplary diagrams that would also contribute to the asymptotic perturbation field with the same radial dependency as the Love numbers, i.e., the diagrams 
in Fig.~(b). The gray shaded area is a placeholder for any allowed interactions between the perturbations and the five couplings to the worldline.}
\label{fig::higher_order_scattering}
\end{subfigure}
\caption{Various diagrams that contribute to the 
asymptotic expansion of the background and 
perturbation fields.}
\label{fig:enter-label}
\end{figure}

Using Eq.~(\ref{Spp1}) together with the bulk action in Eq.~\eqref{Sbulk3}, we could now calculate the contributions of $M$ and $Q$ to the asymptotic expansion of 
$\hat{\varphi}$ and $\hat{\phi}$. 
For example, by considering the diagram depicted in the right panel of Fig.~\ref{fig::eft_background_LO_diagrams},
we can compute the leading-order contribution 
of the scalar charge to $\hat{\phi}$. 
We find that
\begin{align}
\langle \hat{\phi}(x) \rangle_\mathrm{Fig.\,\ref{fig::eft_background_LO_diagrams}} &= -i \frac{Q}{\mpl} \int_\mathbb{R} \rd t \, \langle \hat{\phi}(x) \hat{\phi}(0) \rangle \notag\\
&= -\frac{Q}{\mpl} \int_\mathbb{R} \rd t \int_{\mathbb{R}^4} \frac{\rd^4 k}{(2 \pi)^4} \frac{e^{ik x}}{k^2 + i \epsilon} \notag\\
&= -\frac{Q}{\mpl} \int_{\mathbb{R}^3} 
\frac{\rd^3 {\bm k}}{(2 \pi)^3} \frac{e^{i{\bm k}\cdot {\bm x}}}{{\bm k}^2 
+ i \epsilon}\,,
\end{align}
where, in the last equality, we first solved the 
$t$ integral to obtain a factor of 
$2\pi \delta(k_0)$ and subsequently 
performed the $k_0$ integral. 
Changing to spherical coordinates with the 
radial distance $\hat{r}$, solving the angular 
integrals, and extending the integration limits 
to the entire real line, we further obtain
\begin{align}
\langle \hat{\phi}(x) \rangle_\mathrm{Fig.\,\ref{fig::eft_background_LO_diagrams}} 
&= \frac{i\,Q}{4 \pi^2 \mpl \hat{r}} \int_\mathbb{R} 
\rd \hat{k} \, \hat{k} \frac{e^{i\hat{k} \hat{r}}}
{\hat{k}^2 + i \epsilon} \,.
\end{align}
Then, we leverage the residue theorem by closing the integration contour in the upper complex plane. Picking up the residual at $\hat{k} = \sqrt{\epsilon}\, e^{i3\pi / 4}$ and 
taking the limit $\epsilon \rightarrow 0$, 
we arrive at
\begin{align}
\langle \hat{\phi}(x) \rangle_\mathrm{Fig.\,\ref{fig::eft_background_LO_diagrams}} = -\frac{Q}{4 \pi \mpl \hat{r}}\,.
\label{phiexpe}
\end{align}
As we will see later in Sec.~\ref{sec:minimally_matching}, 
this allows 
us to identify the scalar charge $\phi_1$ in Eq.~\eqref{phiLexpan} to be 
$\phi_1 = -Q/(4 \pi \mpl)$.\footnote{This calculation has been done in isotropic coordinates and is thus not directly comparable to the expansion in Eq.~\eqref{eq::horndeski_bg_equations}, which uses Schwarzschild coordinates. However, we will see in Sec.~\ref{sec:minimally_matching} (Eq.~\eqref{eq:iso-to-Schw} specifically) that we can identify the leading-order terms in both coordinate systems to be equal.}

Likewise, we can calculate the leading-order contribution 
to the gravitational perturbation $\hat{\varphi}$. 
The left panel of Fig.~\ref{fig::eft_background_LO_diagrams} 
results in
\begin{align}
\langle \hat{\varphi}(x) \rangle_\mathrm{Fig.\,\ref{fig::eft_background_LO_diagrams}} = -\frac{\sqrt{2} M}{8 \pi \mpl \hat{r}}\,.
\label{varphiexpe}
\end{align}
Inserting this into Eq.~\eqref{eq::eft_metric_decomposition} and expanding up to linear order, we find that
\begin{equation}
\langle \hat{g}_{00}(x) \rangle_\mathrm{Fig.\,\ref{fig::eft_background_LO_diagrams}} = -1 + \frac{2 M}{8 \pi \mpl^2 \hat{r}} = -1 + \frac{2 M}{\hat{r}}\,.
\end{equation}
This corresponds to the leading-order expansion of $-f$ 
given in Eq.~(\ref{eq::minimally_coupled_bg_expansion}), 
with the transformation to the Schwarzschild coordinate 
(which will be discussed in Sec.~\ref{sec:minimally_matching}).
We could similarly now construct higher-order diagrams that would allow us to calculate the same expansion as was presented above in Eq.~\eqref{eq::minimally_coupled_bg_expansion}. 
We will next turn to the treatment of perturbations 
in the EFT framework.

%%%%%%%%%%%%%%%%%%%%%%%%%%%%%%%%
%%%%%%%%%%%%%%%%%%%%%%%%%%%%%%%%%%%%%%%%%%%%%

\paragraph*{\bf Perturbations.}

Let us discuss the tidal perturbation sector.
When using the metric decomposition in 
Eq.~\eqref{eq::eft_metric_decomposition}, 
the relevant parts of the tidal PP action yield
\begin{align}
\label{Sppl}
{\cal S}_\mathrm{pp}^\lambda = \int & 
\rd t \, \bigg[ \frac{\lambda_{hh}}{8 \mpl^2} \partial_i \partial_j \hat{\varphi} \partial^i \partial^j \hat{\varphi}
+\frac{\lambda_{h \phi}}{2 \sqrt{2} \mpl^2} \partial_i \partial_j \hat{\varphi} \partial^i \partial^j \hat{\phi} \nonumber\\
&~~~+\frac{\lambda_{\phi\phi}}{4 \mpl^2} \partial_i \partial_j \hat{\phi} \partial^i \partial^j \hat{\phi} \bigg] \,, 
\end{align}
where we have only kept the leading-order contributions after expanding $\hat{\sigma}$ and $\hat{\varphi}$ around zero. 
By the same reasoning given around Eq.~\eqref{Spp1} for the truncation of the background PP action, only terms that are quadratic in the second derivatives of fields can contribute to the classical observables that 
arise due to an externally applied quadrupolar perturbation.
To study the behavior of quadrupolar perturbations, we now follow Ref.~\cite{Kol:2011vg} and introduce perturbations into $\hat{\varphi}$ and $\hat{\phi}$, such that
\begin{alignat}{2}
\hat{\varphi} & \rightarrow \hat{\varphi} + \delta \hat{\varphi}\,,\qquad &\delta \hat{\varphi} {}={}& \hat{\varphi}_{ij} \hat{x}^i \hat{x}^j\,, \\
\hat{\phi} & \rightarrow \hat{\phi} 
+ \delta \hat{\phi}\,,\qquad &\delta \hat{\phi} 
{}={}& \hat{\phi}_{ij} \hat{x}^i \hat{x}^j\,, 
\end{alignat}
where $\hat{\varphi}_{ij}$ and $\hat{\phi}_{ij}$ 
are traceless and symmetric tensors. Since in Eq.~\eqref{Sppl} we observe that -- at leading-order -- only $\hat{\phi}$ and $\hat{\varphi}$ contribute to the externally sourced field, we do not consider perturbations to $\hat{\sigma}$ at this stage. 
We will later confirm that this is indeed 
unnecessary, as all relevant information is encoded in $\delta \hat{\varphi}$ and $\delta \hat{\phi}$.
To investigate the impact of tidal Love numbers on the 
asymptotic behavior of perturbations, we insert the above 
split into the tidal part of the PP action and only keep 
terms linear in perturbations, such that 
the relevant part of the tidal action becomes
\begin{align}
\hspace{-0.2cm}
{\cal S}_\mathrm{pp}^\lambda = &\int \rd t \, \bigg[ \frac{\lambda_{hh}}{2 \mpl^2} \hat{\varphi}_{ij} \partial^i \partial^j \hat{\varphi} + \frac{\lambda_{h \phi}}{\sqrt{2} \mpl^2} \hat{\varphi}_{ij} \partial^i \partial^j \hat{\phi} \notag \\
\hspace{-0.2cm}
&~~~~~~~
+ \frac{\lambda_{h \phi}}{\sqrt{2} \mpl^2} \hat{\phi}_{ij} \partial^i \partial^j \hat{\varphi} 
+ \frac{\lambda_{\phi\phi}}{\mpl^2} \hat{\phi}_{ij} \partial^i \partial^j \hat{\phi} \bigg].
\label{Spp}
\end{align}
With these interactions, it is now straightforward to 
calculate the contribution of tidal Love numbers to the asymptotic expansion of $\hat{\varphi}$ and $\hat{\phi}$. 
The relevant diagrams are shown in Fig.~\ref{fig::tidalLOcontribution} 
and evaluate to
\begin{align}
\label{eq::LO_tidal_contribution}
\delta \hat{\varphi} &\supset \left( \frac{3 \lambda_{hh}}{8 \pi \mpl^2} \hat{\varphi}_{ij} + \frac{3 \lambda_{h\phi}}{4 \sqrt{2} \pi \mpl^2} \hat{\phi}_{ij} \right) 
\frac{\hat{x}^i \hat{x}^j}{\hat{r}^5}, \\
\delta \hat{\phi} &\supset \left( \frac{3 \lambda_{h\phi}}{4 \sqrt{2} \pi \mpl^2} \hat{\varphi}_{ij} + \frac{3 \lambda_{\phi\phi}}{4 \pi \mpl^2} \hat{\phi}_{ij} \right) \frac{\hat{x}^i \hat{x}^j}{\hat{r}^5}\,.
\label{eq::LO_tidal_contribution2}
\end{align}
Next, we want to project the above terms onto the 
$l=2$ spherical harmonic $Y_{2,m}(\theta, \varphi)$. 
Following Ref.~\cite{Thorne:1980ru}, we rewrite $\hat{\varphi}_{ij}$ and $\hat{\phi}_{ij}$, as
\begin{align}
\begin{split}
\hat{\varphi}_{ij}=\sum_{m=-2}^{2} \hat{\varphi}_m \mathcal{Y}_{ij}^{2,m}, \qquad
\hat{\phi}_{ij}= \sum_{m=-2}^{2} \hat{\phi}_m \mathcal{Y}_{ij}^{2,m},
\end{split}
\end{align}
where 
\begin{align}
Y_{2,m}(\theta, \varphi) = 
\mathcal{Y}_{ij}^{2,m} 
\frac{\hat{x}^i \hat{x}^j}{\hat{r}^2}\,.
\end{align}
With this decomposition, we can now identify
\begin{align}
\hat{\varphi}_{ij} \hat{x}^i \hat{x}^j &= 
\sum_{m=-2}^{2} \hat{\varphi}_m \hat{r}^2 
Y_{2,m}(\theta, \varphi)\,, \\
\hat{\phi}_{ij} \hat{x}^i \hat{x}^j &= 
\sum_{m=-2}^{2} \hat{\phi}_m \hat{r}^2 
Y_{2,m}(\theta, \varphi)\,.
\end{align}
Given that Eqs.\eqref{perH0} and\eqref{perdp} are independent of $m$, 
we will, without loss of generality, 
restrict our analysis to the $m=0$ mode.
Accordingly, we relabel 
$\hat{\varphi}_{m=0}$ as $\delta \hat{\varphi}_{-2}$ 
and $\hat{\phi}_{m=0}$ as $\delta \hat{\phi}_{-2}$.
Then, terms in Eqs.~\eqref{eq::LO_tidal_contribution} 
and \eqref{eq::LO_tidal_contribution2} yield 
\begin{align}
\label{eq::tidal_LO_spherical}
\delta \hat{\varphi} &\supset \left( \frac{3 \lambda_{hh}}{8 \pi \mpl^2} \delta \hat{\varphi}_{-2} + \frac{3 \lambda_{h\phi}}{4 \sqrt{2} \pi \mpl^2} \delta \hat{\phi}_{-2} \right) 
\frac{1}{\hat{r}^3}, \\
\label{eq::tidal_LO_spherical2}
\delta \hat \phi &\supset \left( \frac{3 \lambda_{h\phi}}{4 \sqrt{2} \pi \mpl^2} \delta \hat{\varphi}_{-2} + \frac{3 \lambda_{\phi\phi}}{4 \pi \mpl^2} \delta \hat{\phi}_{-2} \right) \frac{1}{\hat{r}^3}\,.
\end{align}
Here and in the following, we absorb 
the factor $Y_{2,0}$ into 
the definitions of $\delta \hat{\varphi}_{-2}$ and 
$\delta \hat{\phi}_{-2}$.
We thus successfully connected tidal Love numbers 
to the asymptotic expansion in isotropic coordinates. 
In Sec.~\ref{sec:minimally_matching}, we will explain how this result 
is related to the asymptotic expansion in Schwarzschild 
coordinates. However, we first highlight 
a problem usually overlooked when extracting tidal Love numbers 
in theories beyond GR (see Ref.~\cite{Katagiri:2024fpn} 
for a recent related work).

\paragraph*{\bf Contamination.}
We have calculated the contribution of tidal Love numbers 
to the $1/\hat{r}^3$ term in the asymptotic 
expansion of the perturbation fields. 
So far, however, we have not checked whether 
there are other diagrams that do not contain the Love numbers 
but still contribute to the same order in the expansion. 
In other words, there might be additional terms proportional to $1/\hat{r}^3$ 
in the expansion of $\delta \hat{\varphi}$ and 
$\delta \hat{\phi}$ that do not contain the Love numbers. 
From dimensional reasoning, we can conclude that such a diagram would need five couplings to the worldline 
(to either the mass or scalar charge) and would thus be of the form that is shown in Fig.~\ref{fig::higher_order_scattering}.

In the following, we will see that such diagrams 
cannot contribute to the $1/\hat{r}^3$ term when 
using isotropic coordinates. However, we will see 
that, after transforming our result to Schwarzschild coordinates, other terms in the expansion will transform into a $1/\hat{r}^3$ contribution, which we will explicitly compute. We stress that, while this ambiguity was already discussed 
in pure GR \cite{Kol:2011vg}, it has not yet been considered 
in scalar-tensor theories~\cite{Pani:2014jra,Brown:2022kbw,Creci:2023cfx}. 

\paragraph*{\bf A neat symmetry.}
As was done in Ref.~\cite{Kol:2011vg}, we could now expand the total action and collect all relevant diagrams that would contribute at a certain order in the asymptotic expansion and then directly evaluate their effect. 
However, it turns out that we can sufficiently constrain the expansion 
without calculating a single diagram. 
For this purpose, let us consider the 
action given by 
\begin{equation}
{\cal S} = {\cal S}_\mathrm{bulk} 
+{\cal S}_\mathrm{pp}^\mathrm{(bg)}\,,
\label{Sta}
\end{equation}
where ${\cal S}_\mathrm{bulk}$ and 
${\cal S}_\mathrm{pp}^\mathrm{(bg)}$ are given, 
respectively, by Eqs.~(\ref{Sbulk3}) and (\ref{Spp1}). 
This represents the action with vanishing Love numbers. 
Then, we find that Eq.~(\ref{Sta}) is invariant under the inversion\footnote{In fact, there are additional symmetries present in this action. For example, inverting only the $\hat{\varphi}$ sector, such that $\hat{\varphi} \rightarrow - \hat{\varphi}$ together with $Q \rightarrow -Q$ and $\lambda_{h\phi} \rightarrow - \lambda_{h \phi}$, or also an exchange symmetry $\hat{\varphi} \leftrightarrow \hat{\phi}$ with $M \leftrightarrow \sqrt{2}\, Q$. While these symmetries impose additional structures on $\delta \hat{\varphi}$ and $\delta \hat{\phi}$, these additional structures 
are unnecessary for the analysis we perform in the subsequent sections.}
\ba
\label{eq::minimal_coupled_symmetry}
& &
\hat{\varphi} \rightarrow - \hat{\varphi}\,,\qquad 
\hat{\phi} \rightarrow -\hat{\phi}\,, \nonumber \\
& &
Q \rightarrow -Q\,,\qquad 
M \rightarrow -M\,.
\ea
Because of this symmetry, we can infer that, in the expansion of $\delta \hat{\varphi}$ and $\delta \hat{\phi}$, only terms where the combined 
powers of $M$ and $Q$ are even can appear. In other words, only diagrams with an even number of couplings to the worldline contribute to the asymptotic expansion of the perturbation fields. From dimensional analysis, we can further conclude that a diagram with an even number of worldline couplings needs to result in a term that carries an even power of the radial coordinate, and thus, only terms of even power in the radial coordinate are allowed in the asymptotic expansion of $\delta \hat{\varphi}$ and $\delta \hat{\phi}$. 
More explicitly, for an object that has vanishing Love numbers, $\delta \hat{\varphi}$ and $\delta \hat{\phi}$ need to behave as
\begin{align}
\label{eq::eft_asymp_proto_expansion}
\delta \hat{\varphi} = \sum_{i=-1}^\infty \delta \hat{\varphi}_{2i}\, \hat{r}^{-2i}\,, \qquad
\delta \hat{\phi} = \sum_{i=-1}^\infty \delta \hat{\phi}_{2i}\, \hat{r}^{-2i}\,,
\end{align}
with the $\delta \hat{\varphi}_{2i}$ and $\delta \hat{\phi}_{2i}$ being real 
coefficients.\footnote{Note that the tidal part of 
the PP action, Eq.~\eqref{eq::pp_action_tidal_part}, 
is also invariant under the symmetry in Eq.~\eqref{eq::minimal_coupled_symmetry}. 
While it is not relevant for further considerations, 
we note that one can deduce from a dimensional analysis that the tidal part of the PP action would result in terms that contain an odd power of $\hat{r}$, 
starting at $\mathcal{O}(\hat{r}^{-3})$.}

We will see in Sec.~\ref{sec:minimally_matching} that this information alone is enough to calculate the contribution to the $1/r^3$ term in Schwarzschild coordinates that is independent of the Love numbers.

\subsection{Matching 
\label{sec:minimally_matching}} 

While the expansion in Sec.~\ref{sec:minimally_asymp_expansion} was done in Schwarzschild coordinates for the background fields and the Regge-Wheeler gauge for the perturbations, the expansion in Sec.~\ref{sec:minimally_eft} utilized isotropic coordinates for the background fields and 
did not specify a gauge for the perturbed fields. 
To relate the two obtained asymptotic expansions, 
we could now either explicitly construct the necessary 
coordinate transformations or consider 
coordinate-independent quantities. 
We find it most convenient to transform the EFT expansion 
to Schwarzschild coordinates and then relate the asymptotic 
expansion of the perturbed fields by using 
gauge-invariant quantities.

First, we transform the spatial part of the 
metric decomposition in Eq.~\eqref{eq::eft_metric_decomposition} 
to spherical coordinates, thus arriving at 
a line element of the form
\be
\rd s^2=-\hat{f}(\hat{r}) \rd t^2+
\hat{h}^{-1}(\hat{r}) \left( \rd \hat{r}^2
+\hat{r}^2 \rd \Omega^2 \right)\,,
\label{isoco}
\ee
where $\hat{f}$ and $\hat{h}$ are functions of $\hat{r}$. 
Compared to the Schwarzschild coordinate 
(\ref{metric}), the metric components in 
the isotropic coordinate (\ref{isoco}) 
are related to each other, as 
$f(r)=\hat{f}(\hat{r})$, $h^{-1}(r) \rd r^2
=\hat{h}^{-1}(\hat{r}) \rd \hat{r}^2$, and 
$r^2=\hat{r}^2 \hat{h}^{-1}(\hat{r})$. 
From the latter two equations, we obtain 
$\rd \hat{r}=\hat{r}/(r\sqrt{h}) \rd r$, 
where we have chosen the sign $\rd \hat{r}>0$ 
for $\rd r>0$. This is integrated to give  
$\ln \hat{r}=\int {\rm d}r/(r\sqrt{h})$. 
We choose the integration constant 
to have the leading-order behavior 
$\hat{r} \simeq r$ at spatial infinity.
Using the large-distance solution (\ref{hlexpan}) 
of $h$, it follows that\footnote{Because of the unspecified integration constant, there is a potential ambiguity in this definition. To avoid this ambiguity, one might be tempted to replace the indefinite integral with a definite one and set one of the boundaries to either zero or infinity. However, the integral diverges in both of these limits. Instead, we fix the integration constant by demanding that $\hat{r} \rightarrow r$ 
at spatial infinity, which is 
a necessary condition since the metric would 
otherwise not reduce to the Minkowski metric at $r \rightarrow \infty$.}
\begin{align}
\label{eq:iso-to-Schw}
\hat{r}=r - M + \frac{1}{r} 
\left( \frac{\phi_1^2}{8 \mpl^2} - \frac{M^2}{4} 
\right) + \mathcal{O}(r^{-2})\,,
\end{align}
which is valid for large $r$.
Equipped with this coordinate transformation, 
we are now ready to relate the background fields. 
At the background level, the $(0,0)$ component of the 
metric is given by 
\begin{align}
\label{eq::schwarzschild_isotropic_radius}
f(r) = \hat{f}(\hat{r})= e^{\sqrt{2} \hat{\varphi}(\hat{r})/\mpl}\,.
\end{align}

Relating the perturbations is more difficult since we also would need to perform an infinitesimal transformation to the Regge-Wheeler gauge. 
Instead, we will first only identify the $(0,0)$ 
component of the metric perturbations in isotropic coordinates, denoted by $\hat{H}_0$, and then relate it afterward to $H_0$ by using a gauge-invariant quantity. We write the $(0,0)$ component 
of $\hat{g}_{\mu \nu}$ as
\begin{align}
\label{eq::varphi_to_h00}
\hat{g}_{00}=-\hat{f}+\hat{h}_{00}
=-\hat{f} \left( 1-\hat{H}_0 \right)\,.
\end{align}
From Eq.~(\ref{eq::eft_metric_decomposition}), 
we can also express $\hat{g}_{00}$ in the form 
\begin{align}
\label{eq::varphi_to_h002}
\hat{g}_{00} = - e^{\sqrt{2} (\hat{\varphi} 
+ \delta \hat{\varphi}) / \mpl} 
= -\hat{f} e^{\sqrt{2} \delta \hat{\varphi} / \mpl}\,.
\end{align}
Comparing Eq.~(\ref{eq::varphi_to_h00}) with 
Eq.~(\ref{eq::varphi_to_h002}), 
it follows that 
\be
\label{eq::metric_perturb_definition_H0}
1-\hat{H}_0=
e^{\sqrt{2} \delta \hat{\varphi} / \mpl}\,.
\ee
Since we are working at linear order in 
perturbation theory, we can safely drop all of
the nonlinear terms of order $\delta \hat{\varphi}^2$ in Eq.~\eqref{eq::varphi_to_h00}.
From Eq.~\eqref{eq::metric_perturb_definition_H0}, 
we can relate $\hat{H}_0$ with 
$\delta \hat{\varphi}$, as
\begin{align}
\hat{H}_0(\hat{r})= 
-\frac{\sqrt{2}}{\mpl} \delta 
\hat{\varphi}(\hat{r})\,,
\label{hH0}
\end{align}
for linear perturbations.
For the growing-mode part, i.e., 
the $\hat{r}^2$ term, we have 
\begin{equation}
\hat{H}_{0,-2} = -\frac{\sqrt{2}}{\mpl} \delta \hat{\varphi}_{-2}\,.
\end{equation}
Then, Eqs.~(\ref{eq::tidal_LO_spherical}) 
and (\ref{eq::tidal_LO_spherical2}) yield 
\begin{align}
\label{eq::LO_tidal_terms_isotropic_EFT}
\hat{H}_0 &\supset \left( \frac{3 \lambda_{hh}}{8 \pi \mpl^2} \hat{H}_{0,-2} - \frac{3 \lambda_{h\phi}}{4 \pi \mpl^3} \delta \hat{\phi}_{-2} \right) \frac{1}{\hat{r}^3}\,, \\
\delta \hat{\phi} &\supset \left( -\frac{3 \lambda_{h\phi}}{8 \pi \mpl} \hat{H}_{0,-2} + \frac{3 \lambda_{\phi\phi}}{4 \pi \mpl^2} \delta \hat{\phi}_{-2} \right) \frac{1}{\hat{r}^3}\,.
\end{align}

As mentioned above, we would now need to perform an 
infinitesimal transformation to transform $\hat{H}_0$ and 
$\delta \hat{\phi}$ to $H_0$ and $\delta \phi$. 
Instead, we will construct a gauge-invariant quantity 
and use it to relate the perturbed quantities. 
To this end, we consider an infinitesimal coordinate 
transformation of the form $x_\mu \rightarrow x_\mu + \xi_\mu$,
where
\begin{align}
\label{eq::infinitesimal_gauge_transformation}
\xi_t &= \sum_{l,m} \mathcal{T}(t, r) Y_{l m}(\theta, \varphi), \\
\xi_r &= \sum_{l,m} \mathcal{R}(t, r) Y_{l m}(\theta, \varphi), \\
\xi_a &= \sum_{l,m} \Theta(t, r) \nabla_a Y_{l m}(\theta, \varphi)\,,
\label{eq::infinitesimal_gauge_transformation3}
\end{align}
with $a$ either $\theta$ or $\varphi$. 
Then, the perturbations $H_0$, $\delta \phi$, 
$h_0$, and $G$ transform, respectively, as~\cite{Kobayashi:2014wsa,Kase:2021mix}
\begin{align}
H_0 &\rightarrow H_0 + \frac{2}{f} 
\Dot{\mathcal{T}} -\frac{f' h}{f} \mathcal{R}, \\
\delta \phi &\rightarrow \delta \phi - \phi' h \mathcal{R}\,,\\
h_0 &\rightarrow h_0+\mathcal{T}
+\dot{\Theta}\,,\\
G &\rightarrow G+\frac{2}{r^2}\Theta\,.
\end{align}
We can construct the following 
gauge-invariant perturbation
\be
\Psi \equiv H_0-\frac{f'}{f\phi'}\delta \phi
-\frac{2}{f}\dot{h}_0+\frac{r^2}{f} \ddot{G}\,.
\label{Psidef}
\ee
Up to this point, we have included the time dependence of the gauge transformations of the perturbed fields for generality. However, since we are only interested in static perturbations, we omit the time dependence in the following discussion.
Then, the combination (\ref{Psidef}) reduces 
to $\Psi=H_0-f' \delta \phi/(f\phi')$.
Then, among the two different coordinates, 
there is the following relation 
\be
-\frac{\sqrt{2}}{\mpl} \delta 
\hat{\varphi}(\hat{r})-\frac{f'(r)}
{f(r) \phi'(r)}
\delta \hat{\phi} (\hat{r})=
H_0(r)-\frac{f'(r)}
{f(r) \phi'(r)} \delta \phi(r)\,,
\label{eq:gauge-inv}
\ee
where, on the left-hand side, we used 
the properties (\ref{eq::schwarzschild_isotropic_radius}) 
and (\ref{hH0}).
To carry out the matching procedure between 
the two coordinate systems, we find it convenient 
to split the $H_{0,3}$ and $\delta \phi_3$ coefficients 
of Eqs.~\eqref{eq::minimal_asymp_expansion_H0} and \eqref{eq::minimal_asymp_expansion_dphi} into
\begin{align}
H_{0,3} &= H_{0,3}^\lambda + H_{0,3}^0\,, \\
\delta \phi_{3} &= \delta \phi_{3}^\lambda + \delta \phi_{3}^0\,.
\end{align}
Here, the superscript $\lambda$ denotes the tidal part, which only contains the Love numbers, and the superscript $0$ denotes the non-tidal part, which only contains powers of 
$\phi_1$ and and $M$.\footnote{Since $\phi_1$ 
is proportional to $Q$, we will use $\phi_1$ 
and $Q$ synonymously here.}

To determine the non-tidal part first, we consider the limit $\lambda \rightarrow 0$ and use the relation in Eq.~\eqref{eq:gauge-inv} to match Eq.~\eqref{eq::eft_asymp_proto_expansion} with its unknown 
coefficients $\delta \hat{\varphi}_{2i}$ and $\delta \hat{\phi}_{2i}$ 
to the expansions in Eqs.~\eqref{eq::minimal_asymp_expansion_H0} and \eqref{eq::minimal_asymp_expansion_dphi}. 
For this purpose, we substitute 
the relation (\ref{eq:iso-to-Schw}) into 
Eq.~\eqref{eq::eft_asymp_proto_expansion} 
to express the left-hand side of 
Eq.~(\ref{eq:gauge-inv}) in terms of $r$.
We also exploit the large-distance background solutions 
(\ref{eq::minimally_coupled_bg_expansion}) and (\ref{phiLexpan}) 
in Eq.~(\ref{eq:gauge-inv}) and expand both 
sides of the latter equation in the series 
of $r$ about infinity. 
We can then solve it order by order: 
\begin{enumerate}
\item $\mathcal{O}(r^2)$: At this leading order, 
we can solve the equation for $\delta \hat{\varphi}_{-2}$.
\item $\mathcal{O}(r^1)$: The equation at this order is 
automatically satisfied if we use the above solution for $\delta \hat{\varphi}_{-2}$.
\item $\mathcal{O}(r^0)$: We can solve the equation for $\delta \hat{\varphi}_{0}$.
\item $\mathcal{O}(r^{-1})$: The equation at this order is automatically satisfied if we use the above solutions for $\delta \hat{\varphi}_{-2}$ and $\delta \hat{\varphi}_{0}$. 
\item $\mathcal{O}(r^{-2})$:  We solve for  $\delta \hat{\varphi}_{2}$.
\item $\mathcal{O}(r^{-3})$: At this order, the above pattern breaks. 
The coefficients $\delta \hat{\varphi}_{2i}$ and $\delta \hat{\phi}_{2i}$ do not appear after using the solutions at previous orders, but we still find an equation which we can solve for $H_{0,3}$ in terms of $H_{0,-2}$, $\delta \phi_{-2}$, and $\delta \phi_3$. 
\item $\mathcal{O}(r^{-4})$: We solve for $\delta \hat{\varphi}_{4}$.
\item $\mathcal{O}(r^{-5})$: Finally, at this order, we find an equation similar to that at $\mathcal{O}(r^{-3})$, which we can solve for $\delta \phi_{3}$ in terms of $\delta \phi_{-2}$. Using this solution for the equation of 
$H_{0,3}$ derived in process 6 above, 
we find that the dependence on 
$\delta \phi_{-2}$ cancels.
\end{enumerate}
As a result, while the matching does not allow us to infer every coefficient of the EFT expansion in Eq.~\eqref{eq::eft_asymp_proto_expansion}, 
we find that the coefficients $H_{0,3}^0$ and $\delta \phi_3^0$ are 
uniquely determined to be
\ba
H_{0, 3}^0 &=& H_{0, -2} \left( \frac{3 M^3 \phi_1^2}{5 \mpl^2} - \frac{3 M \phi_1^4}{80 \mpl^4} \right)\,, 
\label{Nonti1}
\\
\delta \phi_3^0 &=& \delta \phi_{-2} \left( \frac{3 M^3 \phi_1^2}{5 \mpl^2} - \frac{3 M \phi_1^4}{80 \mpl^4} \right)\,.
\label{Nonti2}
\ea
Hence we find a contribution to the $r^{-3}$ term 
that is independent of the Love numbers.

Next, we move on to relate the tidal 
contributions in both frames. 
For this purpose, we use the property that, 
in the limits $M \rightarrow 0$ and 
$Q \rightarrow 0$, the nonvanishing 
terms in the expansion of $\hat{H}_0$ and 
$\delta \hat{\phi}$ are given by
\begin{align}
\hat{H}_0 &= \hat{H}_{0,-2} \hat{r}^2 
+ \left( \frac{3 \lambda_{hh}}{8 \pi \mpl^2} \hat{H}_{0,-2} - \frac{3 \lambda_{h\phi}}{4 \pi \mpl^3} \delta \hat{\phi}_{-2} \right) \frac{1}{\hat{r}^3}\,, \\
\delta \hat{\phi} &= \delta \hat{\phi}_{-2} \hat{r}^2 + \left( -\frac{3 \lambda_{h\phi}}{8 \pi \mpl} \hat{H}_{0,-2} + \frac{3 \lambda_{\phi\phi}}{4 \pi \mpl^2} \delta \hat{\phi}_{-2} \right) \frac{1}{\hat{r}^3}\,.
\end{align}
Furthermore, the coordinate transformation in 
Eq.~\eqref{eq:iso-to-Schw} reduces to the identity 
in the limits $M \to 0$ and $\phi_1 \to 0$, 
such that $\hat{r} \equiv r$. 
Then, static infinitesimal transformations in Eqs.~\eqref{eq::infinitesimal_gauge_transformation}-\eqref{eq::infinitesimal_gauge_transformation3} leave $\hat{H}_0$ and $\delta \hat{\phi}$ unchanged. Hence, in such limits, we have 
\begin{align}
\hat{H}_0(\hat{r}) = H_0(r) \quad \text{and} 
\quad \delta \hat{\phi}(\hat{r}) = \delta \phi(r),
\end{align}
which allows us to deduce the tidal 
contributions in the Regge-Wheeler gauge as\footnote{Note that these expressions 
are the same as in Eq.~\eqref{eq::LO_tidal_terms_isotropic_EFT},
just with the hats dropped.}
\begin{align}
\label{eq::tidal_part_SS_RW_coords}
H_{0, 3}^\lambda &= \frac{3 \lambda_{hh}}{8 \pi \mpl^2} H_{0,-2} - \frac{3 \lambda_{h\phi}}{4 \pi \mpl^3} \delta \phi_{-2}\,, \\
\delta \phi_3^\lambda &= -\frac{3 \lambda_{h\phi}}{8 \pi \mpl} H_{0,-2} + \frac{3 \lambda_{\phi\phi}}{4 \pi \mpl^2} \delta \phi_{-2} \,.
\label{eq::tidal_part_SS_RW_coords2}
\end{align}
To summarize, we find that the $1/r^3$ terms in the expansion 
of the perturbated fields are uniquely determined to be
\ba
\label{eq::minimal_1o3_coefficients}
H_{0, 3} &=& H_{0, -2} \left( \frac{3 M^3 \phi_1^2}{5 \mpl^2} - \frac{3 M \phi_1^4}{80 \mpl^4} \right) \nonumber \\
&&+ \frac{3 \lambda_{hh}}{8 \pi \mpl^2} H_{0,-2} - \frac{3 \lambda_{h\phi}}{4 \pi \mpl^3} \delta \phi_{-2}\,, \\
\label{eq::minimal_1o3_coefficients2}
\delta \phi_3 &=&  \delta \phi_{-2} \left( \frac{3 M^3 \phi_1^2}{5 \mpl^2} - \frac{3 M \phi_1^4}{80 \mpl^4} \right) \nonumber \\
& & - \frac{3 \lambda_{h\phi}}{8 \pi \mpl} H_{0,-2} + \frac{3 \lambda_{\phi\phi}}{4 \pi \mpl^2} \delta \phi_{-2} \,.
\ea

At this point, we want to stress that this result is universal for any scalar-tensor theories 
reducing to Einstein gravity with a minimally coupled scalar field in the vacuum outside 
a compact body.
More specifically, it can be applied to any 
model that is related via a transformation to a minimally coupled scalar field, as is possible for the DEF model, which we will investigate in the next section.

Also, we note that the additional contributions in Eqs.~\eqref{eq::minimal_1o3_coefficients}~and~\eqref{eq::minimal_1o3_coefficients2} only appear in this form for a theory that contains a \textit{massless} scalar field. If the scalar field were to possess a mass, all of its contributions to the asymptotic expansion would receive an additional exponential suppression 
(see Ref.~\cite{Diedrichs:2023foj}). The additional terms could thus not mix with the $1/r^3$ term, as they would have a different functional dependence. If one wants to extract $\lambda_{hh}$ in such a case, then it would suffice to numerically integrate the perturbation equations up to a sufficiently large radius since all of the scalar contributions will decay exponentially (see, e.g., Refs.~\cite{Diedrichs:2023trk, Sennett:2017etc}).

%%%%%%%%%%%%%%%%%%%%%%%%%%%%%%%%%%%%%%
\section{Damour-Esposito-Farèse Model \label{sec:DEF_model}} 
%%%%%%%%%%%%%%%%%%%%%%%%%%%%%%%%%%%%%%

We now apply the results in the previous section to the DEF model \cite{Damour:1993hw}. 
In the Jordan frame, the action for such a model is given by \cite{Kase:2020yhw,Kase:2020qvz}
\ba
\hspace{-0.8cm}
{\cal S}_{\rm H} &=& \int \rd^4 x \sqrt{-g} 
\bigg[ \frac{\mpl^2}{2} F(\phi) R \nonumber \\
\hspace{-0.8cm}
& &~~~~~~~~~
-\frac{1}{2} \left( 1 - \frac{3 \mpl^2 F_{,\phi}^2}{2 F^2} \right) 
F(\phi) \nabla^\mu \phi \nabla_\mu \phi 
\bigg],
\label{DEFaction}
\ea
where
\begin{equation}
F = e^{-\beta \phi^2 / ( 2 \mpl^2)}\,,
\end{equation}
with $\beta$ being a coupling constant.
This belongs to a subclass of Horndeski 
theories by choosing the coupling functions 
in Eq.~(\ref{eq::horndeski_action}) to be 
\ba
G_2 &=& \left( 1 - \frac{3 \mpl^2 F_{,\phi}^2}{2 F^2} \right) 
F(\phi) X\,, \qquad 
G_3=0\,,\nonumber \\
G_4 &=& \frac{\mpl^2}{2}\,,
\qquad 
G_5 = 0\,.
\ea
Note that we can retrieve the action of 
a minimally coupled scalar field, which 
we studied in the previous section, when 
setting $\beta$ to zero. 

An important property of the DEF model is that, while $\phi = 0$ 
is always a solution to the equations of motion, there exist 
solutions with a non-trivial scalar field configuration. 
Since the solution that fulfills $\phi = 0$ reduces exactly to 
the general relativistic one, we will denote them as 
the GR branch, while the one with $\phi \neq 0$ is referred 
to as the scalarized branch. On the strong gravitational 
background like the vicinity of NSs, 
the GR branch can be unstable for 
negative $\beta$ due to the tachyonic mass 
of the scalar field. This can trigger 
the approach to the scalarized branch 
in high-curvature regimes.
This phenomenon, which is dubbed spontaneous scalarization, 
occurs for $\beta \le -4.35$ \cite{Harada:1998ge,Novak:1998rk}.

Before presenting our calculations, let us note that 
the action (\ref{DEFaction}) is invariant 
under the parity transformation
\begin{align}
\phi \rightarrow -\phi\,.
\end{align}
By enforcing this symmetry in the PP action, most importantly 
in its tidal part in Eq.~\eqref{eq::pp_action_tidal_part}, 
we necessarily have that $\lambda_{h \phi}$ transforms as
\begin{align}
\lambda_{h \phi} \rightarrow - \lambda_{h \phi}
\end{align}
under the above transformation. Further, since $\phi = 0$ is 
always a solution to the above action, we thus have that, 
if $\phi = 0$ for a particular solution, then it necessarily 
follows that $\lambda_{h \phi}$ vanishes. 
Hence the GR branch needs to 
fulfill $\lambda_{h \phi} = 0$.

For simplicity, in the following, we will assume that 
the background scalar field vanishes at radial infinity, 
i.e.,~$\phi(r) \to 0$ as $r \to \infty$. 
Also, we will still mostly drop the overhead bar that would denote the background fields, unless a clear distinction is crucial.

\subsection{Expansion in Jordan Frame \label{sec:DEF_jordan}} 

In the Jordan frame given by the action (\ref{DEFaction}), 
the matter fields are 
minimally coupled to gravity. 
To perform an asymptotic expansion at 
spatial infinity in the Jordan frame, 
we can resort to the same ansatz that 
was given in Eq.~\eqref{eq::bg_ansatz} 
for the background.
The resulting coefficients are provided 
in App.~\ref{asymp_coeffs_DEF} up to order $r^{-9}$.
The first few terms in the large-distance expansion 
of the background fields are given by 
\ba
\phi(r) &=& \frac{\phi_1}{r} 
+ \frac{M \phi_1}{r^2} + \mathcal{O}(r^{-3})\,, 
\label{phiLJ}
\\
h(r) &=& 1 - \frac{2 M}{r} 
+\frac{\phi_1^2 (1 - 2 \beta)}{2 \mpl^2 r^2} 
+ \mathcal{O}(r^{-3})\,, 
\\
f(r) &=& 1 - \frac{2 M}{r} 
+\frac{\beta \phi_1^2}{2 \mpl^2 r^2} 
+ \mathcal{O}(r^{-3}) \,.
\ea
Notice how the contribution of $\beta$ enters 
at the next-to-leading order in the expansion 
of $h(r)$ and $f(r)$, i.e., at order $r^{-2}$. 

For perturbations, we find that the expressions for 
individual coefficients become much more involved 
due to additional terms that depend on $\beta$, 
which appear from order $r^0$ onward. 
The expansion still takes the following forms
\begin{align}
\label{eq::DEF_pert_asymp_expansion}
H_0(r) = \sum_{i=-2}^\infty H_{0, i} r^{-i}\,,
\qquad 
\delta \phi(r)=
\sum_{i =-2}^\infty \delta \phi_i r^{-i}\,,
\end{align}
where the coefficients $H_{0, i}$ and 
$\delta \phi_i$ are presented 
in App.~\ref{asymp_coeffs_DEF} up to order $r^{-8}$.

We also require the expansion of 
the background and perturbations around 
$r = 0$ for the numerical integration. 
For this purpose, we take into account
a perfect fluid with the background
density $\rho$ and the pressure $P$ 
and take the following ansatz 
\ba
P(r) &=& P_c + \sum_{i = 1}^\infty P_i r^i\,,
\qquad
\rho(r) = \rho_c + \sum_{i = 1}^\infty \rho_i r^i\,, \nonumber \\
\phi(r) &=& \phi_c + \sum_{i = 1}^\infty \phi_i r^i\,,\qquad
h(r) = h_c + \sum_{i = 1}^\infty h_i r^i\,, 
\nonumber \\
f(r) &=& f_c + \sum_{i = 1}^\infty f_i r^i\,,
\ea
where $P_c$, $P_i$, $\rho_c$, $\rho_i$, 
$\phi_c$, $\phi_i$, $h_c$, $h_i$, $f_c$, 
and $f_i$ are constants.
Enforcing regularity at the origin, we need to 
set $P'(0) = \rho'(0) = \phi'(0) = h'(0) = f'(0) = 0$. Upon using the background equations of motion, 
we obtain the same result that was already presented 
in the Appendix of Ref.~\cite{Kase:2020yhw}:
\ba
\label{eq::DEF_initial_condition}
\hspace{-0.7cm}
P(r) &=& P_c - \frac{\rho_c + P_c}{F(\phi_c)} 
\bigg[ \frac{\rho_c + 3 P_c}{12 \mpl^2} 
\nonumber \\
\hspace{-0.7cm}
& &+ \frac{\beta^2 \phi_c^2 (\rho_c - 3 P_c)}{24 \mpl^4} \bigg] r^2
+\mathcal{O}(r^4)\,, \\
\hspace{-0.7cm}
\phi(r) &=& \phi_c 
+\frac{\beta \phi_c (\rho_c - 3 P_c)}{12 \mpl^2 F(\phi_c)} r^2 + \mathcal{O}(r^4)\,, \\
\hspace{-0.7cm}
h(r) &=& 1 - \frac{2 \mpl^2 \rho_c - \beta^2 \phi_c^2 (\rho_c - 3 P_c)}{6 \mpl^4 F(\phi_c)} 
r^2 + \mathcal{O}(r^4)\,, \\    
\hspace{-0.7cm}
f(r) &=& f_c + \frac{f_c}{F(\phi_c)} \bigg[ \frac{\rho_c + 3 P_c}{6 \mpl^2} \nonumber \\
& &+ \frac{\beta^2 \phi_c^2 (\rho_c - 3 P_c)}{12 \mpl^4} \bigg] r^2 + \mathcal{O}(r^4)\,.
\label{eq::DEF_initial_condition4}
\ea
Likewise, we find that the perturbations behave 
at leading order around $r = 0$, as
\ba
H_0(r) &=& H_{0, c} r^2 + \mathcal{O}(r^4)\,, \\
\delta \phi(r) &=& 
\delta \phi_c r^2 + \mathcal{O}(r^4)\,,
\ea
where $H_{0, c}$ and $\delta \phi_c$ 
are constants.

\subsection{Expansion in Einstein Frame  
\label{sec:DEF_einstein}} 

After performing a conformal rescaling
of the metric with
\begin{equation}
\label{eq::conformal_rescaling}
\tilde{g}_{\mu\nu} = F(\phi) g_{\mu\nu}\,,
\end{equation}
we arrive at the so-called Einstein 
frame -- a frame in which there is a
minimally coupled scalar field. 
In the following, we denote quantities with 
a tilde as those in the Einstein frame. 
Explicitly, the Einstein-frame action
with a matter source is given by
\ba
\tilde{{\cal S}} &=&
\int \rd^4 x \sqrt{-\tilde{g}} \left[ \frac{\mpl^2}{2} \tilde{R} - \frac{1}{2} \tilde{g}^{\mu\nu} \nabla_\mu \phi 
\nabla_\nu \phi \right] \nonumber \\
& &+ {\cal S}_m(F^{-1}(\phi) \tilde{g}_{\mu\nu})\,.
\ea
After splitting the action into the bulk and PP 
actions, we thus arrive at the same system 
that was already investigated in Sec.~\ref{sec:minimally_coupled_scalar}. 
Hence we can directly reuse the results 
derived there, most importantly, Eqs.~\eqref{eq::minimal_1o3_coefficients} 
and \eqref{eq::minimal_1o3_coefficients2}.

\subsection{Matching \label{sec:DEF_matching}} 

We will now match the asymptotic expansion 
obtained in the Jordan frame 
to that in the Einstein frame to determine the coefficient $H_{0, 3}$ for the DEF model. To this end, we will explicitly construct the transformation that relates the perturbation fields in both frames.

\paragraph*{\bf Background transformation.}
To transform the background fields, we can directly use the relation given in Eq.~\eqref{eq::conformal_rescaling}. 
However, if we start with background fields that were 
calculated in Schwarzchild coordinates in the Jordan frame, 
then after performing the rescaling depicted in Eq.~\eqref{eq::conformal_rescaling}, 
we are no longer in Schwarzschild coordinates. 
We thus have to perform an additional coordinate transformation, 
which is given by\footnote{Since the coordinate transformation is designed to ensure that the background fields are asymptotically flat, we are also only using the background field $\bar{\phi}$ to define the transformation instead of the full field $\phi$.}
\begin{align}
\label{eq::DEF_to_minimal_coord_transform}
\tilde{r} = r \sqrt{F(\bar{\phi})} \,,
\end{align}
where $\bar{\phi}$ is the background value of the scalar field, 
and $\tilde{r}$ is the radial coordinate corresponding to 
Schwarzschild coordinates in the Einstein frame. 
With this additional relation, we can now relate all background fields in the two different frames to each other.\footnote{Note 
that we do not need to perform an additional coordinate 
transformation to ensure asymptotic flatness, as was done in, 
e.g., Ref.~\cite{Creci:2023cfx}. Since we only consider the case where $\phi \rightarrow 0$ for $r \rightarrow \infty$, it automatically 
follows that the rescaled metric is 
asymptotically flat.}

\paragraph*{\bf Perturbation transformation.}
We do not have to perform any additional operations 
to relate the scalar perturbation 
in both frames to each other. 
These are directly related via the 
coordinate transformation given in Eq.~\eqref{eq::DEF_to_minimal_coord_transform}, which explicitly results in
\begin{align}
\label{eq::DEF_phi_relation}
\widetilde{\delta \phi}(\tilde{r}) = \delta \phi(r)\,.
\end{align}
However, when relating the gravitational perturbation, 
we caution that the conformal rescaling also depends on 
the scalar perturbation. Explicitly writing out the metric 
and scalar field decomposition in 
Eq.~\eqref{eq::conformal_rescaling}, we have
\begin{align}
\tilde{\bar{g}}_{\mu\nu} + \tilde{h}_{\mu\nu} 
= F(\bar{\phi} + \delta \phi) (\bar{g}_{\mu\nu} 
+ h_{\mu\nu})\,.
\end{align}
Taylor expanding $F$ and only keeping terms that are 
linear in the perturbation, we obtain
\begin{align}
\tilde{h}_{\mu\nu}=F_{,\phi}(\bar{\phi}) \bar{g}_{\mu\nu} \delta \phi + F(\bar{\phi}) h_{\mu\nu}\,.
\label{hmunu}
\end{align}
Since $\tilde{h}_{00}
=\tilde{f} \tilde{H}_0$, 
$\bar{g}_{00}=-f$, and $h_{00}=fH_0$, 
the $(0,0)$ component of Eq.~(\ref{hmunu}) gives
\begin{align}
\tilde{f} \tilde{H}_0 = 
-F_{,\phi}(\bar{\phi})f
\delta \phi + F(\bar{\phi}) f H_0\,.
\end{align}
Since there is a relation 
$\tilde{f}=F(\bar{\phi})f$ between the $(0,0)$ 
background components of metrics in 
the Einstein and Jordan frames, 
it follows that
\begin{equation}
\tilde{H}_0(\tilde{r}) = 
-\frac{F_{,\phi}(\bar{\phi}(r))}{F(\bar{\phi}(r))} 
\delta \phi(r) + H_0(r)\,.
\label{H0JE}
\end{equation}
The large-distance solutions to $\tilde{H}_0(\tilde{r})$ and 
$\widetilde{\delta \phi}(\tilde{r})$ 
in the Einstein frame are the same as the ones derived for the minimally coupled scalar field 
in Sec.~\ref{sec:minimally_asymp_expansion}, 
which are specifically given in 
Eqs.~(\ref{eq::minimal_asymp_expansion_H0}) 
and (\ref{eq::minimal_asymp_expansion_dphi}), 
respectively--- one only needs to add 
an overhead tilde to them.
In the Jordan frame, 
the corresponding expanded solutions 
of $H_0(r)$ and $\delta \phi (r)$
are of the form (\ref{eq::DEF_pert_asymp_expansion}), 
where the coefficients
are shown in App.~\ref{asymp_coeffs_DEF}.
We also exploit the background 
solution (\ref{phiLJ}) of $\phi(r)$ 
expanded at large distances as well as 
the relation (\ref{eq::DEF_to_minimal_coord_transform}) 
between the radial distances $\tilde{r}$ and $r$.
By using the correspondences (\ref{eq::DEF_phi_relation}) and 
(\ref{H0JE}) and comparing the coefficients of 
the large-distance expansions with respect 
to $r$ at each order, we can express $H_{0,3}$ 
and $\delta \phi_3$ in terms of $\tilde{H}_{0,3}$, 
$\widetilde{\delta \phi}_{3}$, and the corrections 
arising from the coupling $\beta$. 
Here, $\tilde{H}_{0,3}$ and 
$\widetilde{\delta \phi}_{3}$
are equivalent to the right-hand sides of 
Eqs.~(\ref{eq::minimal_1o3_coefficients}) and (\ref{eq::minimal_1o3_coefficients2}), 
respectively, where we simply replaced $\delta \phi_{-2}$ with 
$\widetilde{\delta \phi}_{-2}$ and $H_{0,-2}$ with
$\tilde{H}_{0,-2}$.
Then, it follows that
\begin{widetext}
\begin{align}
\label{eq::DEF_1o3_coeffs}
H_{0,3} &= \beta \delta\phi_{-2} \phi_{1} \frac{- 128 M^{4} \mpl^{4} + M^{2} \phi_{1}^{2} \mpl^{2} \left(1320 \beta - 208\right) + \phi_{1}^{4} \left(135 \beta^{2} - 30 \beta - 27\right)}{1440 \mpl^{6}} \\
&\phantom{{}={}} + H_{0,-2} M \beta \phi_{1}^{2} \frac{\phi_{1}^{2} \left(50 - 69 \beta\right) - 152 M^{2} \mpl^{2}}{144 \mpl^{4}} + H_{0, -2} \left( \frac{3 M^3 \phi_1^2}{5 \mpl^2} - \frac{3 M \phi_1^4}{80 \mpl^4} \right) + \frac{3 \lambda_{hh}}{8 \pi \mpl^2} H_{0,-2} - \frac{3 \lambda_{h\phi}}{4 \pi \mpl^3} \delta \phi_{-2}\,, \nonumber\\
\delta \phi_3 &= 3\delta\phi_{-2}  M \beta \phi_{1}^{2}  \frac{\phi_{1}^{2} (2 - 3 \beta) - 8 M^{2} \mpl^{2}}{16 \mpl^{4}} + \delta\phi_{-2} \left( \frac{3 M^3 \phi_1^2}{5 \mpl^2} - \frac{3 M \phi_1^4}{80 \mpl^4} \right) - \frac{3 \lambda_{h\phi}}{8 \pi \mpl} H_{0,-2} + \frac{3 \lambda_{\phi\phi}}{4 \pi \mpl^2} \delta \phi_{-2}\,.
\label{eq::DEF_1o3_coeffs2}
\end{align}
\end{widetext}

%%%%%%%%%%%%%%%%%%%%%%%%%%%%%%%
\begin{figure*}[t]
\centering
\includegraphics[width=0.49\linewidth]{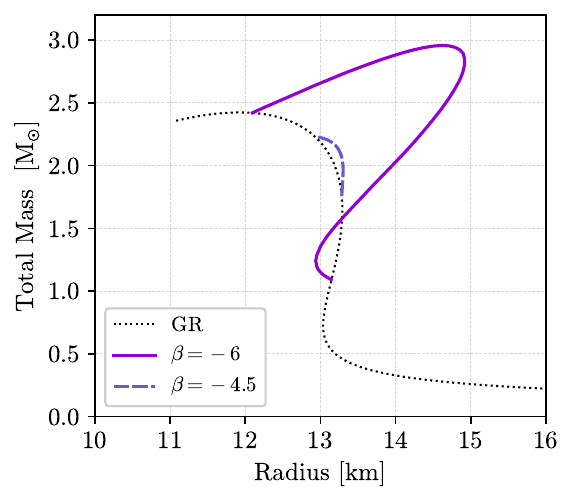}
\includegraphics[width=0.49\linewidth]{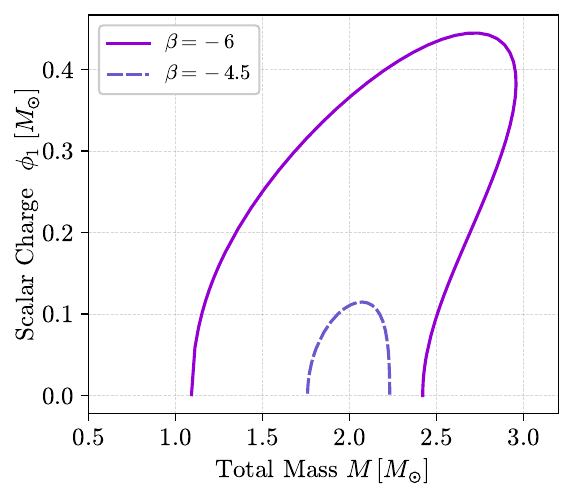}
\caption{\textbf{Left:} 
the mass-radius curve for NSs in the DEF model with 
$\beta = -4.5$ and $\beta = -6$, where $M_{\odot}$ 
is the solar mass. We choose the DD2 equation of state for  
describing the property of perfect fluids inside the NS.
The dotted black line shows the non-scalarized solutions, which are identical to the GR solutions, while the purple-colored lines 
represent the scalarized solutions. 
\textbf{Right:} scalar charge $\phi_1$ 
versus the total gravitational mass.}
\label{DEF::mass_rad_charge}
\end{figure*}
%%%%%%%%%%%%%%%%%%%%%%%%%%%%%%
Using these relations, we are now ready to extract tidal Love numbers for specific NS solutions, which will be done in Sec.~\ref{Numesec}.

\subsection{Numerical Analysis}
\label{Numesec}

In the following, we will describe the numerical construction
of background solutions and, afterward, the construction 
of perturbation profiles. Since the background solutions have 
already been well-studied (see e.g., Refs.~\cite{Damour:1993hw, Damour:1996ke}), we will only briefly review their features 
and then directly move to the Love numbers. For numerics, 
we choose two coupling constants $\beta = -4.5$ and $\beta = -6$. 
The former is close to observational bounds, while the latter 
is already excluded from binary pulsar measurements \cite{Freire:2012mg,Shao:2017gwu,Freire:2024adf}.
We have included the latter case in our analysis, 
since we would like to demonstrate 
the behavior of resulting tidal Love numbers in general.

\paragraph*{\bf Background solutions.}
To obtain the tidal Love numbers, it is first necessary to 
construct the background solutions numerically, 
which are then used to integrate the perturbation equations 
of motion. We resort to the fourth-order Runge-Kutta method 
to integrate the equations given in Eqs.~\eqref{eq::horndeski_bg_equations}-\eqref{back5} 
outward. For the boundary conditions around $r=0$, 
we exploit Eqs.~\eqref{eq::DEF_initial_condition}-\eqref{eq::DEF_initial_condition4} and retrieve 
the total mass and scalar charge of the object 
by matching the integrated profiles at large 
radii to the asymptotic expansions given in App.~\ref{asymp_coeffs_DEF}. 
In Fig.~\ref{DEF::mass_rad_charge}, we show 
the masses, radii, and scalar charges 
obtained by using the DD2 equation of state~\cite{Typel:2009sy, Hempel:2011mk}. 
Here, we can observe the well-known feature of this model: Below a certain central pressure, we only 
have non-scalarized solutions identical to those 
in GR; from a certain pressure onward, 
there is another scalarized branch. 
Only NSs on the scalarized branch
develop a non-vanishing scalar profile, 
and hence, a non-zero scalar charge. 

\paragraph*{\bf Extracting the Love numbers.} 

The relevant information for each integrated stellar profile is stored in the values $\delta \phi_{-2}$, $\delta \phi_{3}$, $H_{0,-2}$, and $H_{0,3}$. We obtain these values by performing a 
least-squares fit of the expansion given in Eq.~\eqref{eq::DEF_pert_asymp_expansion} to $H_0''$ and $\delta \phi''$ outside the star. 
Here, we use the second derivative of the 
perturbed fields to extract the information, 
since they asymptote to a constant value and are 
thus numerically more convenient to use. For the fitting, we integrate each profile up to a radius of $r_\mathrm{max} \approx 75$\,km and then 
exploit $\approx 100$ data points in the range of 
$r \approx 60$\,km up to $r = r_\mathrm{max}$ 
for the least-squares fit. The advantage of adopting a least-squares fit is that the numerically integrated profiles can be plagued by subtle numerical errors, which we smooth out in this way. 
We present results regarding the robustness of this approach further below. 

The tidal Love numbers could now be obtained by performing a shooting procedure to fix either $H_{0,-2}$ or $\delta \phi_{-2}$ 
to zero, such that only one Love number appears 
in the asymptotic series of each perturbed field, 
which can then be easily extracted. 
However, we find it more convenient to exploit the linear nature of the perturbation equations of motion: For each background configuration, we numerically prepare two linearly independent perturbation profiles denoted as 
$H_0^{(1)}$, $\delta \phi^{(1)}$ for the first set profiles 
and $H_0^{(2)}$, $\delta \phi^{(2)}$ 
for the second profiles. Next, as noted in Eq.~\eqref{eq::perturb_coeff_naiv_H03}, 
the coefficients of $H_{0,3}$ can be written as 
\begin{align}
\begin{split}
H_{0,3}^{(1)} &= c_{h\phi} \delta \phi_{-2}^{(1)} + c_{hh} H_{0,-2}^{(1)}\,, \\
H_{0,3}^{(2)} &= c_{h\phi} \delta \phi_{-2}^{(2)} + c_{hh} H_{0,-2}^{(2)}\,.
\end{split}
\end{align}
We can restate the above equation
in a matrix form, such that
\begin{equation}
\begin{pmatrix}
\delta \phi_{-2}^{(1)} & H_{0,-2}^{(1)} \\
\delta \phi_{-2}^{(2)} & H_{0,-2}^{(2)}
\end{pmatrix}
\begin{pmatrix}
c_{h\phi} \\
c_{hh}
\end{pmatrix} =
\begin{pmatrix}
H_{0,3}^{(1)} \\
H_{0,3}^{(2)}
\end{pmatrix}.
\end{equation}
We thus find that the coefficients $c_{h\phi}$ and $c_{hh}$ can be directly obtained via\footnote{Note that the matrix inverse is guaranteed to exist by the linear nature of the perturbation equations, 
so long as one chooses linearly independent initial conditions for the two integrated profiles.}
\begin{equation}
\label{eq::tidal_extract_1}
\begin{pmatrix}
c_{h\phi} \\
c_{hh}
\end{pmatrix} =
\begin{pmatrix}
\delta \phi_{-2}^{(1)} & H_{0,-2}^{(1)} \\
\delta \phi_{-2}^{(2)} & H_{0,-2}^{(2)}
\end{pmatrix}^{-1}
\begin{pmatrix}
H_{0,3}^{(1)} \\
H_{0,3}^{(2)}
\end{pmatrix}.
\end{equation}
Likewise, by considering the coefficients 
$\delta \phi_3^{(1)}$ and $\delta \phi_3^{(2)}$ 
in Eq.~(\ref{eq::perturb_coeff_naiv_dphi03}), 
we can derive an expression for $c_{\phi\phi}$ 
and $c_{\phi h}$, as 
\begin{equation}
\label{eq::tidal_extract_2}
\begin{pmatrix}
c_{\phi\phi} \\
c_{\phi h}
\end{pmatrix} =
\begin{pmatrix}
\delta \phi_{0,-2}^{(1)} & H_{0,-2}^{(1)} \\
\delta \phi_{0,-2}^{(2)} & H_{0,-2}^{(2)}
\end{pmatrix}^{-1}
\begin{pmatrix}
\delta \phi_3^{(1)} \\
\delta \phi_3^{(2)}
\end{pmatrix}.
\end{equation}

%%%%%%%%%%%%%%%%%%%%%%%%%%%%%%%
\begin{figure}[t]
\centering
\includegraphics[width=0.99\textwidth]{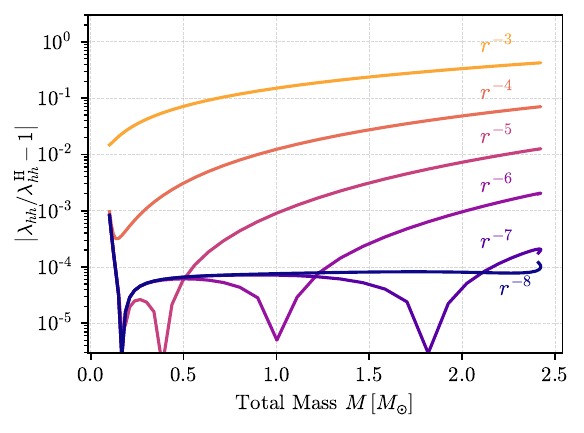}
\caption{Absolute relative difference of the quadrupolar Love numbers $\lambda_{hh}$ on the GR branch when extracted using the exact formula derived by Hinderer~\cite{Hinderer:2007mb}, to the ones obtained by the matching procedure described in the main text. The text inset denotes up to which order the asymptotic expansion is used for the matching, e.g., the dark blue curve corresponds to the expansion up to (including) order $r^{-8}$.}
\label{fig:tidals_matching_error}
\end{figure}
After having computed the coefficients $c_{\phi\phi}$, $c_{h\phi}$, $c_{\phi h}$, and $c_{h h}$, we can resort to
Eqs.~\eqref{eq::DEF_1o3_coeffs}-\eqref{eq::DEF_1o3_coeffs2} to 
extract the tidal coefficients $\lambda_{hh}$, $\lambda_{h \phi}$ and $\lambda_{\phi \phi}$. 
Note that there are two possibilities for 
obtaining $\lambda_{h \phi}$: One could either use Eq.~\eqref{eq::tidal_extract_1} and thus the coefficient $c_{h \phi}$, or one could use 
$c_{\phi h}$, which is instead derived from Eq.~\eqref{eq::tidal_extract_2}. 
Independent of which equation is chosen, 
the resulting $\lambda_{h \phi}$ should be identical.

In what follows, we will distinguish between \textit{corrected} Love numbers, denoted by $\lambda_{x}$, and \textit{non-corrected} Love numbers, denoted by $\lambda_{x, \mathrm{nc}}$. 
The corrected Love numbers are extracted via Eqs.~\eqref{eq::DEF_1o3_coeffs}~and~\eqref{eq::DEF_1o3_coeffs2} and thus take the additional coefficients of the $r^{-3}$ term into account, while the \textit{uncorrected} ones 
simply use Eqs.~\eqref{eq::tidal_part_SS_RW_coords}~and~\eqref{eq::tidal_part_SS_RW_coords2} as the sole contribution to the $r^{-3}$ term. 
More explicitly, we obtain the non-corrected 
Love numbers as
\ba
\label{eq::non_corrected_love_numbers_def}
\hspace{-0.7cm}
\lambda_{hh, \mathrm{nc}} 
&=& \frac{8\pi \mpl^2}{3} c_{hh}\,,\qquad
\lambda_{h\phi, \mathrm{nc}} 
= -\frac{4\pi \mpl^3}{3} c_{h\phi}, \nonumber \\
\hspace{-0.7cm}
\hat{\lambda}_{h\phi, \mathrm{nc}} 
&=& -\frac{8\pi \mpl}{3} c_{\phi h}\,,\qquad 
\lambda_{\phi\phi, \mathrm{nc}}
= \frac{4\pi \mpl^2}{3} c_{\phi\phi} \,,
\ea
where we have also introduced $\hat{\lambda}_{h\phi, \mathrm{nc}}$, which is obtained from $c_{\phi h}$, in contrast to $\lambda_{h\phi, \mathrm{nc}}$, which is obtained from $c_{h \phi}$. As noted in Sec.~\ref{sec:pp_action}, both values should be identical for the corrected Love numbers.

%%%%%%%%%%%%%%%%%%%%%%%%%%%%%%%%%%%
\begin{figure*}[t]
\centering
\includegraphics[width=0.49\linewidth]{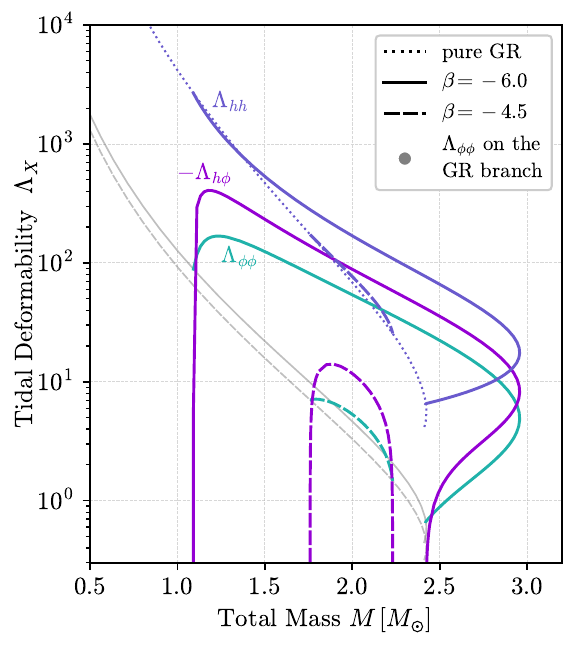}
\includegraphics[width=0.49\linewidth]{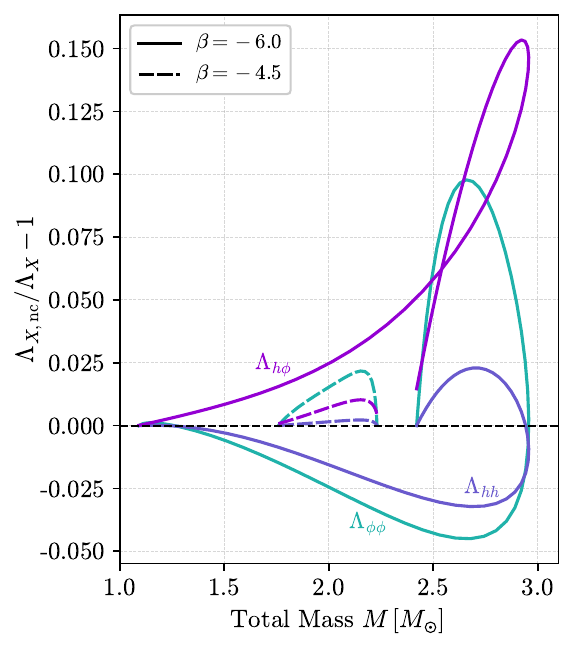}
\caption{\textbf{Left:} the quadrupolar even-parity dimensionless Love numbers for NSs in the DEF model with $\beta \in \{-6, -4.5\}$ and the DD2 equation of state. We obtain the dimensionless Love numbers via $\Lambda_x \equiv \lambda_x / M^5$, where $x \in \{ hh, h\phi, \phi\phi \}$. $\Lambda_{\phi\phi}$ is shown for both in the scalarized (colored) and GR (gray) branches.
Note that $\Lambda_{h\phi}$ drops to zero on the GR 
branch, which is due to the parity symmetry explained 
at the beginning of Sec.~\ref{sec:DEF_model}. 
\textbf{Right:} the relative difference between the non-corrected ($\Lambda_{x,nc}$) and corrected ($\Lambda_x$) Love numbers.}
\label{DEF::tidal_love_numbers}
\end{figure*}
%%%%%%%%%%%%%%%%%%%%%%%%%%%%%%%%%%

%%%%%%%%%%%%%%%%%%%%%%%%%%%%%%%%%%%
\begin{figure}[t]
\centering
\includegraphics[width=0.98\linewidth]{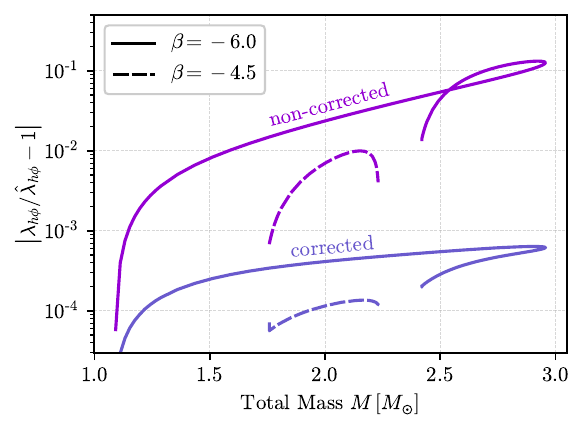}
\caption{Comparison of the extracted tidal Love numbers $\lambda_{h\phi}$ when obtained via $c_{h\phi}$ from Eq.~\eqref{eq::tidal_extract_1} to the ones obtained via $c_{\phi h}$ from Eq.~\eqref{eq::tidal_extract_2}. The blue line displays the relative difference for the \textit{corrected} Love numbers, as defined in Eqs.~\eqref{eq::corrected_love_numbers_def}--\eqref{eq::corrected_love_numbers_def4}, while the purple line demonstrates the relative difference for the \textit{non-corrected} Love numbers, defined in Eq.~\eqref{eq::non_corrected_love_numbers_def}. If the Love numbers are correctly extracted, then the relative difference should approach zero. Notably, the curves of non-corrected Love numbers show significant deviations, with a relative difference of up to $13$\,\%.}
\label{fig:hphi_comparison}
\end{figure}
Further, the corrected Love numbers are obtained 
from Eqs.~\eqref{eq::DEF_1o3_coeffs}~and~\eqref{eq::DEF_1o3_coeffs2}, 
which results in 
\begin{widetext}
\ba
\label{eq::corrected_love_numbers_def}
\lambda_{hh} 
&=& \frac{8 \pi \mpl^2}{3} \bigg[ c_{hh} - M \beta \phi_{1}^{2} \frac{\phi_{1}^{2} \left(50 - 69 \beta\right) - 152 M^{2} \mpl^{2}}{144 \mpl^{4}} - \frac{3 M^3 \phi_1^2}{5 \mpl^2} + \frac{3 M \phi_1^4}{80 \mpl^4} \bigg]\,,\\
\lambda_{h\phi} 
&=& -\frac{4 \pi \mpl^3}{3} \bigg[ c_{h\phi} - \beta \phi_{1} \frac{- 128 M^{4} \mpl^{4} + M^{2} \phi_{1}^{2} \mpl^{2} \left(1320 \beta - 208\right) + \phi_{1}^{4} \left(135 \beta^{2} - 30 \beta - 27\right)}{1440 \mpl^{6}} \bigg]\,,
\\
\hat{\lambda}_{h\phi} 
&=& -\frac{8\pi \mpl}{3} c_{\phi h}\,,
\\
\lambda_{\phi\phi} &=& 
\frac{4\pi \mpl^2}{3} \bigg[ c_{\phi\phi} - 3 M \beta \phi_{1}^{2}  \frac{\phi_{1}^{2} (2 - 3 \beta) - 8 M^{2} 
\mpl^{2}}{16 \mpl^{4}} - \frac{3 M^3 \phi_1^2}{5 \mpl^2} 
+ \frac{3 M \phi_1^4}{80 \mpl^4} \bigg]\,.
\label{eq::corrected_love_numbers_def4}
\ea
\end{widetext}

\paragraph*{\bf Numerical accuracy.}
Before presenting our results for the tidal Love numbers, 
we briefly discuss the inaccuracy in our numerical results 
due to the matching procedure mentioned earlier. 
While there does not seem to be a closed-form analytic 
solution for the perturbed fields in the exterior region 
in scalar-tensor theories considered here, 
we can compare our matching procedure in the case of 
pure GR to the exact formula derived 
by Hinderer~\cite{Hinderer:2007mb}:
\begin{align}
\lambda_{hh}^\mathrm{H} = &\frac{16}{15} M^5 (1 - 2 C)^2 
\big[ 2 + 2 C (y - 1) - y\big]  \\
&\times \big\{ 2C \big[ 6 - 3y + 3C (5y - 8) \big] \notag\\
&+ 4C^3 \big[ 13 - 11 y + C(3y - 2) + 2 C^2 (1 + y) \big] \notag\\
&+ 3(1 - 2 C)^2 \big[2 - y + 2 C (y - 1)\big] \log (1 - 2 C) \big\}^{-1}
\,.\notag
\end{align}
We use this formula to extract the tidal Love number 
at the surface of the star by setting $C=M/R$ and 
$y=R H_0'(R)/H_0(R)$, with $R$ being the star radius.
We thus prepare a range of NS solutions 
and obtain the Love numbers once by fitting the 
asymptotic expansion of $H_0$ -- given in App.~\ref{asymp_coeffs_DEF} 
with $\beta = 0$ -- and once 
by using the above exact formula.

These results are presented in Fig.~\ref {fig:tidals_matching_error}, demonstrating the absolute relative difference between both approaches for various orders of the asymptotic expansion. We first note that the relative difference increases for larger values of the total mass, which is to be expected since each higher order in the expansion carries along a higher power of the mass, thus rendering the neglected higher-order terms more significant. We further observe that the relative difference exponentially decreases with each additional order. However, this improvement seems to stagnate for higher orders, where the relative difference approaches a plateau. 
This is also expected because of the inherent numerical uncertainties that arise. Since this plateau is roughly constant for the highest order considered here, we conclude that including even higher-order terms in the expansion would not improve the fit, as the uncertainties are already dominated by the numerics. Observing that the expansion of order $r^{-8}$ results in a relative difference that is less than $0.01\,\%$ for the pure GR tidal Love numbers, we will assume that the fitting error is 
on a similar order of magnitude for the perturbations 
in the DEF model. Hence we use this fitting order 
for all of the following numerical results.

\paragraph*{\bf Numerical results.}
The left panel in Fig.~\ref{DEF::tidal_love_numbers} displays 
the corrected dimensionless tidal Love numbers, 
defined as $\Lambda_x = \lambda_x / M^5$, where 
$x \in \{ hh, h\phi, \phi\phi \}$.
We generally observe that the magnitude of tidal Love numbers 
decreases for larger values of $\beta$, 
which we expect since we approach pure GR in the limit that $\beta = 0$. 
We further note that, as expected from the discussion at the beginning of Sec.~\ref{sec:DEF_model}, 
the mixed Love number $\Lambda_{h \phi}$ drops to zero 
(within numerical accuracy) on the GR branch. Further, the purely scalar 
Love number $\Lambda_{\phi\phi}$ is non-zero even on the GR branch 
but gets enhanced on the scalarized branch. 
While this enhancement is quite significant for $\beta = -6$, 
in which case it partly increases by more than one order 
of magnitude, it is much less significant for $\beta = -4.5$. 

The right panel in Fig.~\ref{DEF::tidal_love_numbers} shows the relative difference between the corrected and non-corrected Love numbers. 
For $\beta = -6$ and considering only stable solutions, the discrepancy is largest for $\Lambda_{h\phi}$, reaching up to $15\,\%$, while $\Lambda_{hh}$ deviates at most by about $3\,\%$, 
and $\Lambda_{\phi\phi}$ by at most about $5\,\%$. For $\beta = -4.5$, the deviation is much less 
significant: Up to $2.2\,\%$ for $\Lambda_{\phi\phi}$, $1\,\%$ for $\Lambda_{h\phi}$, and $0.23\,\%$ for $\Lambda_{hh}$. 
The fact that the relative difference generally decreases as one chooses smaller values of 
$|\beta|$ can also be expected from 
the definition of the corrected Love numbers in Eqs.~\eqref{eq::corrected_love_numbers_def}-\eqref{eq::corrected_love_numbers_def4}. 
The additional terms involve positive powers of 
the scalar charge (proportional to $\phi_1$), whose 
contributions tend to decrease for smaller $|\beta|$.

Additionally, in Fig.~\ref{fig:hphi_comparison}, we present 
the relative difference of $\lambda_{h\phi}$ to $\hat{\lambda}_{h \phi}$. 
If both the analytical aspects and the numerical implementation 
are correct, then we expect both values to be equal. 
Indeed, we find that the relative difference of corrected 
Love numbers is less than $0.07\,\%$ for $\beta = -6$ 
and around $0.01\,\%$ for $\beta = -4.5$. 
As indicated in Fig.~\ref{fig:tidals_matching_error}, 
we expect a relative difference caused by numerical 
inaccuracies on the order of $0.01\,\%$, meaning that -- 
within the numerical accuracy -- we can verify that 
$\lambda_{h\phi} = \hat{\lambda}_{h \phi}$.
Let us stress that this is a non-trivial test of the 
EFT considerations that ultimately lead to 
Eqs.~\eqref{eq::DEF_1o3_coeffs}-\eqref{eq::DEF_1o3_coeffs2}, 
as well as the robustness of the numerical implementation. 
In the same figure, we also show the relative difference of 
the non-corrected values, for which the relative difference 
increases up to $13\,\%$ for $\beta = -6$, and $1\,\%$ 
for $\beta = -4.5$. Since they exhibit notable 
deviations from zero, we should use corrected values
for the proper estimation of tidal Love numbers.

\section{Scalar-Gauss-Bonnet Gravity \label{sec:SGB}} 

After successfully describing the extraction procedure of 
tidal Love numbers in the DEF model, we wish to discuss 
the case of scalar-Gauss-Bonnet gravity. 
This is characterized by considering the scalar field $\phi$ 
coupled to the Gauss-Bonnet curvature invariant 
$R_\mathrm{GB}^2$. The corresponding action is now given by
\begin{align}
{\cal S}_{\rm H} = \int \rd^4 x \sqrt{-g} 
\bigg[ \frac{\mpl^2}{2} R 
-\frac{1}{2} \nabla^\mu \phi \nabla_\mu \phi 
+ \xi(\phi) R_\mathrm{GB}^2 \bigg]\,,
\label{SsGB}
\end{align}
where
\begin{equation}
R_\mathrm{GB}^2 = R^2 - 4 R_{\alpha \beta} R^{\alpha \beta} 
+ R_{\alpha \beta \mu \nu} R^{\alpha \beta \mu \nu}\,,
\end{equation}
with $R_{\alpha \beta}$ and 
$R_{\alpha \beta \mu \nu}$ being the 
Ricci and Riemann tensors, respectively.
We can express the theory (\ref{SsGB}) in terms of 
the Horndeski action by choosing \cite{Kobayashi:2011nu}
\ba
G_2 &=& X + 8 \xi^{(4)} X^2 
(3 - \ln |X|)\,,\nonumber \\
G_3 &=& 4 \xi^{(3)} X (7 - 3 \ln |X|)\,,
\nonumber \\
G_4 &=& \frac{\mpl^2}{2} 
+ 4 \xi^{(2)} X (2 - \ln |X|)\,,
\nonumber \\
G_5 &=& -4 \xi^{(1)} \ln |X|\,,
\ea
where $\xi^{(n)}(\phi) \equiv 
\rd^n \xi (\phi)/\rd \phi^n$.
The crucial difference between this model and the DEF model 
is that here we cannot arrive at a frame where the scalar 
field is only minimally coupled by performing a conformal rescaling. 
This property in the DEF model was crucial for identifying 
the $r^{-3}$ term in the asymptotic expansion of 
$H_0$ and $\delta \phi$, in which case we could correctly 
extract tidal Love numbers.

In scalar-Gauss-Bonnet theories, we will 
consider the case in which $\xi$ 
contains even power-law functions 
of $\phi$, i.e., 
\begin{align}
\xi(\phi) = \sum_{n = 1}^\infty c_{2n} \phi^{2n}\,,
\end{align}
where $c_{2n}$'s are constants. 
Note that this type of coupling was 
chosen to describe the phenomenon of spontaneous 
scalarization of BHs and NSs \cite{Doneva:2017bvd,Silva:2017uqg,Antoniou:2017acq,Doneva:2017duq}. 
We will first look at the asymptotic expansion, as it can be 
inferred from the equations of motion. 
Afterward, we go over how the point-particle action 
changes compared to the minimally coupled case discussed in Sec.~\ref{sec:minimally_coupled_scalar}.

%%%%%%%%%%%%%%%%%%%
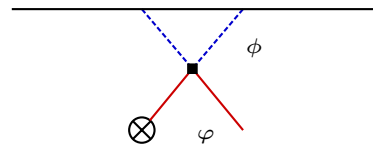
\begin{figure}[t]
    \centering
    \begin{subfigure}[t]{\textwidth}
    \begin{tikzpicture}[thick, scale = 1.6, every node/.style={transform shape}]
  \begin{feynman}
    % set up the worldline
    \vertex (a1);
    \vertex [right=of a1] (a2);
    \vertex [right=of a2] (a3);
    
    \vertex (a11) at ($(a2)!\twoVertexDistance!(a1)$);
    \vertex (a12) at ($(a2)!\twoVertexDistance!(a3)$);

    % set up some vertices for the bottom (invisible) line
    \vertex [below=of a1] (b1);
    \vertex [right=of b1] (b2);
    \vertex [right=of b2] (b3);

    \node[crossed dot, minimum size=6pt, white] (b11) at ($(b2)!\twoVertexDistance!(b1)$);
    \vertex (b12) at ($(b2)!\twoVertexDistance!(b3)$);

    \vertex (c1) at ($(b2)!0.33!(a2)$);
    \vertex (c2) at ($(a2)!0.33!(b2)$);

    \node[crossed dot, minimum size=6pt] (c3) at ($(b11)!0.33!(a11)$);
    \vertex (c4) at ($(b12)!0.33!(a12)$);
    
    \diagram* {
      (a11) -- [\phistyle] (c2),
      (a12) -- [\phistyle, edge label={\tiny$\phi$}, near start] (c2),
      (c3) -- [\varphistyle] (c2),
      (c4) -- [\varphistyle, edge label={\tiny$\varphi$}, near start] (c2),
      (a1) -- [thick] (a3)
    };
    \node[square dot, minimum size=2pt] (c4) at (c2);
  \end{feynman}
\end{tikzpicture}
\end{subfigure}
\caption{Leading-order diagram arising from the scalar field coupled to a 
Gauss-Bonnet term. Since we consider theories in which $\xi$ 
is an even function of $\phi$, the above diagram corresponds to 
the leading-order effect from $\xi_2$. We denote the corresponding 
operator, which is given in Eq.~\eqref{eq::GB_LO_operator}, 
with a black square.}
\label{fig:Gauss-Bonnet_LO}
\end{figure}
%%%%%%%%%%%%%%%%%%%%%%%%%%%%

%
\subsection{Asymptotic Expansion}

On the static and spherically symmetric 
background given by the line element 
(\ref{metric}), we first derive 
the large-distance solutions to $f$, $h$, and 
$\phi$ by using the expansions (\ref{eq::bg_ansatz}). 
To this end, it is convenient also to expand the function $\xi$ 
around the asymptotic field value $\phi_0$ \cite{Minamitsuji:2022tze}.
Setting $\phi_0=0$, the asymptotic expansion 
is given by 
\begin{align}
\xi(\phi) = \sum_{n = 2}^\infty \frac{\xi_n}{n!} 
\phi^n\,,
\end{align}
where $\xi_n = \xi^{(n)}(0)$. Note that we can neglect 
the terms corresponding to $\xi_0$ and $\xi_1$, 
as the former would lead to a surface term 
in the action and the latter would inhibit 
spontaneous scalarization \cite{Doneva:2022ewd}. 
To derive the asymptotic behavior of the perturbation fields, 
we use the same ansatz as in Eq.~\eqref{eq::minimally_coupled_perturb_expansion} 
and solely focus on external gravitational perturbations, 
thus setting $\delta \phi_{-2} = 0$. 
We find that the asymptotic expansions 
of the perturbation fields now take 
the forms of
\begin{align}
H_0(r) &= H_{0,-2} r^2 - 2 H_{0,-2} M r 
+ \frac{H_{0,-2} \phi_1^2}{3 \mpl^2} \nonumber\\
&\phantom{{}={}} + \frac{H_{0,-2} M \phi_1^2}{6 \mpl^2 r} 
+\frac{H_{0,-2} \phi_1^2 (M^2 + 48 \xi_2)}{3 \mpl^2 r^2} \nonumber \\
&\phantom{{}={}} + \frac{H_{0,3}}{r^3} 
+\mathcal{O}\left( r^{-4} \right)\,, \\
\delta \phi(r) &= - \frac{H_{0,-2} M \phi_1}{3} 
+ \frac{12 H_{0,-2} M \phi_1 \xi_2}{r^2}  \nonumber\\
&\phantom{{}={}}+ \frac{\delta \phi_3}{r^3} 
+ \mathcal{O}\left( r^{-4} \right).
\end{align}
Notice how $\xi_2$ enters at order $r^{-2}$, 
and we thus expect that this parameter also 
contributes to the $r^{-3}$ order.

\subsection{EFT Side}

Even though we do not have a transformation that results 
in the scalar field being minimally coupled, if the same symmetry 
that was presented in Eq.~\eqref{eq::minimal_coupled_symmetry} is 
also manifest here, then we might be able to leverage it 
again to obtain the correct coefficients $H_{0,3}$ and $\delta \phi_3$. 
To this end, we expand the scalar-Gauss-Bonnet coupling in terms of 
the metric ansatz specified in Eq.~\eqref{eq::eft_metric_decomposition}. 
We find that
\begin{widetext}
\begin{align}
\int \rd^4 x \sqrt{-g}\,\xi(\phi) R_\text{GB}^2 = 
&\int \rd^4 x \,\xi(\hat{\phi})\,
\partial_i \bigg[ e^{\sqrt{2} \hat{\varphi}/\mpl} \bigg\{ 2 \sqrt{2} \frac{    (\partial_{j}\partial^{j}\hat{\varphi}) \partial^{i}\hat{\psi}{}}{\mpl{} \
\hat{\psi}^{3/2}} + 2 \frac{(\partial_{j}\hat{\varphi} \partial^{j}\hat{\varphi}) \
\partial^{i}\hat{\psi}{}}{\mpl{}^2 \hat{\psi}{}^{3/2}} +
\sqrt{2} \frac{(\partial_{j}\hat{\psi}{} \partial^{j}\hat{\psi}{}) \
\partial^{i}\hat{\varphi}}{\mpl{} \hat{\psi}{}^{5/2}} \nonumber \\
&- 6 \frac{
(\partial_{j}\hat{\varphi} \partial^{j}\hat{\psi}{}) \
\partial^{i}\hat{\varphi}}{\mpl{}^2 \hat{\psi}{}^{3/2}} - 4 \frac{
(\partial_{j}\partial^{j}\hat{\varphi}) \partial^{i}\hat{\varphi}}{\mpl{}^2 \
\hat{\psi}{}^{1/2}} + 2 \sqrt{2} \frac{(\partial_{j}\hat{\varphi} \partial^{j}\hat{\varphi}) \
\partial^{i}\hat{\varphi}}{\mpl{}^3 \hat{\psi}{}^{1/2}} - 2\sqrt{2} \frac{
\partial_{j}\partial^{i}\hat{\varphi} \partial^{j}\hat{\psi}{}}{\mpl{} \
\hat{\psi}{}^{3/2}} + 4 \frac{\partial^{i}\partial_{j}\hat{\varphi} \
\partial^{j}\hat{\varphi}}{\mpl{}^2 \hat{\psi}{}^{1/2}}\bigg\} \bigg]\,. 
\end{align}
\end{widetext}

Notably, independent of the function $\xi$, the scalar-Gauss-Bonnet term 
breaks the symmetry specified in Eq.~\eqref{eq::minimal_coupled_symmetry} 
and, hence, we cannot apply the same reasoning as was done for the minimally coupled model. We further expand the above expression in $\sigma$ and $\hat{\varphi}$ around the Minkowski metric and $\hat{\phi}$ around its asymptotic value $\hat{\phi}_0=0$. 
We find that the leading-order operator due to 
the scalar-Gauss-Bonnet term is then given by
\begin{equation}
\label{eq::GB_LO_operator}
{\cal S}_\mathrm{GB}^\mathrm{(LO)} = \frac{4 \xi_2}{\mpl^2} 
\int \rd^4 x\, \hat{\phi}\, \partial_i \hat{\phi} \Big( 
\partial_j \partial^j \hat{\varphi} \partial^i \hat{\varphi} - \partial^i \partial_j \hat{\varphi} \partial^j \hat{\varphi} \Big)\,,
\end{equation}
where we performed an integration by parts and dropped a surface term. 
From this leading-order action, we find that the diagram shown in 
Fig.~\ref{fig:Gauss-Bonnet_LO} gives the leading-order contribution to the asymptotic expansion. 
Again, using dimensional analysis, it follows 
that this diagram contributes to the expansion of $H_0$ with a term proportional to $\xi_2 / \hat{r}^2$, exactly what we found in the previous section.

Since there is neither a useful symmetry in the EFT action nor a transformation to the Einstein frame, it seems necessary to calculate every diagram that contributes to the asymptotic expansion due to the scalar-Gauss-Bonnet term for determining the correct form of $H_{0,3}$ and $\delta \phi_3$, which we leave for future work.

%%%%%%%%%%%%%%%%%%%%%%%%%
\section{Conclusions 
\label{sec:conclusions}} 
%%%%%%%%%%%%%%%%%%%%%%%%%

We have investigated the ambiguity that arises 
when extracting quadrupolar tidal coefficients in scalar-tensor theories. 
We have resolved this ambiguity for the case of a minimally coupled scalar field and the DEF model by leveraging an EFT approach starting from the PP action of a compact object. In doing so, we have found that the calculation of tidal Love numbers is 
non-trivial, even in the deceivingly simple case of a scalar field that is minimally coupled to gravity in the absence of matter.

The crucial point for the correct extraction of tidal Love numbers is the introduction of isotropic coordinates given by the metric components (\ref{eq::eft_metric_decomposition}). 
When using this coordinate system for a minimally coupled 
scalar field in GR, 
the symmetry of the bulk and background 
PP actions imposes that the perturbed fields 
$\delta \hat{\varphi}$ and $\delta \hat{\phi}$ 
for an object with vanishing Love numbers
should only contain even powers of the radial coordinate 
$\hat{r}$, see 
Eq.~(\ref{eq::eft_asymp_proto_expansion}).
On the other hand, the PP action relevant 
to tidal Love numbers is given by 
Eq.~(\ref{Spp}), whose leading-order asymptotic contributions 
are in the forms (\ref{eq::tidal_LO_spherical}) and (\ref{eq::tidal_LO_spherical2}). 
To obtain the tidal and non-tidal contributions in the Schwarzschild 
coordinate (\ref{metric}) with the radial distance $r$, 
we performed the transformation between the two coordinates 
and exploited the gauge-invariant relation (\ref{eq:gauge-inv}). 
This allows us to extract the non-tidal contributions (\ref{Nonti1})-(\ref{Nonti2}) besides the tidal contributions 
(\ref{eq::tidal_part_SS_RW_coords}) and (\ref{eq::tidal_part_SS_RW_coords2}).
The formulas (\ref{eq::minimal_1o3_coefficients}) and (\ref{eq::minimal_1o3_coefficients2}) are valid for scalar-tensor theories in which a conformal transformation leads to the 
Einstein-frame action with a minimally 
coupled scalar field.

The DEF model belongs to a subclass of Horndeski theories 
in which the Jordan-frame action is given by 
Eq.~(\ref{DEFaction}) with a nonminimal 
coupling $F(\phi)=e^{-\beta \phi^2/(2\mpl^2)}$.
To obtain the tidal Love numbers in the 
Jordan frame, we used the transformation 
properties (\ref{eq::DEF_phi_relation}) 
and (\ref{H0JE}) for 
the perturbed fields $\delta \phi$ 
and $H_0$ in the Einstein 
and Jordan frames. Then, we derived 
the formulas (\ref{eq::DEF_1o3_coeffs}) and (\ref{eq::DEF_1o3_coeffs2}) 
in the Jordan frame, which allows us to 
extract the tidal Love numbers as 
Eqs.~(\ref{eq::corrected_love_numbers_def})-(\ref{eq::corrected_love_numbers_def4}).
For NSs with the DD2 equation 
of state, we numerically solved the background 
and perturbation equations 
of motion up to a sufficiently large 
distance and computed the tidal 
Love numbers in the Jordan frame.
If the Love numbers are not extracted correctly, 
we showed that the mismatch between the corrected 
and non-corrected ones can become quite noticeable 
with a deviation of up to $15$\,\% for the coupling 
$\beta = -6$.
This indicates the necessity of performing these calculations properly. Otherwise, the predictive power of the resulting GW signal template will be limited, hindering the extraction of model constraints.

While we have used the fact that the DEF model in the Jordan frame 
can be related to the Einstein frame to extract Love numbers properly, 
the same procedure cannot be applied to models where this relation 
does not exist. This is the case for the scalar 
field coupled to a Gauss-Bonnet term, 
where a conformal rescaling to the 
Einstein frame is not possible.
Instead, it will be necessary to explicitly calculate the Feynman diagrams that contribute to the asymptotic expansion, which is an endeavor that we leave for future work.

We have focused on the quadrupolar Love numbers, but studying this ambiguity for the other classes of Love numbers, most notably the dipolar ones~\cite{Bernard:2019yfz}, will be important.
We note that there have been recent advancements in the study of tidal heating of NSs 
within general relativity~\cite{HegadeKR:2023glb,Ripley:2023qxo,Ripley:2023lsq,HegadeKR:2024agt}.
Moreover, several recent results have shown that the dynamical tidal Love numbers of black 
holes (rotating or non-asymptotically flat) can be 
non-zero \cite{Saketh:2023bul,Perry:2023wmm,Chakrabarti:2013lua,Chia:2024bwc}.
It would be interesting to apply our formulation to these scenarios and their generalizations (e.g., involving a scalar field).
We leave such investigations for future work.

%%%%%%%%%%%%%%%%%%%%%%%%
\begin{acknowledgments}
%%%%%%%%%%%%%%%%%%%%%%%%

The authors thank Ryotaro Kase and Siddarth Ajith for useful discussions.  
R.F.D. acknowledges support by the Deutsche Forschungsgemeinschaft (DFG, German Research Foundation) through the CRC-TR 211 `Strong-interaction matter under extreme conditions'– project number 315477589 – TRR 211. R.F.D. is also partially supported by the State of Hesse within the Research Cluster ELEMENTS (Project ID 500/10.006) and the Helmholtz Graduate School for Hadron and Ion Research (HGS-HIRe). R.F.D. thanks the group of S.~Tsujikawa at Waseda University for their hospitality during his stay.
K.Y. acknowledges support from NSF Grant No. PHY-2309066, No. PHYS-2339969, and the Owens Family Foundation. 
S.T. was supported by the Grant-in-Aid 
for Scientific Research Fund of the JSPS No.~22K03642 and Waseda University Special Research Project No.~2024C-474. 

\end{acknowledgments}

%%%%%%%%%%%%%%%%%%%%%%%%%%%%%%%%%%%%%%%%
\onecolumngrid
\appendix

\cftlocalchange{toc}{309pt}{0cm}% change settings to suppress dots
\cftaddtitleline{toc}{section}{\textbf{Appendices}}{}
\cftlocalchange{toc}{1.55em}{2.55em}% restore original settings

\section{Background Equations \label{app::backgroundEquations}}

We present the definition of the functions that are used in the background equations of motion of the Horndeski action in Eq.~\eqref{eq::horndeski_bg_equations}.
\begin{align}
A_1&=-h^2 (G_{3,X}-2 G_{4,\phi X} ) \phi'^2-2 G_{4,\phi} h\,,\notag\\
A_2&=2 h^3 ( 2 G_{4,XX}-G_{5,\phi X} ) \phi'^3-4 h^2 ( G_{4,X}-G_{5,\phi} ) \phi'\,,\notag\\
A_3&=-h^4G_{5,XX} \phi'^4+h^2G_{5,X}  ( 3 h-1 ) \phi'^2\,,\notag\\
A_4&=h^2 ( 2 G_{4,XX}-G_{5,\phi X} ) \phi'^4+h ( 3 G_{5,\phi}-4 G_{4,X} ) \phi'^2-2 G_4\,,\notag\\
A_5&=-\frac12 \left[G_{5,XX} h^3{\phi'}^{5}- hG_{5,X}  ( 5 h-1 ) \phi'^3\right]\,,\notag\\
A_6&=h ( G_{3,\phi}-2 G_{4,\phi\phi} ) \phi'^2+G_2\,,\notag\\
A_7&=-2 h^2 ( 2 G_{4,\phi X}-G_{5,\phi\phi} ) \phi'^3-4 G_{4,\phi} h\phi'\,,\notag\\
A_8&=G_{5,\phi X} h^3\phi'^4-h ( 2 G_{4,X} h-G_{5,\phi} h-G_{5,\phi} ) \phi'^2-2 G_4  ( h-1 )\,,\notag\\
A_9&=-h ( G_{2,X}-G_{3,\phi} ) \phi'^2-G_2\,,\notag\\
A_{10}&=\frac12 G_{5,\phi X} h^3\phi'^4-\frac12 h^2 ( 2 G_{4,X}-G_{5,\phi} ) \phi'^2-G_4 h\,.
\end{align}

\section{Coefficients of perturbation 
equations of motion
\label{app::evenParityPerturbations}}

The coefficients appearing in the perturbation 
equations of motion (\ref{pereq1})-(\ref{pereq6}) 
are given by
\ba
a_1&=&\sqrt{fh} \left[  \left\{ G_{4,\phi}+\frac12 h ( G_{3,X}-2 G_{4,\phi X} ) \phi'^2 \right\} r^2
+2 h \phi' \left\{ G_{4,X}-G_{5,\phi}-\frac12h ( 2 G_{4,XX}-G_{5,\phi X} ) \phi'^2 \right\} r
\right.
\notag\\
&&
\left.
+\frac12 G_{5,XX} h^3\phi'^4-\frac12 G_{5,X} h ( 3 h-1 ) \phi'^2 \right]\,, \notag\\
a_2&=&\sqrt{fh}\left( {\frac {a_1}{\sqrt{fh}}} \right)' 
- \left( {\frac {\phi''}{\phi'}}-\frac12 {\frac {f'}{f}} \right) a_1
+{\frac {r}{\phi'} \left( {\frac {f'}{f}}-{\frac {h'}{h}} \right) a_4}
-\frac12 {\frac { \sqrt {f}r^2( \rho+P )}{\sqrt {h}\phi'}}\,,\notag\\
a_3&=&-\frac12 \phi' a_1-ra_4\,,\qquad 
a_4=\frac{\sqrt{fh}}{2} {\cal H}\,,\qquad
a_5=a_2'-a_1''\,,\qquad
a_6=- {\frac {\sqrt {f}}{2\sqrt {h}\phi'} 
\left( {\cal H}' + \frac{{\cal H}}{r}-{\frac {{\cal F}}{r}} \right) }\,,
\notag\\
a_7 &=&a_3'+\frac14 {\frac {\sqrt {f}r^2( \rho+P )}{\sqrt {h}}}\,,\qquad
a_8=-\frac12 {\frac {a_4}{h}}\,,\qquad
a_9= a_4'+ \left( \frac1r-\frac12 
{\frac{f'}{f}} \right) a_4\,,\qquad 
b_1=\frac{1}{4} \sqrt{\frac{h}{f}}{\cal H}\,,
\notag\\
c_2&=&\sqrt{fh} \left[  \left\{  
\frac{1}{2f}\left( -\frac12 h ( 3 G_{3,X}-8 G_{4,\phi X} ) \phi'^2
+\frac12 h^2 ( G_{3,XX}-2 G_{4,\phi XX} ) \phi'^4
-G_{4,\phi} \right) r^2
\right.\right.
\notag\\
&&
\left.\left.
-{\frac {h\phi'}{f}} \left( 
\frac12 {h^2 ( 2 G_{4,XXX}-G_{5,\phi XX} ) \phi'^4}
-\frac12 {h ( 12 G_{4,XX}-7 G_{5,\phi X} ) \phi'^2}
+3 ( G_{4,X}-G_{5,\phi} ) \right) r
\right.\right.
\notag\\
&&
\left.\left.
+\frac{h\phi'^2}{4f}\left(
G_{5,XXX} h^3\phi'^4
- G_{5,XX} h ( 10 h-1 ) \phi'^2
+3 G_{5,X}  ( 5 h-1 ) 
\right) \right\} f'
\right.
\notag\\
&&
\left.
+\phi' \left\{ \frac12G_{2,X}-G_{3,\phi}
-\frac12 h ( G_{2,XX}-G_{3,\phi X} ) \phi'^2 \right\} r^2
\right.
\notag\\
&&
\left.
+ 2\left\{ -\frac12h ( 3 G_{3,X}-8 G_{4,\phi X} ) \phi'^2
+\frac12h^2 ( G_{3,XX}-2 G_{4,\phi XX} ) \phi'^4
-G_{4,\phi} \right\} r
\right.
\notag\\
&&
\left.
-\frac12 h^3 ( 2 G_{4,XXX}-G_{5,\phi XX} ) \phi'^5
+\frac12 h \left\{ 2\left(6 h-1\right) G_{4,XX}+\left(1-7 h\right)G_{5,\phi X} \right\} \phi'^3
- ( 3 h-1 )  ( G_{4,X}-G_{5,\phi} ) \phi' \right] \,,\notag\\
c_3&=&-\frac12 {\frac {\sqrt {f} r^2}{\sqrt {h}}}\frac{\partial{\cal E}_{11}}{\partial\phi}\,,\notag\\
c_4&=&\frac14 \frac {\sqrt {f}}{\sqrt {h}} 
\left[ {\frac {h\phi'}{f} \left\{ 
2 G_{4,X}-2 G_{5,\phi}
-h ( 2 G_{4,XX}-G_{5,\phi X} ) \phi'^2
-{\frac {h\phi'  ( 3 G_{5,X}-G_{5,XX} \phi'^2h ) }{r}} \right\}}f'
\right.
\notag\\
&&
\left.
+4 G_{4,\phi}
+2 h ( G_{3,X}-2 G_{4,\phi X} ) \phi'^2
+{\frac {4 h ( G_{4,X}-G_{5,\phi} ) \phi'-2 h^2 ( 2 G_{4,XX}-G_{5,\phi X} ) \phi'^3}{r}} \right] \,,\notag\\
c_5&=&-h \phi'c_4-\frac12 {\frac {\sqrt{fh}}{r}}{\cal G}-\frac12 {\frac {f'}{f}}a_4\,,\notag\\
c_6&=&\frac18 {\frac {f' \phi' }{f}}a_1+\frac12 {\frac {f' r}{f}}a_4-\frac14 \phi' c_2+\frac12 h\phi' rc_4
+\frac14 \sqrt{fh}\,{\cal G}
\,,\qquad d_2=2 hc_4\,,\notag\\
d_3&=&
-{\frac {1}{r^2} \left( {\frac {2\phi''}{\phi'}}+{\frac {h'}{h}} \right) }a_1
+{\frac {2f}{ ( f' r-2 f ) \phi'} \left( 
{\frac {2\phi''}{h\phi' r}}
+ {\frac {{f'}^{2}}{f^2}}
- {\frac {f' h'}{fh}}
-{\frac {2f'}{fr}}
+{\frac {2h'}{hr}}
+ {\frac {h'}{h^2r}} \right) }a_4
\notag\\
&&
+{\frac {f' r-2 f}{fr}}\frac{\partial a_4}{\partial\phi}
+{\frac {\sqrt {f}}{\phi' \sqrt {h}r^2}}{\cal F}
-{\frac {{f}^{3/2}}{\sqrt {h} ( f' r-2 f ) \phi'} 
\left( {\frac {f'}{fr}}+{\frac {2\phi''}{\phi' r}}+{\frac {h'}{hr}}-\frac{2}{r^2} \right) }{\cal G}
-{\frac {\sqrt {f} ( \rho+P ) }{\phi' \sqrt {h}}}\,,\notag\\
e_2&=&-\frac{1}{2\phi'} \left(\frac{f'}{f}a_1+2 c_2+4 hrc_4\right)\,,\qquad
e_3=\frac14 {\frac {\sqrt {f}r^2}{\sqrt {h}}}\frac{\partial{\cal E}_{\phi}}{\partial\phi}\,,\notag\\
e_4&=&{\frac {1}{\phi'}}c_4'-\frac12 {\frac {f' }{f\phi'^2h}}a_4'
-\frac12 {\frac {\sqrt {f}}{\phi'^2\sqrt {h}r}}{\cal G}'
+{\frac {1}{h\phi' r^2} \left( {\frac {\phi''}{\phi'}}+\frac12 {\frac {h'}{h}} \right) }a_1
\notag\\
&&
+{\frac {1}{4h\phi'^2} \left[ {\frac { ( f' r-6 f ) f'}{f^2r}}
+\frac {h'  ( f' r+4 f ) }{hrf}
-{\frac {4f ( 2 \phi'' h+h' \phi' ) }{\phi' h^2r ( f' r-2 f ) }} \right] }a_4
+\frac12 {\frac {h'}{h\phi'}}c_4
-\frac12 {\frac {f' r-2 f}{fhr\phi'}}\frac{\partial a_4}{\partial \phi}
\notag\\
&&
+\frac12 {\frac {f' hr-f}{r^2\sqrt {f}\phi'^2{h}^{3/2}}}{\cal F}
+\frac12 {\frac {\sqrt {f}}{r\phi'^2{h}^{3/2}} 
\left[ {\frac {f ( 2 \phi'' h+h' \phi' ) }{h\phi'  ( f' r-2 f ) }}+\frac12 {\frac {2 f-f' hr}{fr}} \right] }{\cal G}
+\frac12 {\frac {\sqrt {f} ( \rho+P ) }{{h}^{3/2}\phi'^2}}
\,,\notag\\
g_2 &=&-\frac12 r^2a_4\,,\qquad
g_4=\frac18 r^2\sqrt{fh}{\cal G}\,,\qquad
g_{12}=-hr^2c_4\,,\qquad
g_{13}=\frac14 {\frac {r^2f'}{f}}a_4-\frac12 r^2a_4'-\frac32 ra_4\,,\notag\\
g_{14}&=&-\frac12 r^2c_5\,,\qquad
g_{15}=-\frac12 r^2d_3\,,\qquad
g_{16}=-\frac12 \sqrt{fh}\,{\cal G}\,,\notag\\
k_1&=&-\frac14 {\frac { \sqrt {f}}{\sqrt {h}}}{\cal F}\,,\qquad
k_2=\frac14 {\frac {\sqrt {f}}{\sqrt {h}}}{\cal G}\,,\qquad
k_3=\frac{2f{\cal G}'+f'({\cal G}-{\cal F})}{4\sqrt{fh}\,\phi'}\,,
\ea
with 
\ba
{\cal F} &\equiv& 2G_4+h \phi'^2 G_{5,\phi}
-h \phi'^2 \left( \frac{1}{2}h' \phi'
+h \phi'' \right) G_{5,X}\,,\nonumber \\
{\cal G} &\equiv& 2G_4+2h \phi'^2 G_{4,X}
-h \phi'^2 \left( G_{5,\phi}
+\frac{f'h \phi' G_{5,X}}{2f} \right)\,,\nonumber \\
{\cal H} &\equiv& 2G_4+2h \phi'^2 G_{4,X}
-h \phi'^2 G_{5,\phi}-\frac{h^2 \phi'^3 G_{5,X}}{r}\,.
\ea
Note that ${\cal E}_{11}$ and ${\cal E}_{\phi}$ are defined, 
respectively, by Eqs.~(\ref{back2}) and (\ref{back5}).

\section{Asymptotic Expansions \label{app::asymp_expansions}}

\subsection{Minimally Coupled Scalar Field}
\label{asymp_coeffs_minimally_coupled}

For the model studied in Sec.~\ref{sec:minimally_coupled_scalar}, 
we present the coefficients for the asymptotic expansion of the background fields and then show the coefficients for the perturbation fields.

\subsubsection{Background}

\ba
& &
h_{0} = 1\,,\qquad h_{1}= - 2 M\,,\qquad
h_{2} = \frac{\phi_1^{2}}{2 \mpl^{2}}\,,\qquad
h_{3} = \frac{M \phi_1^{2}}{2 \mpl^{2}}\,,\qquad
h_{4} = \frac{2 M^{2} \phi_1^{2}}{3 \mpl^{2}}\,,
\qquad
h_{5} = \frac{M^{3} \phi_1^{2}}{\mpl^{2}} 
- \frac{M \phi_1^{4}}{48 \mpl^{4}}\,,\nonumber \\
& &
h_{6} = \frac{8 M^{4} \phi_1^{2}}{5 \mpl^{2}} 
- \frac{M^{2} \phi_1^{4}}{10 \mpl^{4}}\,,\qquad
h_{7} = \frac{8 M^{5} \phi_1^{2}}{3 \mpl^{2}} - \frac{59 M^{3} \phi_1^{4}}{180 \mpl^{4}} 
+ \frac{M \phi_1^{6}}{320 \mpl^{6}}\,,\qquad
h_{8} = \frac{32 M^{6} \phi_1^{2}}{7 \mpl^{2}} - \frac{32 M^{4} \phi_1^{4}}{35 \mpl^{4}} 
+ \frac{8 M^{2} \phi_1^{6}}{315 \mpl^{6}}\,,
\nonumber \\
& &
h_{9} = \frac{8 M^{7} \phi_1^{2}}{\mpl^{2}} - \frac{327 M^{5} \phi_1^{4}}{140 \mpl^{4}} 
+ \frac{141 M^{3} \phi_1^{6}}{1120 \mpl^{6}} 
-\frac{5 M \phi_1^{8}}{7168 \mpl^{8}}\,,
\ea
\ba
& &
f_0=1\,,\qquad f_1=-2M\,,\qquad 
f_2=0\,,\qquad f_3=\frac{M \phi_1^{2}}{6 \mpl^{2}}\,,
\qquad f_4=\frac{M^{2} \phi_1^{2}}{3 \mpl^{2}}\,,
\qquad f_5=\frac{3 M^{3} \phi_1^{2}}{5 \mpl^{2}} - \frac{3 M \phi_1^{4}}{80 \mpl^{4}}\,,\nonumber\\
& &
f_{6} = \frac{16 M^{4} \phi_1^{2}}{15 \mpl^{2}} - \frac{8 M^{2} \phi_1^{4}}{45 \mpl^{4}}\,,\qquad
f_{7} = \frac{40 M^{5} \phi_1^{2}}{21 \mpl^{2}} 
-\frac{145 M^{3} \phi_1^{4}}{252 \mpl^{4}} + \frac{5 M \phi_1^{6}}{448 \mpl^{6}}\,,\qquad
f_{8} = \frac{24 M^{6} \phi_1^{2}}{7 \mpl^{2}} - \frac{111 M^{4} \phi_1^{4}}{70 \mpl^{4}} + \frac{3 M^{2} \phi_1^{6}}{35 \mpl^{6}}\,,\nonumber\\
& &
f_{9}=\frac{56 M^{7} \phi_1^{2}}{9 \mpl^{2}} - \frac{721 M^{5} \phi_1^{4}}{180 \mpl^{4}} + \frac{5257 M^{3} \phi_1^{6}}{12960 \mpl^{6}} - \frac{35 M \phi_1^{8}}{9216 \mpl^{8}}\,,
\ea
\ba
& &
\phi_{0}=0\,,\qquad 
\phi_{1}=\phi_1\,,\qquad
\phi_{2}= M \phi_1\,,\qquad
\phi_{3}=\frac{4 M^{2} \phi_1}{3}
-\frac{\phi_1^{3}}{12 \mpl^{2}}\,,\qquad
\phi_{4} = 2 M^{3} \phi_1 
-\frac{M \phi_1^{3}}{3 \mpl^{2}}\,, \nonumber \\
& &
\phi_{5}= \frac{16 M^{4} \phi_1}{5} - \frac{29 M^{2} \phi_1^{3}}{30 \mpl^{2}} + \frac{3 \phi_1^{5}}{160 \mpl^{4}}\,,\qquad
\phi_{6} = \frac{16 M^{5} \phi_1}{3} 
-\frac{37 M^{3} \phi_1^{3}}{15 \mpl^{2}} 
+ \frac{2 M \phi_1^{5}}{15 \mpl^{4}}\,,\nonumber \\
& &
\phi_{7} = \frac{64 M^{6} \phi_1}{7} - \frac{206 M^{4} \phi_1^{3}}{35 \mpl^{2}} + \frac{751 M^{2} \phi_1^{5}}{1260 \mpl^{4}} - \frac{5 \phi_1^{7}}{896 \mpl^{6}}\,,\qquad
\phi_{8} = 16 M^{7} \phi_1 - \frac{472 M^{5} \phi_1^{3}}{35 \mpl^{2}} + \frac{676 M^{3} \phi_1^{5}}{315 \mpl^{4}} - \frac{2 M \phi_1^{7}}{35 \mpl^{6}}
\,,\nonumber \\
& &
\phi_{9} = \frac{256 M^{8} \phi_1}{9} - \frac{9476 M^{6} \phi_1^{3}}{315 \mpl^{2}} + \frac{5717 M^{4} \phi_1^{5}}{840 \mpl^{4}} - \frac{6959 M^{2} \phi_1^{7}}{20160 \mpl^{6}} + \frac{35 \phi_1^{9}}{18432 \mpl^{8}}\,.
\ea

\subsubsection{Perturbations}

\ba
\hspace{-0.3cm}
& &
\delta \phi_{-2} = \delta\phi_{-2}\,,\qquad
\delta \phi_{-1} = - 2 M \delta\phi_{-2}\,,\qquad
\delta \phi_{0} = - \frac{H_{0,-2} M \phi_{1}}{3} 
+ \frac{2 M^{2} \delta\phi_{-2}}{3}\,,\qquad
\delta \phi_{1} = \frac{M \delta\phi_{-2} \phi_{1}^{2}}{6 \mpl^{2}}\,,\nonumber \\
\hspace{-0.3cm}
& &
\delta \phi_{2} = \frac{M^{2} \delta\phi_{-2} \phi_{1}^{2}}{3 \mpl^{2}}\,,\qquad
\delta \phi_{3} = \delta\phi_{3}\,,\qquad
\delta \phi_{4} = 3 M \delta\phi_{3} + \delta\phi_{-2} \left(- \frac{11 M^{4} \phi_{1}^{2}}{15 \mpl^{2}} -\frac{47 M^{2} \phi_{1}^{4}}{720 \mpl^{4}}\right)\,,\nonumber\\
\hspace{-0.3cm}
& &
\delta \phi_{5} = H_{0,-2} \left(- \frac{3 M^{4} \phi_{1}^{3}}{35 \mpl^{2}} + \frac{3 M^{2} \phi_{1}^{5}}{560 \mpl^{4}}\right) + \frac{H_{0,3} M \phi_{1}}{7} + \delta\phi_{-2} \left(- \frac{232 M^{5} \phi_{1}^{2}}{105 \mpl^{2}} - \frac{19 M^{3} \phi_{1}^{4}}{90 \mpl^{4}} + \frac{M \phi_{1}^{6}}{224 \mpl^{6}}\right) \nonumber\\
\hspace{-0.3cm}
&&\phantom{{} = {}}~~~+ \delta\phi_{3} \left(\frac{48 M^{2}}{7} - \frac{5 \phi_{1}^{2}}{28 \mpl^{2}}\right)\,,\nonumber\\
\hspace{-0.3cm}
& &
\delta \phi_{6} = H_{0,-2} \left(- \frac{3 M^{5} \phi_{1}^{3}}{7 \mpl^{2}} + \frac{3 M^{3} \phi_{1}^{5}}{112 \mpl^{4}}\right) + \frac{5 H_{0,3} M^{2} \phi_{1}}{7} + \delta\phi_{-2} \left(- \frac{36 M^{6} \phi_{1}^{2}}{7 \mpl^{2}} - \frac{51 M^{4} \phi_{1}^{4}}{140 \mpl^{4}} + \frac{3 M^{2} \phi_{1}^{6}}{70 \mpl^{6}}\right) \nonumber \\
\hspace{-0.3cm}
&&\phantom{{} = {}}~~~+ \delta\phi_{3} \left(\frac{100 M^{3}}{7} - \frac{8 M \phi_{1}^{2}}{7 \mpl^{2}}\right)\,,\nonumber\\
\hspace{-0.3cm}
& &
\delta \phi_{7} = H_{0,-2} \left(- \frac{10 M^{6} \phi_{1}^{3}}{7 \mpl^{2}} + \frac{M^{4} \phi_{1}^{5}}{8 \mpl^{4}} - \frac{M^{2} \phi_{1}^{7}}{448 \mpl^{6}}\right) + H_{0,3} \left(\frac{50 M^{3} \phi_{1}}{21} - \frac{5 M \phi_{1}^{3}}{84 \mpl^{2}}\right) \nonumber \\
\hspace{-0.3cm}
&&\phantom{{} = {}}~~~+ \delta\phi_{-2} \left(- \frac{688 M^{7} \phi_{1}^{2}}{63 \mpl^{2}} - \frac{37 M^{5} \phi_{1}^{4}}{180 \mpl^{4}} + \frac{18493 M^{3} \phi_{1}^{6}}{90720 \mpl^{6}} - \frac{17 M \phi_{1}^{8}}{9216 \mpl^{8}}\right) + \delta\phi_{3} \left(\frac{200 M^{4}}{7} - \frac{191 M^{2} \phi_{1}^{2}}{42 \mpl^{2}} + \frac{5 \phi_{1}^{4}}{96 \mpl^{4}}\right)\,,\nonumber\\
\hspace{-0.3cm}
& &
\delta \phi_{8} = H_{0,-2} \left(- \frac{4 M^{7} \phi_{1}^{3}}{\mpl^{2}} + \frac{15 M^{5} \phi_{1}^{5}}{28 \mpl^{4}} - \frac{M^{3} \phi_{1}^{7}}{56 \mpl^{6}}\right) + H_{0,3} \left(\frac{20 M^{4} \phi_{1}}{3} - \frac{10 M^{2} \phi_{1}^{3}}{21 \mpl^{2}}\right) \nonumber\\
\hspace{-0.3cm}
& &\phantom{{} = {}}~~~ 
-\delta\phi_{-2} \left( \frac{200 M^{8} \phi_{1}^{2}}{9 \mpl^{2}} - \frac{569 M^{6} \phi_{1}^{4}}{450 \mpl^{4}} - \frac{38657 M^{4} \phi_{1}^{6}}{56700 \mpl^{6}} + \frac{139 M^{2} \phi_{1}^{8}}{6300 \mpl^{8}}\right) + \delta\phi_{3} \left(56 M^{5} - \frac{1532 M^{3} \phi_{1}^{2}}{105 \mpl^{2}} + \frac{52 M \phi_{1}^{4}}{105 \mpl^{4}}\right),
\ea
\ba
& &
H_{0, -2} = H_{0,-2}\,,\qquad
H_{0, -1} = - 2 H_{0,-2} M\,,\qquad
H_{0, 0} = \frac{H_{0,-2} \phi_{1}^{2}}{3 \mpl^{2}} 
-\frac{2 M \delta\phi_{-2} \phi_{1}}{3 \mpl^{2}}\,,\qquad
H_{0, 1} = \frac{H_{0,-2} M \phi_{1}^{2}}{6 \mpl^{2}}\,, \nonumber\\
& &
H_{0, 2} = \frac{H_{0,-2} M^{2} \phi_{1}^{2}}
{3 \mpl^{2}}\,,\qquad
H_{0, 3} = H_{0,3}\,,\qquad
H_{0, 4}= -H_{0,-2} \left( \frac{11 M^{4} \phi_{1}^{2}}{15 \mpl^{2}} + \frac{47 M^{2} \phi_{1}^{4}}{720 \mpl^{4}}\right) + 3 H_{0,3} M\,,
\nonumber \\
& &
H_{0, 5} = H_{0,-2} \left(- \frac{50 M^{5} \phi_{1}^{2}}{21 \mpl^{2}} - \frac{289 M^{3} \phi_{1}^{4}}{2520 \mpl^{4}} - \frac{M \phi_{1}^{6}}{1120 \mpl^{6}}\right) + H_{0,3} \left(\frac{50 M^{2}}{7} - \frac{9 \phi_{1}^{2}}{28 \mpl^{2}}\right) + \frac{2 M \delta\phi_{3} \phi_{1}}{7 \mpl^{2}} 
\nonumber \\
& &~~~~
\phantom{{} = {}} + \delta\phi_{-2} \left(- \frac{6 M^{4} \phi_{1}^{3}}{35 \mpl^{4}} + \frac{3 M^{2} \phi_{1}^{5}}{280 \mpl^{6}}\right)\,,
\nonumber \\
& &
H_{0, 6} = H_{0,-2} \left(- \frac{6 M^{6} \phi_{1}^{2}}{\mpl^{2}} + \frac{33 M^{4} \phi_{1}^{4}}{280 \mpl^{4}} + \frac{9 M^{2} \phi_{1}^{6}}{560 \mpl^{6}}\right) + H_{0,3} \left(\frac{110 M^{3}}{7} - \frac{13 M \phi_{1}^{2}}{7 \mpl^{2}}\right) + \frac{10 M^{2} \delta\phi_{3} \phi_{1}}{7 \mpl^{2}} 
\nonumber \\
& &~~~~
\phantom{{} = {}} + \delta\phi_{-2} \left(- \frac{6 M^{5} \phi_{1}^{3}}{7 \mpl^{4}} + \frac{3 M^{3} \phi_{1}^{5}}{56 \mpl^{6}}\right)\,, \nonumber\\
& &
H_{0, 7}= H_{0,-2} \left(- \frac{124 M^{7} \phi_{1}^{2}}{9 \mpl^{2}} + \frac{464 M^{5} \phi_{1}^{4}}{315 \mpl^{4}} + \frac{241 M^{3} \phi_{1}^{6}}{3240 \mpl^{6}} + \frac{25 M \phi_{1}^{8}}{64512 \mpl^{8}}\right) + H_{0,3} \left(\frac{100 M^{4}}{3} - \frac{148 M^{2} \phi_{1}^{2}}{21 \mpl^{2}} + \frac{25 \phi_{1}^{4}}{224 \mpl^{4}}\right) \nonumber \\
& &~~~~
\phantom{{} = {}} + \delta\phi_{-2} \left(- \frac{20 M^{6} \phi_{1}^{3}}{7 \mpl^{4}} + \frac{M^{4} \phi_{1}^{5}}{4 \mpl^{6}} - \frac{M^{2} \phi_{1}^{7}}{224 \mpl^{8}}\right) + \delta\phi_{3} \left(\frac{100 M^{3} \phi_{1}}{21 \mpl^{2}} - \frac{5 M \phi_{1}^{3}}{42 \mpl^{4}}\right)\,,
\nonumber \\
& &
H_{0, 8} = H_{0,-2} \left(- \frac{272 M^{8} \phi_{1}^{2}}{9 \mpl^{2}} + \frac{9979 M^{6} \phi_{1}^{4}}{1575 \mpl^{4}} + \frac{6257 M^{4} \phi_{1}^{6}}{56700 \mpl^{6}} - \frac{53 M^{2} \phi_{1}^{8}}{12600 \mpl^{8}}\right) + H_{0,3} \left(\frac{208 M^{5}}{3} - \frac{2332 M^{3} \phi_{1}^{2}}{105 \mpl^{2}} + \frac{34 M \phi_{1}^{4}}{35 \mpl^{4}}\right) \nonumber \\
& &~~~~
\phantom{{} = {}} + \delta\phi_{-2} \left(- \frac{8 M^{7} \phi_{1}^{3}}{\mpl^{4}} + \frac{15 M^{5} \phi_{1}^{5}}{14 \mpl^{6}} - \frac{M^{3} \phi_{1}^{7}}{28 \mpl^{8}}\right) + \delta\phi_{3} \left(\frac{40 M^{4} \phi_{1}}{3 \mpl^{2}} - \frac{20 M^{2} \phi_{1}^{3}}{21 \mpl^{4}}\right)\,.
\ea

\subsection{DEF Model}
\label{asymp_coeffs_DEF}

For the model discussed in Sec.~\ref{sec:DEF_model}, 
we show the coefficients for the asymptotic expansion of the 
background fields and also present the coefficients for 
the perturbation fields.

\subsubsection{Background}

\ba
& &
h_{0} = 1\,,\qquad h_{1} = - 2M\,,\qquad
h_{2} = \frac{\phi_{1}^{2} \left(1 - 2\beta \right)}{2\mpl^{2}}\,,\qquad
h_{3} = \frac{M \phi_{1}^{2} 
\left(1 - 3 \beta\right)}{2 \mpl^{2}}\,,
\nonumber \\
& &
h_{4} = \frac{M^{2} \phi_{1}^{2} \left(2 - 7 \beta\right)}{3 \mpl^{2}} + \frac{\beta \phi_{1}^{4} \left(1 - 3 \beta\right)}{12 \mpl^{4}}\,,\qquad
h_{5} = \frac{M^{3} \phi_{1}^{2} \left(6 - 23 \beta\right)}{6 \mpl^{2}} + \frac{M \phi_{1}^{4} \left(- 51 \beta^{2} + 18 \beta - 1\right)}{48 \mpl^{4}}\,,\nonumber\\
& &
h_{6} = \frac{2 M^{4} \phi_{1}^{2} \left(12 - 49 \beta\right)}{15 \mpl^{2}} + \frac{M^{2} \phi_{1}^{4} \left(- 195 \beta^{2} + 73 \beta - 6\right)}{60 \mpl^{4}} + \frac{\beta \phi_{1}^{6} \left(- 30 \beta^{2} + 15 \beta - 2\right)}{240 \mpl^{6}}\,,\nonumber\\
& &
h_{7} = \frac{2 M^{5} \phi_{1}^{2} \left(60 - 257 \beta\right)}{45 \mpl^{2}} + \frac{M^{3} \phi_{1}^{4} \left(- 3135 \beta^{2} + 1229 \beta - 118\right)}{360 \mpl^{4}} + \frac{M \phi_{1}^{6} \left(- 875 \beta^{3} + 475 \beta^{2} - 79 \beta + 3\right)}{960 \mpl^{6}}\,,\nonumber\\
& &
h_{8} = \frac{8 M^{6} \phi_{1}^{2} \left(20 - 89 \beta\right)}{35 \mpl^{2}} + \frac{2 M^{4} \phi_{1}^{4} \left(- 1141 \beta^{2} + 463 \beta - 48\right)}{105 \mpl^{4}} + \frac{M^{2} \phi_{1}^{6} \left(- 10605 \beta^{3} + 6132 \beta^{2} - 1152 \beta + 64\right)}{2520 \mpl^{6}} \nonumber\\
& &~~~~\phantom{=} + \frac{\beta \phi_{1}^{8} \left(- 420 \beta^{3} + 315 \beta^{2} - 84 \beta + 8\right)}{5040 \mpl^{8}}\,, \nonumber \\
& &
h_{9} = \frac{M^{7} \phi_{1}^{2} \left(280 -1286 
\beta \right)}{35\mpl^{2}} -\frac{M^{5} \phi_{1}^{4} \left( 87185 \beta^{2} - 36300 \beta + 3924\right)}{1680 \mpl^{4}} - \frac{M^{3} \phi_{1}^{6} \left( 105525 \beta^{3} - 64036 \beta^{2}+ 12975 \beta - 846\right)}
{6720 \mpl^{6}} \nonumber\\
& &~~~~\phantom{=} + \frac{M \phi_{1}^{8} \left(- 93555 \beta^{4} + 74340 \beta^{3} - 21686 \beta^{2} + 2556 \beta - 75\right)}{107520 \mpl^{8}}\,,
\ea
\ba
& &
f_{0} = 1\,,\qquad
f_{1} = - 2 M\,,\qquad
f_{2} = \frac{\beta \phi_{1}^{2}}{2 \mpl^{2}}\,,\qquad
f_{3} = \frac{M \phi_{1}^{2} \left(1 - 3 \beta\right)}{6 \mpl^{2}}\,,
\nonumber \\
& &
f_{4} = \frac{M^{2} \phi_{1}^{2} \left(2 - 7 \beta\right)}{6 \mpl^{2}} + \frac{\beta \phi_{1}^{4} \left(9 \beta - 2\right)}{24 \mpl^{4}}\,,\qquad
f_{5} = \frac{M^{3} \phi_{1}^{2} \left(18 - 65 \beta\right)}{30 \mpl^{2}} + \frac{M \phi_{1}^{4} \left(105 \beta^{2} + 10 \beta - 9\right)}{240 \mpl^{4}}\,,\nonumber \\
& &
f_{6} = \frac{16 M^{4} \phi_{1}^{2} \left(3 - 11 \beta\right)}{45 \mpl^{2}} + \frac{M^{2} \phi_{1}^{4} \left(- 15 \beta^{2} + 34 \beta - 8\right)}{45 \mpl^{4}} + \frac{\beta \phi_{1}^{6} \left(30 \beta^{2} - 15 \beta + 2\right)}{90 \mpl^{6}}\,,\nonumber \\
& &
f_{7} = \frac{2 M^{5} \phi_{1}^{2} \left(100 - 371 \beta\right)}{105 \mpl^{2}} + \frac{M^{3} \phi_{1}^{4} \left(- 8505 \beta^{2} + 7567 \beta - 1450\right)}{2520 \mpl^{4}} + \frac{M \phi_{1}^{6} \left(25515 \beta^{3} - 10395 \beta^{2} + 259 \beta + 225\right)}{20160 \mpl^{6}}\,,\nonumber \\
& &
f_{8}= \frac{6 M^{6} \phi_{1}^{2} \left(4 - 15 \beta\right)}{7 \mpl^{2}} + \frac{M^{4} \phi_{1}^{4} \left(- 10255 \beta^{2} + 7572 \beta - 1332\right)}{840 \mpl^{4}} + \frac{M^{2} \phi_{1}^{6} \left(4375 \beta^{3} - 560 \beta^{2} - 660 \beta + 144\right)}{1680 \mpl^{6}} \nonumber \\
& &\phantom{{} = {}}~~+ \frac{\beta \phi_{1}^{8} \left(4375 \beta^{3} - 3500 \beta^{2} + 980 \beta - 96\right)}{13440 \mpl^{8}}\,, \nonumber \\
& &
f_{9}= \frac{2 M^{7} \phi_{1}^{2} \left(980 - 3713 \beta\right)}{315 \mpl^{2}} + \frac{M^{5} \phi_{1}^{4} \left(- 530145 \beta^{2} + 361244 \beta - 60564\right)}{15120 \mpl^{4}}  \nonumber \\
& &\phantom{{} = {}}~~~+ \frac{M^{3} \phi_{1}^{6} \left(343035 \beta^{3} + 701316 \beta^{2} - 473979 \beta + 73598\right)}{181440 \mpl^{6}} \nonumber \\
& &\phantom{{} = {}}~~~~+ \frac{M \phi_{1}^{8} \left(2096325 \beta^{4} - 1575420 \beta^{3} + 369138 \beta^{2} - 12916 \beta - 3675\right)}{967680 \mpl^{8}}\,,
\ea
\ba
& &
\phi_{0} = 0\,,\qquad
\phi_{1} = \phi_{1}\,,\qquad
\phi_{2} = M \phi_{1}\,,\qquad
\phi_{3} = \frac{4 M^{2} \phi_{1}}{3} + \frac{\phi_{1}^{3} \left(3 \beta - 1\right)}
{12 \mpl^{2}}\,,\qquad 
\phi_{4} = 2 M^{3} \phi_{1} + \frac{M \phi_{1}^{3} \left(3 \beta - 1\right)}{3 \mpl^{2}}\,,\nonumber \\
& &
\phi_{5} = \frac{16 M^{4} \phi_{1}}{5} + \frac{M^{2} \phi_{1}^{3} \left(175 \beta - 58\right)}{60 \mpl^{2}} + \frac{\phi_{1}^{5} \left(75 \beta^{2} - 50 \beta + 9\right)}{480 \mpl^{4}}\,, \nonumber \\
& &
\phi_{6}= \frac{16 M^{5} \phi_{1}}{3} + \frac{M^{3} \phi_{1}^{3} \left(225 \beta - 74\right)}{30 \mpl^{2}} + \frac{M \phi_{1}^{5} \left(135 \beta^{2} - 90 \beta + 16\right)}{120 \mpl^{4}}\,, \nonumber \\
& &
\phi_{7} = \frac{64 M^{6} \phi_{1}}{7} + \frac{2 M^{4} \phi_{1}^{3} \left(2842 \beta - 927\right)}{315 \mpl^{2}} + \frac{M^{2} \phi_{1}^{5} \left(25725 \beta^{2} - 17101 \beta + 3004\right)}{5040 \mpl^{4}} + \frac{\phi_{1}^{7} \left(5145 \beta^{3} - 5145 \beta^{2} + 1813 \beta - 225\right)}{40320 \mpl^{6}}
\,, \nonumber \\
& &
\phi_{8} = 16 M^{7} \phi_{1} + \frac{4 M^{5} \phi_{1}^{3} \left(3283 \beta - 1062\right)}{315 \mpl^{2}} + \frac{4 M^{3} \phi_{1}^{5} \left(1470 \beta^{2} - 973 \beta + 169\right)}{315 \mpl^{4}} + \frac{M \phi_{1}^{7} \left(140 \beta^{3} - 140 \beta^{2} + 49 \beta - 6\right)}{105 \mpl^{6}}
\,, \nonumber \\
& &
\phi_{9} = \frac{256 M^{8} \phi_{1}}{9} + \frac{M^{6} \phi_{1}^{3} \left(29531 \beta - 9476\right)}{315 \mpl^{2}} + \frac{M^{4} \phi_{1}^{5} \left(202041 \beta^{2} - 133060 \beta + 22868\right)}{3360 \mpl^{4}} \nonumber \\
& &\phantom{{} = {}}
~~~+ \frac{M^{2} \phi_{1}^{7} \left(331695 \beta^{3} - 331128 \beta^{2} + 115077 \beta - 13918\right)}{40320 \mpl^{6}}\nonumber \\
& &\phantom{{} = {}}
~~~
+ \frac{\phi_{1}^{9} \left(76545 \beta^{4} - 102060 \beta^{3} + 53298 \beta^{2} - 12916 \beta + 1225\right)}{645120 \mpl^{8}}\,.
\ea

\subsubsection{Perturbations}

\ba
& &
\delta \phi_{-2} = \delta\phi_{-2}\,,\qquad
\delta \phi_{-1} = - 2 M \delta \phi_{-2}\,,\qquad
\delta \phi_{0} = -\frac{H_{0,-2} M \phi_{1}}{3} + \delta\phi_{-2} \left( \frac{2 M^{2}}{3} - \frac{\beta \phi_{1}^{2}}{2 \mpl^{2}} \right)\,,\nonumber \\
& &
\delta \phi_{1} = \frac{M \delta\phi_{-2} \phi_{1}^{2} \left(1 - 3 \beta\right)}{6 \mpl^{2}}\,,\qquad
\delta \phi_{2} = \delta\phi_{-2} 
\left[ \frac{M^{2} \phi_{1}^{2} \left(2 - 5 \beta\right)}{6 \mpl^{2}} + \frac{\beta \phi_{1}^{4} \left(2 - 3 \beta\right)}{24 \mpl^{4}}
\right]\,,\qquad
\delta \phi_{3} = \delta\phi_{3}\,, \nonumber \\
& &
\delta \phi_{4} = 3 M \delta\phi_{3} + \delta\phi_{-2} \left[ \frac{M^{4} \phi_{1}^{2} \left(157 \beta - 66\right)}{90 \mpl^{2}} + \frac{M^{2} \phi_{1}^{4} \left(- 105 \beta^{2} + 86 \beta - 47\right)}{720 \mpl^{4}} + \frac{\beta \phi_{1}^{6} \left(- 15 \beta^{2} + 15 \beta - 4\right)}{180 \mpl^{6}}\right]\,, \nonumber \\
& &
\delta \phi_{5} = H_{0,-2} \left[ \frac{M^{4} \phi_{1}^{3} \left(95 \beta - 54\right)}{630 \mpl^{2}} + \frac{M^{2} \phi_{1}^{5} \left(345 \beta^{2} - 250 \beta + 27\right)}{5040 \mpl^{4}} \right]
+ \frac{H_{0,3} M \phi_{1}}{7}\,, \nonumber \\
& &\phantom{{} = {}}~~~+ \delta\phi_{-2} 
\left[ \frac{2 M^{5} \phi_{1}^{2} \left(817 \beta - 348\right)}{315 \mpl^{2}} + \frac{M^{3} \phi_{1}^{4} \left(- 450 \beta^{2} + 379 \beta - 266\right)}{1260 \mpl^{4}} + \frac{M \phi_{1}^{6} \left(- 815 \beta^{3} + 915 \beta^{2} - 319 \beta + 15\right)}{3360 \mpl^{6}}\right] \nonumber \\
& &\phantom{{} = {}}~~~
+ \delta\phi_{3} \left[ \frac{48 M^{2}}{7} + \frac{\phi_{1}^{2} \left(21 \beta - 5\right)}{28 \mpl^{2}}\right]\,,\nonumber \\
& &
\delta \phi_{6}= H_{0,-2} \left[ \frac{M^{5} \phi_{1}^{3} \left(95 \beta - 54\right)}{126 \mpl^{2}} + \frac{M^{3} \phi_{1}^{5} \left(345 \beta^{2} - 250 \beta + 27\right)}{1008 \mpl^{4}}\right]
+ \frac{5 H_{0,3} M^{2} \phi_{1}}{7} \nonumber \\
& &\phantom{{} = {}}~~~
+ \delta\phi_{-2} \left[ \frac{4 M^{6} \phi_{1}^{2} \left(941 \beta - 405\right)}{315 \mpl^{2}} + \frac{M^{4} \phi_{1}^{4} \left(1149 \beta^{2} - 712 \beta - 918\right)}{2520 \mpl^{4}} \right. 
\nonumber \\
  & &{}\hspace{2.2cm} \left. + \frac{M^{2} \phi_{1}^{6} \left(- 1215 \beta^{3} + 1486 \beta^{2} - 666 \beta + 72\right)}{1680 \mpl^{6}} + \frac{\beta \phi_{1}^{8} \left(- 315 \beta^{3} + 420 \beta^{2} - 196 \beta + 32\right)}{4480 \mpl^{8}}\right] \nonumber \\
& &\phantom{{} = {}}~~~ + \delta\phi_{3} \left[ \frac{100 M^{3}}{7} + \frac{M \phi_{1}^{2} \left(63 \beta - 16\right)}{14 \mpl^{2}}\right]\,, 
\nonumber \\
& &
\delta \phi_{7}= H_{0,-2} \left[ \frac{5 M^{6} \phi_{1}^{3} \left(95 \beta - 54\right)}{189 \mpl^{2}} + \frac{M^{4} \phi_{1}^{5} \left(2010 \beta^{2} - 1507 \beta + 189\right)}{1512 \mpl^{4}} + \frac{M^{2} \phi_{1}^{7} \left(1035 \beta^{3} - 1095 \beta^{2} + 331 \beta - 27\right)}{12096 \mpl^{6}}\right]
\nonumber \\
& &\phantom{{} = {}}~~~
+ H_{0,3} \left[ \frac{50 M^{3} \phi_{1}}{21} + \frac{5 M \phi_{1}^{3} \left(3 \beta - 1\right)}{84 \mpl^{2}} \right] \nonumber \\
& &\phantom{{} = {}}~~~ 
+ \delta\phi_{-2} \left[ \frac{2 M^{7} \phi_{1}^{2} \left(11887 \beta - 5160\right)}{945 \mpl^{2}} + \frac{M^{5} \phi_{1}^{4} \left(29187 \beta^{2} - 21556 \beta - 1036\right)}{5040 \mpl^{4}} \right. 
\nonumber \\
  & &
  {}\hspace{2.1cm} \left. + \frac{M^{3} \phi_{1}^{6} \left(- 392715 \beta^{3} + 499068 \beta^{2} - 258741 \beta + 36986\right)}{181440 \mpl^{6}} \right. \nonumber \\
  & &{}\hspace{2.1cm} \left. + \frac{M \phi_{1}^{8} \left(- 415305 \beta^{4} + 589740 \beta^{3} - 303066 \beta^{2} + 57796 \beta - 1785\right)}{967680 \mpl^{8}}\right] \nonumber \\
& &\phantom{{} = {}}~~~~ +\delta\phi_{3} \left[ \frac{200 M^{4}}{7} + \frac{M^{2} \phi_{1}^{2} \left( 1455 \beta - 382\right)}{84 \mpl^{2}} + \frac{\phi_{1}^{4} \left(441 \beta^{2} - 234 \beta + 35\right)}{672 \mpl^{4}} \right]\,,\nonumber \\
& &
\delta \phi_{8} = H_{0,-2} \left[
\frac{M^{7} \phi_{1}^{3}(190 \beta-108)}{27\mpl^{2}} + \frac{M^{5} \phi_{1}^{5} \left(1185 \beta^{2} - 926 \beta + 135\right)}{252 \mpl^{4}} + \frac{M^{3} \phi_{1}^{7} \left(1035 \beta^{3} - 1095 \beta^{2} + 331 \beta - 27\right)}{1512 \mpl^{6}}\right] \nonumber \\
& &\phantom{{} = {}}~~~+ H_{0,3} 
\left[ \frac{20 M^{4} \phi_{1}}{3} + \frac{10 M^{2} \phi_{1}^{3} \left(3 \beta - 1\right)}{21 \mpl^{2}}\right] \nonumber \\
& &
\phantom{{} = {}}~~~+ \delta\phi_{-2} \left[
\frac{4 M^{8} \phi_{1}^{2} \left(60037 \beta - 26250\right)}{4725 \mpl^{2}} + \frac{M^{6} \phi_{1}^{4} \left(240155 \beta^{2} - 180038 \beta + 11949\right)}{9450 \mpl^{4}}\right. \nonumber \\
  & & {}\hspace{2.0cm} \left. + \frac{M^{4} \phi_{1}^{6} \left(- 294480 \beta^{3} + 389205 \beta^{2} - 232266 \beta + 38657\right)}{56700 \mpl^{6}} \right. \nonumber \\
  & &{}\hspace{2.0cm} \left.+ \frac{M^{2} \phi_{1}^{8} \left(- 141375 \beta^{4} + 211560 \beta^{3} - 119745 \beta^{2} + 26914 \beta - 1668\right)}{75600 \mpl^{8}} \right. \nonumber \\
  & &{}\hspace{2.0cm} \left.+ \frac{\beta \phi_{1}^{10} \left(- 1260 \beta^{4} + 2100 \beta^{3} - 1365 \beta^{2} + 410 \beta - 48\right)}{18900 \mpl^{10}}\right] \nonumber \\
& &\phantom{{} = {}}~~~+ \delta\phi_{3} 
\left[ 56 M^{5} + \frac{2 M^{3} \phi_{1}^{2} \left( 2865 \beta - 766\right)}{105 \mpl^{2}} + \frac{M \phi_{1}^{4} \left(630 \beta^{2} - 345 \beta + 52\right)}{105 \mpl^{4}}\right]\,,
\ea
\ba
& &
H_{0, -2} = H_{0,-2}\,,\qquad 
H_{0, -1} = - 2 H_{0,-2} M - \frac{\beta \delta\phi_{-2} \phi_{1}}{\mpl^{2}}\,,\qquad
H_{0, 0} = \frac{H_{0,-2} \phi_{1}^{2} \left(2 - 3 \beta\right)}{6 \mpl^{2}} + \frac{M \delta\phi_{-2} \phi_{1} \left(3 \beta - 2\right)}{3 \mpl^{2}}\,,
\nonumber \\
& &
H_{0, 1} = \frac{H_{0,-2} M \phi_{1}^{2} \left(1 - \beta\right)}{6 \mpl^{2}} + \frac{\beta \delta\phi_{-2} \phi_{1}^{3} \left(3 \beta + 1\right)}{12 \mpl^{4}}\,,\nonumber \\
& &
H_{0, 2} = H_{0,-2} \left[ \frac{M^{2} \phi_{1}^{2} \left(2 - 3 \beta\right)}{6 \mpl^{2}} + \frac{\beta \phi_{1}^{4} \left(2 - 3 \beta\right)}{24 \mpl^{4}}\right] + \frac{M \beta^{2} \delta\phi_{-2} \phi_{1}^{3}}{2 \mpl^{4}}\,,\qquad
H_{0, 3} = H_{0,3}\,,\nonumber \\
& &
H_{0, 4} = H_{0,-2} \left[ \frac{M^{4} \phi_{1}^{2} \left(97 \beta - 66\right)}{90 \mpl^{2}} + \frac{M^{2} \phi_{1}^{4} \left(- 45 \beta^{2} + 66 \beta - 47\right)}{720 \mpl^{4}} + \frac{\beta \phi_{1}^{6} \left(- 15 \beta^{2} + 15 \beta - 4\right)}{180 \mpl^{6}}\right]+ 3 H_{0,3} M 
\nonumber \\
& &\phantom{{} = {}}
~~~~
- \frac{\beta \delta\phi_{3} \phi_{1}}{\mpl^{2}} + \delta\phi_{-2} \left[ \frac{M^{3} \beta \phi_{1}^{3} \left(38 - 155 \beta\right)}{60 \mpl^{4}} + \frac{M \beta \phi_{1}^{5} \left(- 495 \beta^{2} + 390 \beta - 37\right)}{1440 \mpl^{6}}\right]\,,\nonumber \\
& &
H_{0, 5} = H_{0,-2} \left[\frac{M^{5} \phi_{1}^{2} \left(367 \beta - 250\right)}{105 \mpl^{2}} + \frac{M^{3} \phi_{1}^{4} \left(- 55 \beta^{2} - 56 \beta - 289\right)}{2520 \mpl^{4}} + \frac{M \phi_{1}^{6} \left(- 2415 \beta^{3} + 2035 \beta^{2} - 471 \beta - 9\right)}{10080 \mpl^{6}}\right] 
\nonumber \\
& & \phantom{{} = {}}~~~~
+ H_{0,3} \left[ \frac{50 M^{2}}{7} + \frac{3 \phi_{1}^{2} \left(7 \beta - 3\right)}{28 \mpl^{2}}\right] + \frac{2 M \delta\phi_{3} \phi_{1} \left(1 - 14 \beta\right)}{7 \mpl^{2}} 
\nonumber \\
& &\phantom{{} = {}}~~~~
+ \delta\phi_{-2} \left[
\frac{8 M^{6} \beta \phi_{1}}{315 \mpl^{2}} + \frac{M^{4} \phi_{1}^{3} \left(- 5987 \beta^{2} + 1880 \beta - 108\right)}{630 \mpl^{4}} + \frac{M^{2} \phi_{1}^{5} \left(- 9255 \beta^{3} + 7434 \beta^{2} - 1073 \beta + 54\right)}{5040 \mpl^{6}}  \right. \nonumber \\
& &{}\hspace{2.2cm} \left. + \frac{\beta \phi_{1}^{7} \left(- 105 \beta^{3} + 135 \beta^{2} + 19 \beta - 9\right)}{20160 \mpl^{8}}\right]\,, 
\nonumber \\
& &
H_{0, 6} = H_{0,-2} \left[ 
\frac{M^{6} \phi_{1}^{2} \left(397 \beta - 270\right)}{45 \mpl^{2}} + \frac{M^{4} \phi_{1}^{4} \left(2404 \beta^{2} - 3698 \beta + 297\right)}{2520 \mpl^{4}}  \right. \nonumber \\
& &{}\hspace{2.1cm} \left. + \frac{M^{2} \phi_{1}^{6} \left(- 7305 \beta^{3} + 5516 \beta^{2} - 1557 \beta + 162\right)}{10080 \mpl^{6}} + \frac{\beta \phi_{1}^{8} \left(- 315 \beta^{3} + 420 \beta^{2} - 196 \beta + 32\right)}{4480 \mpl^{8}}\right]
\nonumber \\
& &\phantom{{} = {}}~~~~
+ H_{0,3} \left[ \frac{110 M^{3}}{7} + \frac{M \phi_{1}^{2} \left(61 \beta - 26\right)}{14 \mpl^{2}}\right] \nonumber \\
& &\phantom{{} = {}}~~~~ + \delta\phi_{-2} \left[ \frac{8 M^{7} \beta \phi_{1}}{63 \mpl^{2}} + \frac{M^{5} \phi_{1}^{3} \left(- 1613 \beta^{2} + 589 \beta - 54\right)}{63 \mpl^{4}} + \frac{M^{3} \phi_{1}^{5} \left(- 20190 \beta^{3} + 15503 \beta^{2} - 2529 \beta + 135\right)}{2520 \mpl^{6}}  \right. 
\nonumber \\
& &{}\hspace{2.2cm} \left. + \frac{M \beta \phi_{1}^{7} \left(- 2745 \beta^{3} + 2770 \beta^{2} - 623 \beta + 24\right)}{6720 \mpl^{8}}\right] + \delta\phi_{3} \left[ \frac{5 M^{2} \phi_{1} \left(6 - 47 \beta\right)}{21 \mpl^{2}} + \frac{\beta \phi_{1}^{3} \left(11 - 42 \beta\right)}{42 \mpl^{4}}\right] \,,\nonumber \\
& &
H_{0, 7} = H_{0,-2} \left[ \frac{4 M^{7} \phi_{1}^{2} \left(4801 \beta - 3255\right)}{945 \mpl^{2}} + \frac{M^{5} \phi_{1}^{4} \left(85385 \beta^{2} - 117256 \beta + 22272\right)}{15120 \mpl^{4}} \right. \nonumber \\
& & {}\hspace{2.1cm} \left. + \frac{M^{3} \phi_{1}^{6} \left(- 372105 \beta^{3} + 234366 \beta^{2} - 71415 \beta + 13496\right)}{181440 \mpl^{6}}  \right. \nonumber \\
& & {}\hspace{2.1cm} \left. + \frac{M \phi_{1}^{8} \left(- 410865 \beta^{4} + 502500 \beta^{3} - 216562 \beta^{2} + 31996 \beta + 375\right)}{967680 \mpl^{8}}\right] \nonumber \\
& & \nonumber \phantom{{} = {}}~~~~~+ H_{0,3} \left[ \frac{100 M^{4}}{3} + \frac{M^{2} \phi_{1}^{2} \left(1413 \beta - 592\right)}{84 \mpl^{2}} + \frac{\phi_{1}^{4} \left(147 \beta^{2} - 118 \beta + 25\right)}{224 \mpl^{4}}\right] \nonumber \\
& &\phantom{{} = {}}~~~~~+ \delta\phi_{-2} \left[ \frac{80 M^{8} \beta \phi_{1}}{189 \mpl^{2}} + \frac{M^{6} \phi_{1}^{3} \left(- 57591 \beta^{2} + 23380 \beta - 2700\right)}{945 \mpl^{4}}  \right. \nonumber \\
& &{}\hspace{2.4cm} \left. + \frac{M^{4} \phi_{1}^{5} \left(- 913563 \beta^{3} + 686704 \beta^{2} - 125140 \beta + 7560\right)}{30240 \mpl^{6}}  \right. 
\nonumber \\
& &{}\hspace{2.4cm} \left. + \frac{M^{2} \phi_{1}^{7} \left(- 384795 \beta^{4} + 407238 \beta^{3} - 118241 \beta^{2} + 11540 \beta - 540\right)}{120960 \mpl^{8}}  \right. \nonumber \\
& &{}\hspace{2.4cm} \left. + \frac{\beta \phi_{1}^{9} \left(- 11025 \beta^{4} + 19140 \beta^{3} - 3858 \beta^{2} - 1700 \beta + 375\right)}{1935360 \mpl^{10}}\right] \nonumber \\
& &\phantom{{} = {}}~~~~~~+ \delta\phi_{3} 
\left[ \frac{10 M^{3} \phi_{1} \left(10 - 57 \beta\right)}{21 \mpl^{2}} + \frac{M \phi_{1}^{3} \left(- 294 \beta^{2} + 95 \beta - 5\right)}{42 \mpl^{4}} \right]\,,
\nonumber \\
& &
H_{0, 8} = H_{0,-2} \left[\frac{8 M^{8} \phi_{1}^{2} \left(26431 \beta - 17850\right)}{4725 \mpl^{2}} + \frac{M^{6} \phi_{1}^{4} \left(210825 \beta^{2} - 272558 \beta + 59874\right)}{9450 \mpl^{4}} \right.
\nonumber \\
& &{}\hspace{2.1cm} \left. + \frac{M^{4} \phi_{1}^{6} \left(- 1062045 \beta^{3} + 306210 \beta^{2} - 26379 \beta + 25028\right)}{226800 \mpl^{6}} \right. 
\nonumber \\
& &{}\hspace{2.1cm} \left. + \frac{M^{2} \phi_{1}^{8} \left(- 286425 \beta^{4} + 324195 \beta^{3} - 134025 \beta^{2} + 21683 \beta - 636\right)}{151200 \mpl^{8}} \right. \nonumber \\
& &{}\hspace{2.1cm} \left. + \frac{\beta \phi_{1}^{10} \left(- 1260 \beta^{4} + 2100 \beta^{3} - 1365 \beta^{2} + 410 \beta - 48\right)}{18900 \mpl^{10}}\right]  \nonumber \\
& &\phantom{{} = {}}~~~~~
+ H_{0,3} \left[ \frac{208 M^{5}}{3} + \frac{M^{3} \phi_{1}^{2} \left(5685 \beta - 2332\right)}{105 \mpl^{2}} + \frac{M \phi_{1}^{4} \left(405 \beta^{2} - 325 \beta + 68\right)}{70 \mpl^{4}}\right] \nonumber \\
& &\phantom{{} = {}}~~~~~
+\delta\phi_{-2} \left[ 
\frac{32 M^{9} \beta \phi_{1}}{27 \mpl^{2}} + \frac{M^{7} \phi_{1}^{3} \left(- 214305 \beta^{2} + 94334 \beta - 12600\right)}{1575 \mpl^{4}} \right. 
\nonumber \\
& &{}\hspace{2.5cm} \left. + \frac{M^{5} \phi_{1}^{5} \left(- 760941 \beta^{3} + 572334 \beta^{2} - 115535 \beta + 8100\right)}{7560 \mpl^{6}} \right. \nonumber \\
& &{}\hspace{2.5cm} \left. + \frac{M^{3} \phi_{1}^{7} \left(- 3857625 \beta^{4} + 4157325 \beta^{3} - 1339725 \beta^{2} + 159823 \beta - 8100\right)}{226800 \mpl^{8}} \right. \nonumber \\
& &{}\hspace{2.5cm} \left. + \frac{M \beta \phi_{1}^{9} \left(- 132525 \beta^{4} + 175050 \beta^{3} - 69675 \beta^{2} + 8980 \beta - 72\right)}{302400 \mpl^{10}} \right]  \nonumber \\
& &\phantom{{} = {}}~~~~~ + \delta\phi_{3} \left[ \frac{2 M^{4} \phi_{1} \left(100 - 459 \beta\right)}{15 \mpl^{2}} + \frac{M^{2} \phi_{1}^{3} \left(- 6395 \beta^{2} + 2378 \beta - 200\right)}{210 \mpl^{4}} + \frac{\beta \phi_{1}^{5} \left(- 420 \beta^{2} + 235 \beta - 36\right)}{420 \mpl^{6}}\right]\,.
\ea

%%%%%%%%%%%%%%%%%%%%%%%% Bibliography %%%%%
\cftaddtitleline{toc}{section}{\textbf{References}}{\thepage}

\let\oldaddcontentsline\addcontentsline
\renewcommand{\addcontentsline}[3]{}
\bibliography{biblio.bib}
\let\addcontentsline\oldaddcontentsline
%%%%%%%%%%%%%%%%%%%%%%%%%%%%%%%%%%%%%%%%
\end{document}